\documentclass[final,3p,times,sort&compress]{elsarticle}

\usepackage{epsfig}
\usepackage{amssymb}
\usepackage{amsthm}
\usepackage{color}
\usepackage{graphicx}
\usepackage{dcolumn}
\usepackage[table]{xcolor}
\usepackage{lscape}
\usepackage{caption}
\usepackage{longtable}
\usepackage{soul}
\usepackage{amsmath,amssymb,blkarray}
\usepackage{color}

\journal{Physica A}

\begin{document}

\begin{frontmatter}



\title{Distribution of a bipartite entanglement in a mixed spin-(1/2,1) Heisenberg tetramer }
\author[label1]{Hana Vargov\'a\corref{cor1}}
\ead{hcencar@saske.sk}
\cortext[cor1]{Corresponding author:}
\author[label2]{Jozef Stre\v{c}ka}

\address[label1]{Institute  of  Experimental  Physics,  Slovak   Academy   of Sciences, Watsonova 47, 040 01 Ko\v {s}ice, Slovakia}
\address[label2]{Department of Theoretical Physics and Astrophysics, Faculty of Science, P.~J. \v{S}af\' arik University, Park Angelinum 9, 040 01 Ko\v{s}ice, Slovakia}
\begin{abstract}
The distribution of bipartite entanglement in a mixed spin-(1/2,1) Heisenberg tetramer composed from two spin-1/2 and two spin-1 entities is investigated in detail in presence of an external magnetic field.  Four different negativities measuring a strength of bipartite entanglement are analyzed at zero and non-zero temperatures. Derived rigorous analytic results and respective numerical results are discussed with the  particular emphasis laid on the significance of a strength of the pair spin-spin interactions and spin diversity in the entanglement description. Based on both aforementioned driving forces the regions of parametric space, where the bipartite entanglement can exist  solely for  one type of spin pair or  all four spin pairs, were identified.
\end{abstract}


\end{frontmatter}


\section{Introduction}
\label{introduction}

Quantum many-body systems represent  one of the most fascinating group of materials due to the amount of non-trivial phenomena (e.g., a superconductivity, quantum Hall effect, etc.) inherent to the non-local character of correlations among the system constituents, known as entanglement~\cite{Amico}. Although, the idea of existence of certain hidden force(s) has been  appeared repeatedly since 1930s, only after the publication of  Bell's set of inequalities~\cite{Bell}, the concept of nonlocality has taken a stable place in a quantum mechanics. In a quantum mechanics the many-body system is called {\it entangled} if its quantum state cannot be factorized into a product of  quantum states of its local constituents. Otherwise, the many-body  system is called {\it separable}~\cite{Benneta}.

The new era of entanglement has been connected to a progress in a quantum information science~\cite{Nielsen}. In this area involving the study of quantum teleportation~\cite{Furusawa}, quantum computation~\cite{Loss,Hayashi}, quantum cryptography~\cite{Deutsch,Bechmann}, the entanglement is considered as a major resource  related to quantum processing of information. In addition, it is believed that the entanglement represents a significant mechanism, which leads to the speed-up  in a quantum computation and communication~\cite{Bennet2000}.  

During last decades an enormous effort of researchers has been devoted to deeper understand the underlying physics as  well as to develop the  criterion for a characterization of entangled state, particularly when many-body system is investigated~\cite{Horodecki2009,Amico}. Many of these studies have been focused on  molecular magnets  serving as a  perspective candidate for a quantum information processing~\cite{Leuenberger, Gaita}.  The molecular magnets are  low-dimensional, what is an ideal feature for a miniaturization strategy  as a modern trend in smart devices. Moreover, the weak inter-molecular coupling  allows  to perceive the molecular magnets being composed from  effectively independent clusters of a few interacting spins. As a direct consequence of this simplification, the theoretical calculations of entanglement-related quantities, like concurrence~\cite{Wootters}, negativity~\cite{Vidal} or von Neumann entropy~\cite{Bennett}, may become possible.

Currently, an entanglement of the simplest bipartite spin systems is relatively well understood, nevertheless, the study of entanglement in  multipartite ones is infrequent due to much  higher computational complexity.  After all, the studies of entanglement in the cluster magnetic systems of trimers~\cite{Tribedi,Wang2001,Bose,Pal2009,Pal2011,Cima,Liu2015, Deniz, Ahami, Zhang2021,Najarbashi,Zad2017, Zad2016,Zheng,Sun}, tetramers~\cite{Sun,Tribedi,Irons,Bose,Pal2011,Benabdallah,Szalowski,Torrico,Rojas, Ananikian,Shawish,Ma,Zhad2022,Li,Zheng2017,Abgaryan2015,Hu,Karlova,Ghannadan2021}, pentamers~\cite{Sun,Xu2016}, hexamers~\cite{Deb,Zhad2018} or even larger structures~\cite{Wang2002,Sadiek,Sadiek2} can be found in literature. To the best of our knowledge, only a few of these studies explain the behavior of entanglement in the system of more than three spins with  different magnitudes distributed on a linear chain~\cite{Li}, diamond chain~\cite{Abgaryan2015}, small clusters~\cite{Sun}, tetrahedral lattice~\cite{Hu} and double sawtooth frustrated ladder~\cite{Zhad2018}. As was shown previously, the diversity of spin constituents  is one of the crucial intrinsic factors, which has a non-negligible impact on a degree of quantum and thermal entanglement~\cite{Zheng,Zhang2011,Abgaryan2015,Cenci2020,Vargova2022,Vargova2022a}. 

In the present paper we would like to bring insight into the aforementioned problems through the investigation of a  mixed spin-(1/2,1) Heisenberg tetramer with a geometry of a square plaquette. The square plaquette is constructed in analogy to the previous results~\cite{Guo,Yang,Wang2009,Zhu, Xu,Cenci2020,Vargova2022,Vargova2022a} as a couple of two interacting mixed spin-(1/2,1) Heisenberg dimers. The influence of extrinsic parameters, such as an external magnetic field and temperature, on the stability of entangled state in the model under consideration has been also taken into account. The description of entanglement is realized  through the behavior of the bipartite negativity~\cite{Vidal}. Experimental representatives of the studied mixed spin-(1/2, 1) Heisenberg tetramer can be for instance found in the family of heterobimetallic tetranuclear complexes such as Cu$_2$Ni$_2$~\cite{Ribas, Yonemura} and Fe$_2$Ni$_2$~\cite{Park,Zhuang2015}.

The paper is organized as follows. In Sec.~\ref{Model} we introduce the analyzed model and describe the procedure how to calculate all possible bipartite negativities in a mixed spin-(1/2,1) Heisenberg tetramer. In Sec.~\ref{Results} we present the most interesting results at a zero (Subsec.~\ref{quantum}) and non-zero (Subsec.~\ref{thermal}) temperatures. In Sec.~\ref{Conclusions} we draw our conclusions.  The technical details of calculations are summarized in Appendices~\ref{App A} to \ref{App E}.

\section{Model and Method}
\label{Model}
\begin{figure}[h]
\centering
{\includegraphics[width=.18\textwidth,trim=1.0cm 4.5cm 16cm 20.5cm, clip]{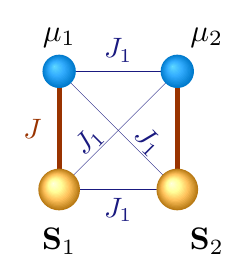}}
\caption{A schematic representation of a mixed spin-(1/2,1) Heisenberg tetramer with a geometry of a square plaquette. Large and small balls correspond to spins with eigenvalues $S_{1}\!=\!S_2\!=\!1$ and $\mu_{1}\!=\!\mu_2\!=\!1/2$, respectively. Four spins interact among themselves via the Heisenberg couplings $J$ (thick vertical lines) and $J_1$ (thin horizontal and diagonal lines).}
\label{fig1}
\end{figure}
The magnetic arrangement of  the mixed spin-(1/2,1) Heisenberg tetramer is schematically illustrated in Fig.~\ref{fig1}. The respective  Hamiltonian of the model under investigation can be mathematically expressed as follows:
\allowdisplaybreaks
\begin{align}
\hat{\cal H}&=J\left(\hat{\bf S}_{1}\cdot\hat{\boldsymbol\mu}_{1}\!+\!\hat{\bf S}_{2}\cdot\hat{\boldsymbol\mu}_{2}\right)\!+\! J_1\left(\hat{\bf S}_{1}\!+\!\hat{\boldsymbol\mu}_{1}\right)\cdot\left(\hat{\bf S}_{2}\!+\!\hat{\boldsymbol\mu}_{2}\right)
\!-\!h\left(\hat{S}^z_{1}\!+\!\hat{S}^z_{2}\!+\!\hat{\mu}^z_{1}\!+\!\hat{\mu}^z_{2}\right).
\label{eq1}
\end{align} 
 Here,  $\hat{\bf S}_{\gamma}$ and  $\hat{\boldsymbol\mu}_{\gamma}$ ($\gamma\!=\!1,2$) are  the  spin angular momentum of two different magnitudes $S\!=\!1$ and $\mu\!=\!1/2$, respectively. The interaction parameter $J$ denotes the Heisenberg interaction between two nearest-neighbor spins of  different sizes. The interaction parameter $J_1$ describes the isotropic Heisenberg interaction between the nearest-neighbor spins with an identical spin size as well as the next-nearest-neighbor spins of different magnitudes (see Fig.~\ref{fig1}). Finally, the parameter $h\!=\!g\mu_BB$ stands for the influence of magnetic field $B$, which affects each spin magnetic moment characterized by the same Land\'e $g$-factor ($\mu_B$ is the Bohr magneton).
 
 The eigenvalues and respective eigenvectors can be easily calculated through a direct diagonalization of the matrix form of Hamiltonian \eqref{eq1} defined in a standard spin basis. The basis of the mixed spin-(1/2,1) Heisenberg tetramer involving in total 36 different basis states is constructed as a tensor product of  basis states of two spin-(1/2,1) Heisenberg dimers, namely $\vert \varphi_k\rangle\!=\!\vert \mu^z_{1},S^z_{1}\rangle \otimes\vert \mu^z_{2},S^z_{2}\rangle$. 
 $Z$-components of both different spin angular momenta $\mu^z_{\gamma}$, $S^z_{\gamma}$ ($\gamma\!=\!1,2$) achieve one of the following allowed values $\mu^z_{\gamma}\!=\!\pm\tfrac{1}{2}$ and $S^z_{\gamma}\!=\!\pm1,0$. The complete list of basis states is given in Tab.~\ref{tab_A1} of ~\ref{App A}. The Hamiltonian \eqref{eq1}  in a matrix form has a block diagonal structure involving seven independent blocks classified according to the value of $z$-projection of the new composite operator $\hat{\boldsymbol\sigma}_T\!=\!\sum_{\gamma=1,2}\hat{\boldsymbol\sigma}_{\gamma}\!=\!\sum_{\gamma=1,2}(\mathbf{\hat{S}}_{\gamma}\!+\!\hat{\boldsymbol\mu}_{\gamma})$. We note that  $\sigma_T^z\!=\!-\sigma_T,\dots,\sigma_T$ and the quantum spin number $\sigma_T$ runs from $|\sigma_{1}\!-\!\sigma_{2}|$ to $\sigma_{1}\!+\!\sigma_{2}$. Because we assume the mixed spin-(1/2,1) Heisenberg dimer with the total spin $\sigma_{\gamma}\!=\!\{1/2,3/2\}$, thus, $\sigma_T^z$ acquires one of seven possible values $\pm 3$, $\pm2 $, $\pm 1$ and 0. Due to a complexity of the computational problem,  the eigenvalues and respective eigenvectors can be calculated with the help of Symbolic Toolbox of MATLAB software~\cite{Matlab}. The complete list of eigenvalues ($\varepsilon_k$) and respective eigenvectors ($\vert \psi_k\rangle$) characterized according to four quantum numbers $\vert \sigma^z_T,\sigma_T,\sigma_{1},\sigma_{2}\rangle$ is given in Tab.~\ref{tab_A2} of~\ref{App A}. 

In order to examine a bipartite entanglement in a quantum mixed spin-(1/2,1) Heisenberg tetramer, we use the concept of negativity elaborated by Peres~\cite{Peres} and Horodecki~\cite{Horodecki}, which is based on an idea that all eigenvalues of a partially transposed  density matrix of a separable state of the bipartite system A-B are always positive. For a quantification, we will therefore exploit the bipartite negativity ${\cal N}_{A|B}$~\cite{Vidal} defined as a sum of absolute values of all negative  eigenvalues $(\lambda_{A|B})_i$ of a partially transposed density matrix $\hat{\rho}_{A|B}^{T_{A}}$. The superscript $T_A$ means that a partial transposition is done with respect to the subsystem A. Thus, the bipartite negativity follows from the formula
\begin{align}
{\cal N}_{A|B}\!=\!{\cal N}(\rho_{{A|B}})\!=\!\sum_{(\lambda_{A|B})_i<0}\left\vert(\lambda_{A|B})_i\right\vert.
\label{eq2}
\end{align}
According to the definition~\eqref{eq2}, the negativity of a separable state of a bipartite quantum system becomes zero (${\cal N}_{A|B}\!=\!0$), while the negativity of an entangled (non-separable) state always reaches non-zero values (${\cal N}_{A|B}\!\neq\!0$).

In order to achieve our main goal, it is necessary to calculate four different bipartite negativities
\begin{align}
{\cal N}_{S_{1}|S_{2}}\!&=\!{\cal N}(\rho_{{S_{1}|S_{2}}})\!=\!\sum_{(\lambda_{S_{1}|S_{2}})_i<0}\left\vert(\lambda_{S_{1}|S_{2}})_i\right\vert,
\hspace*{1.7cm}
{\cal N}_{\mu_{1}|\mu_{2}}\!=\!{\cal N}(\rho_{{\mu_{1}|\mu_{2}}})\!=\!\sum_{(\lambda_{\mu_{1}|\mu_{2}})_i<0}\left\vert(\lambda_{\mu_{1}|\mu_{2}})_i\right\vert,
\label{eq4}\\
{\cal N}_{\mu_{1}|S_{1}}\!=\!{\cal N}_{\mu_{2}|S_{2}}\!&=\!{\cal N}(\rho_{{\mu_{1}|S_{1}}})\!=\!\sum_{(\lambda_{\mu_{1}|S_{1}})_i<0}\left\vert(\lambda_{\mu_{1}|S_{1}})_i\right\vert,
\hspace*{.2cm}\;\;
{\cal N}_{\mu_{1}|S_{2}}\!=\!{\cal N}_{\mu_{2}|S_{1}}\!=\!{\cal N}(\rho_{{\mu_{1}|S_{2}}})\!=\!\sum_{(\lambda_{\mu_{1}|S_{2}})_i<0}\left\vert(\lambda_{\mu_{1}|S_{2}})_i\right\vert,
\label{eq6}
\end{align}
as a sum of negative eigenvalues of respective partially transposed density matrices $\hat{\rho}^{T_{S_{1}}}_{{S_{1}|S_{2}}}$, $\hat{\rho}^{T_{\mu_{1}}}_{\mu_{1}|\mu_{2}}$, $\hat{\rho}^{T_{\mu_{1}}}_{\mu_{1}|S_{1}}$ and $\hat{\rho}^{T_{\mu_{1}}}_{\mu_{1}|S_{2}}$. Each of the bipartite negativities is derived from the overall density operator
\begin{align}
\hat{\rho}\!=\!\frac{1}{\cal Z}{\rm e}^{-\beta \hat{\cal H}}\!=\!\frac{1}{\cal Z}\sum_{k=1}^{36}{\rm e}^{-\beta \varepsilon_k}\vert \psi_k \rangle\langle \psi_k\vert,
\label{eq7}
\end{align}
when tracing out degrees of freedom of two remaining spins.  Thus, one obtains four reduced density operators
\begin{align}
\hat{\rho}_{S_{1}|S_{2}}\!&=\!\frac{1}{\cal Z}\sum_{k=1}^{36} {\rm Tr}_{\mu_{1}}{\rm Tr}_{\mu_{2}}{\rm e}^{-\beta \varepsilon_k}\vert \psi_k\rangle \langle \psi_k\vert,
\hspace*{1.5cm}
\hat{\rho}_{\mu_{1}|\mu_{2}}\!=\!\frac{1}{\cal Z}\sum_{k=1}^{36} {\rm Tr}_{S_{1}}{\rm Tr}_{S_{2}}{\rm e}^{-\beta \varepsilon_k}\vert \psi_k\rangle \langle \psi_k\vert,
\label{eq9}\\
\hat{\rho}_{\mu_{1}|S_{1}}\!&=\!\frac{1}{\cal Z}\sum_{k=1}^{36} {\rm Tr}_{\mu_{2}}{\rm Tr}_{S_{2}}{\rm e}^{-\beta \varepsilon_k}\vert \psi_k\rangle \langle \psi_k\vert,
\hspace*{1.5cm}
\hat{\rho}_{\mu_{1}|S_{2}}\!=\!\frac{1}{\cal Z}\sum_{k=1}^{36} {\rm Tr}_{\mu_{2}}{\rm Tr}_{S_{1}}{\rm e}^{-\beta \varepsilon_k}\vert \psi_k\rangle \langle \psi_k\vert,
\label{eq11}
\end{align}
where $\beta\!=\!1/(k_BT)$, $k_B$ is the Boltzmann's constant, $T$ is absolute temperature and  ${\cal Z}$ is the partition function ${\cal Z}=\sum_{k=1}^{36}{\rm e}^{-\beta\varepsilon_k}$ of the whole mixed spin-(1/2,1) Heisenberg tetramer (for an explicit form of ${\cal Z}$, see \eqref{A_eq1}.). Having the reduced density operator in a matrix representation, we are able to directly perform a partial transpose over  one spin  and finally obtain the respective bipartite negativity as a sum of the absolute values of the negative eigenvalues. All calculation details can be found in Appendices \ref{App B} (for ${\cal N}_{S_{1}|S_{2}}$), \ref{App C} (for ${\cal N}_{\mu_{1}|\mu_{2}}$), \ref{App D} (for ${\cal N}_{\mu_{1}|S_{1}}$) and \ref{App E} (for ${\cal N}_{\mu_{1}|S_{2}}$).

It is worthwhile to remark that the structure of obtained reduced density matrices is identical with the known structure of reduced density matrices of a certain quantum dimer, e.q., (1/2,1/2)-dimer, (1/2,1)-dimer and (1,1)-dimer, which can be found elsewhere~\cite{Hu2008,Cenci2020,Ghannadan2021}. However, individual matrix elements are different and generally have much more complex form. 

\section{Results and discussion}
\label{Results}

Before proceeding to a detailed discussion of the most interesting results we note that our attention will be henceforth restricted to the  physically most intriguing case of the mixed spin-(1/2,1) Heisenberg tetramer with the antiferromagnetic coupling constants $J_1\!\geq\!0$ and $J\!>\!0$. To retain the most general character of presented results all dependencies are normalized with respect to the coupling constant $J$.
\subsection{Bipartite quantum negativity}
\label{quantum}
 In this section we  concentrate our attention to a comprehensive analysis of the bipartite quantum negativity of the mixed spin-(1/2,1) Heisenberg tetramer~\eqref{eq1} at a zero temperature.
The energy spectrum of the analyzed model in absence of the magnetic field $h/J$ involves  two singlet states ($\vert 0,0,\tfrac{1}{2},\tfrac{1}{2}\rangle$, $\vert 0,0,\tfrac{3}{2},\tfrac{3}{2}\rangle$), three triplet states 
($\vert \sigma_T^z,1,\tfrac{3}{2},\tfrac{3}{2}\rangle$, $\vert \sigma_T^z,1,\tfrac{3}{2},\tfrac{1}{2}\rangle$, $\vert \sigma_T^z,1,\tfrac{1}{2},\tfrac{3}{2}\rangle$ for $\sigma_T^z\!=\!-1,0,1$), three quintet states ($\vert \sigma_T^z,2,\tfrac{3}{2},\tfrac{3}{2}\rangle$, $\vert \sigma_T^z,2,\tfrac{3}{2},\tfrac{1}{2}\rangle$, $\vert \sigma_T^z,2,\tfrac{1}{2},\tfrac{3}{2}\rangle$ for $\sigma_T^z\!=\!-2,\dots,2$), and one septet state ($\vert \sigma_T^z,3,\tfrac{3}{2},\tfrac{3}{2}\rangle$ for $\sigma_T^z\!=\!-3,\dots,3$), see Tab.~\ref{tab_A2}. In the limiting case of $J_1/J\!=\!0$ four quantum states $|\pm1,1,1/2,1/2\rangle$, $|0,1,1/2,1/2\rangle$ and $|0,0,1/2,1/2\rangle$ are  degenerate and the ground state of the mixed spin-(1/2,1) Heisenberg tetramer \eqref{eq1} can be interpreted as a product state of two  non-interacting mixed spin-(1/2,1) Heisenberg dimers with the intra-dimer interaction $J$. For this reason, one can intuitively expect the  non-zero bipartite quantum negativity exclusively inside of the mixed spin-(1/2,1) Heisenberg dimers of $\mu_{1}\!-\!S_{1}$ and $\mu_{2}\!-\!S_{2}$, respectively. Indeed, the negativity  ${\cal N}_{S_{1}|S_{2}}$, ${\cal N}_{\mu_{1}|\mu_{2}}$ and ${\cal N}_{\mu_{1}|S_{2}}$ equal zero and the non-zero negativity  ${\cal N}_{\mu_{1}|S_{1}}\!=\!1/3$ perfectly fits the previously detected results for the mixed spin-(1/2,1) Heisenberg dimer~\cite{Cenci2020}. An increasing magnetic field completely lifts the fourfold degeneracy at $J_1/J\!=\!0$ and favors the ground state $\vert 1,1,\tfrac{1}{2},\tfrac{1}{2}\rangle$ with a maximal quantum entanglement  of $\sqrt{2}/3$ until the fully polarized separable state $\vert 3,3,\tfrac{3}{2},\tfrac{3}{2}\rangle$ is achieved at $h/J\!=\!1.5$. Also the  aforementioned observations perfectly coincide with the previous results~\cite{Cenci2020}.

\begin{figure}[t!]
\centering
{\includegraphics[width=.45\textwidth,trim=1.5cm 8.45cm 1.cm 8cm, clip]{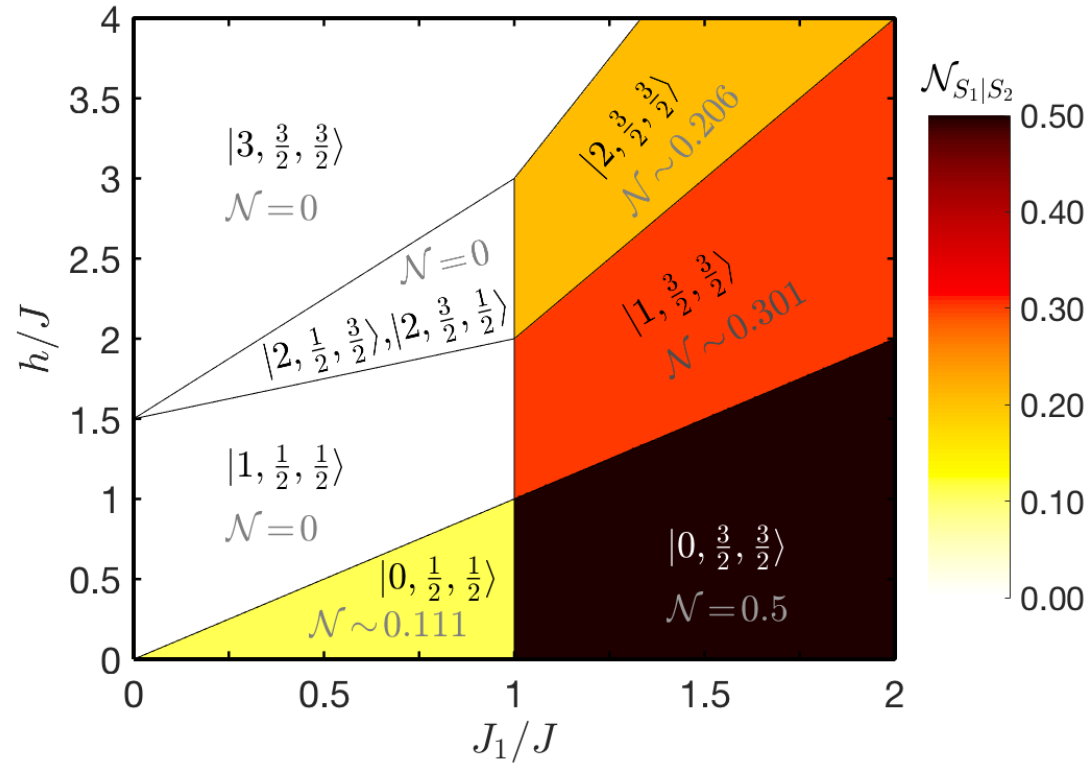}}
{\includegraphics[width=.45\textwidth,trim=1.5cm 8.45cm 1.cm 8cm, clip]{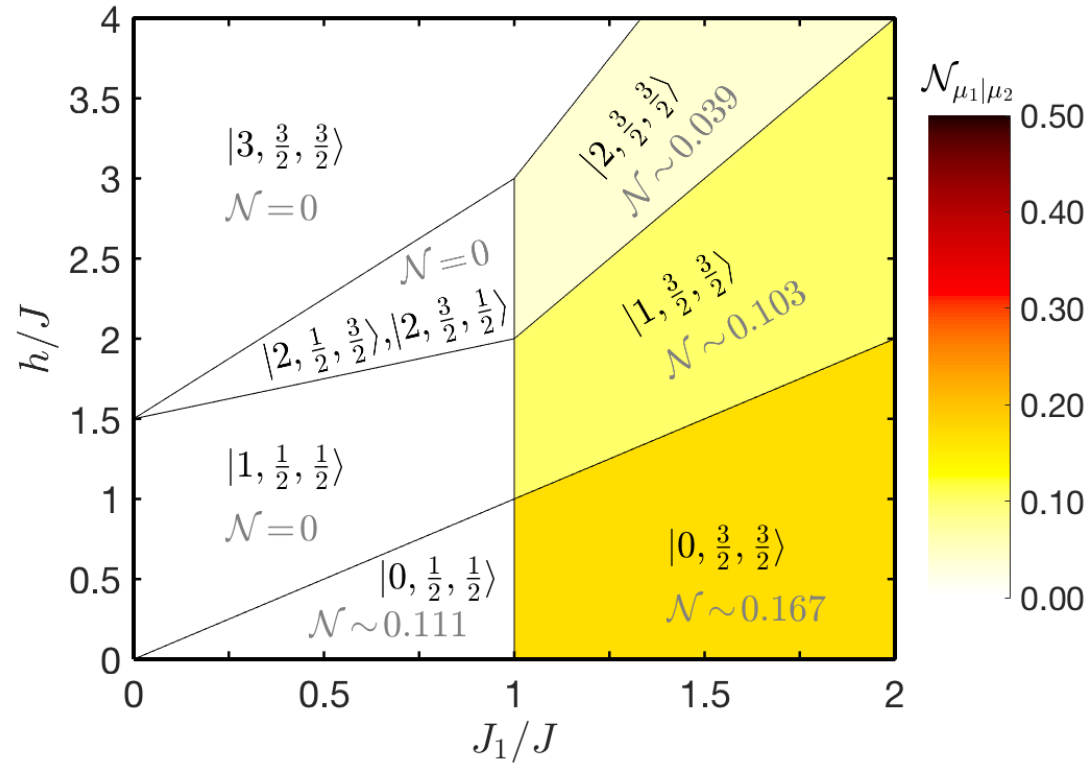}}\\
{\includegraphics[width=.45\textwidth,trim=1.5cm 8.45cm 1.cm 8cm, clip]{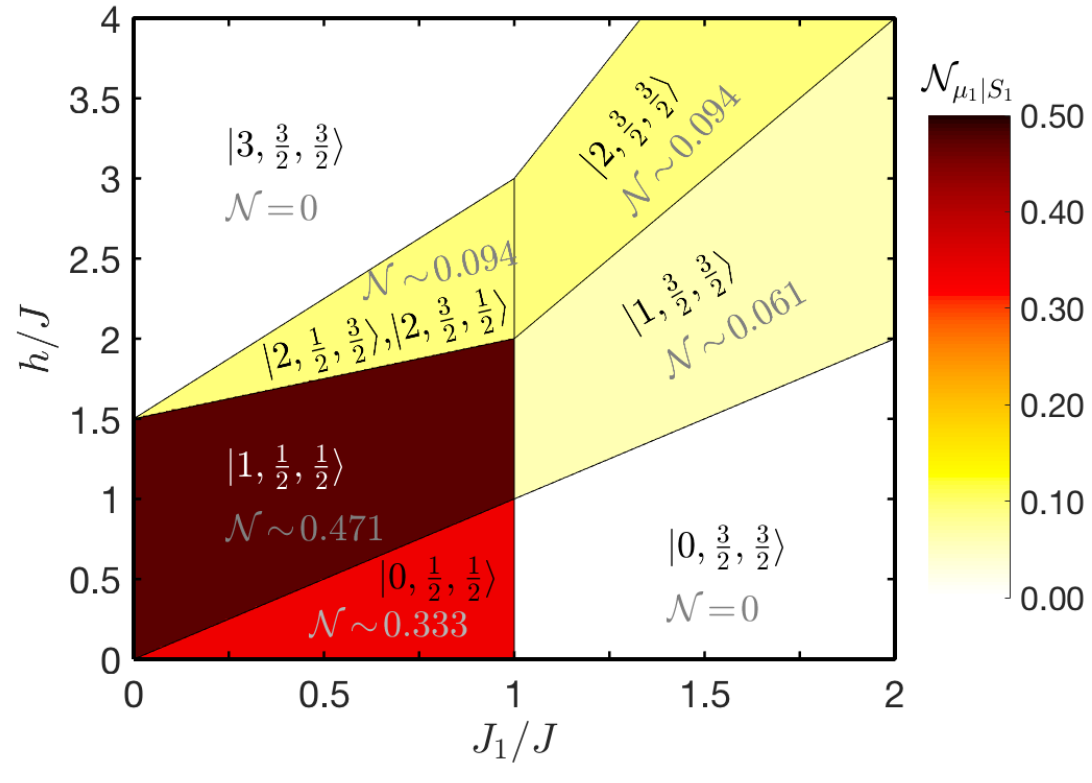}}
{\includegraphics[width=.45\textwidth,trim=1.5cm 8.45cm 1.cm 8cm, clip]{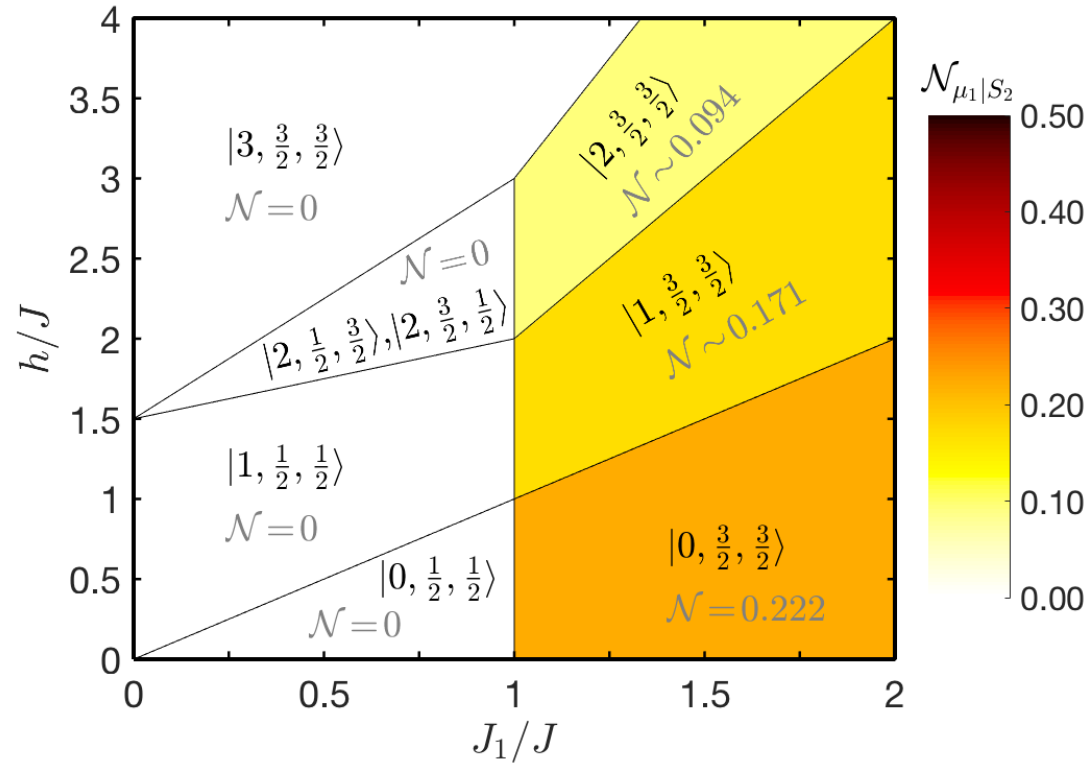}}
\caption{Density plots of four different bipartite quantum  negativities  (${\cal N}_{S_{1}|S_{2}}$, ${\cal N}_{\mu_{1}|\mu_{2}}$, ${\cal N}_{\mu_{1}|S_{1}}$ and ${\cal N}_{\mu_{1}|S_{2}}$ ) of the mixed spin-(1/2,1) Heisenberg tetramer in the $J_1/J-h/J$ plane. Solid black lines visualize borders between different  ground states characterized by a set of three quantum numbers $|\sigma_T^z,\sigma_{1},\sigma_{2}\rangle$.}
\label{fig2}
\end{figure}

For $J_1/J\!>\!0$ and $h/J\!=\!0$ one of the non-degenerate singlet states $\vert 0,0,\tfrac{1}{2},\tfrac{1}{2}\rangle$ and $\vert 0,0,\tfrac{3}{2},\tfrac{3}{2}\rangle$ has the lowest energy depending on whether $J_1/J\!<\!1$ or $J_1/J\!>\!1$, while they both remain energetically most favorable up to the magnetic field $h/J\!=\!J_1/J$. The nonzero magnetic field either completely or partially removes the degeneracy of remaining states and energetically favors only the states with the maximum value of $z$-component of the total spin $\sigma_T^z\!=\!\sigma_T$. In this regard, the reduced set of only three quantum spin numbers $\sigma_T$, $\sigma_{1}$ and $\sigma_{2}$ is needed to fully characterize all possible ground states $\vert \sigma_T^z,\sigma_{1},\sigma_{2}\rangle$ in a regime of $J_1/J\!>\!0$ and $h/J\!>\!0$. As a consequence, the respective ground-state phase diagram involves in total seven different ground states as illustrated in Fig.~\ref{fig2}. Besides the overall structure of the ground-state phase diagram, Fig.~\ref{fig2} simultaneously illustrates a degree of the quantum entanglement between four inequivalent spin pairs.

In a weak interaction limit ($J_1/J\!<\!1$), one could expect that the dominant driving force originates from the exchange interaction $J$ and thus, the higher entanglement is found within the mixed spin-($\mu_{1},S_{1}$) Heisenberg dimer. As illustrated in Fig.~\ref{fig2} and Tab.~\ref{tab1}, the highest value of negativity for $J_1/J\!<\!1$ is detected between  the spins $\mu_{1}\!-\!S_{1}$  with the maximal amplitude of ${\cal N}_{\mu_{1}|S_{1}}\!=\!\sqrt{2}/3$ corresponding to the  ground state $|1,\frac{1}{2},\frac{1}{2}\rangle$. Surprisingly, the negativity of the ground state $|1,\frac{1}{2},\frac{1}{2}\rangle$ is higher than the negativity of the ground state $|0,\frac{1}{2},\frac{1}{2}\rangle$, though it is generally expected that the increasing magnetic field reduces the bipartite quantum entanglement. The explanation of this non-trivial phenomenon could be found in a transcription of respective eigenvectors in a form of tensor product of eigenvectors of two mixed spin-(1/2,1) Heisenberg dimers (the complete list of  eigenvalues and respective eigenvectors of a mixed spin-(1/2,1) Heisenberg dimer is given in Eq.~\eqref{A_eq2} of Appendix~\ref{App A}). After a simple algebra one can obtain that the eigenvector $|1,\frac{1}{2},\frac{1}{2}\rangle$ is a pure state expressed as a tensor product of two maximally entangled ground states $\vert \frac{1}{2},\frac{1}{2}\rangle_{\mu_{\gamma},S_{\gamma}}$ ($\gamma\!=\!1,2$)
\begin{align}
\left\vert 1,\frac{1}{2},\frac{1}{2}\right\rangle\!=\!\left\vert \frac{1}{2},\frac{1}{2}\right\rangle_{\mu_1,S_1}\otimes \left\vert \frac{1}{2},\frac{1}{2}\right\rangle_{\mu_2,S_2},
\label{eq12}
\end{align}
and thus, the relevant negativity ${\cal N}_{\mu_{1}|S_{1}}$  acquires the highest possible value   ${\cal N}_{\mu_{1}|S_{1}}\!=\!\sqrt{2}/3$~\cite{Cenci2020}.   It is worthwhile to remark that 
all three remaining bipartite negativities (${\cal N}_{S_{1}|S_{2}}$, ${\cal N}_{\mu_{1}|\mu_{2}}$ and ${\cal N}_{\mu_{1}|S_{2}}$) are consequently equal to zero, because all non-local quantum correlations are pertinent to $\mu_{1}\!-\!S_{1}$ dimer.
On the other hand, the other ground-state eigenvector $|0,\frac{1}{2},\frac{1}{2}\rangle$ is  given by the quantum superposition of the ground state $\vert \frac{1}{2},\frac{1}{2}\rangle_{\mu_{\gamma},S_{\gamma}}$ and the first excited state $\vert \frac{1}{2},\!-\frac{1}{2}\rangle_{\mu_{\gamma},S_{\gamma}}$ of a  mixed spin-(1/2,1) Heisenberg dimer ($\gamma\!=\!1,2$)
\begin{align}
\left\vert 0,\frac{1}{2},\frac{1}{2}\right\rangle\!=\!\frac{1}{\sqrt{2}}\left[ \left\vert \frac{1}{2},\frac{1}{2}\right\rangle_{\mu_1,S_1}\otimes \left\vert \frac{1}{2},\!-\frac{1}{2}\right\rangle_{\mu_2,S_2}\!-\!\left\vert \frac{1}{2},\!-\frac{1}{2}\right\rangle_{\mu_1,S_1}\otimes \left\vert \frac{1}{2},\frac{1}{2}\right\rangle_{\mu_2,S_2}\right].
\label{eq13}
\end{align}
Subsequently, the corresponding negativity does not reach the maximal value, only the reduced one, ${\cal N}_{\mu_{1}|S_{1}}\!=\!1/3$. Utilizing the same procedure, we rewrite the eigenvector $\frac{1}{2}\left[ |2,\frac{1}{2},\frac{3}{2}\rangle\!+\!|2,\frac{3}{2},\frac{1}{2}\rangle\right]$ as follows
\begin{align}
\frac{1}{2}\left[ |2,\frac{1}{2},\frac{3}{2}\rangle\!+\!|2,\frac{3}{2},\frac{1}{2}\rangle\right]\!=\!-\frac{1}{2}\left[ \left\vert \frac{3}{2},\frac{3}{2}\right\rangle_{\mu_1,S_1}\otimes \left\vert \frac{1}{2},\frac{1}{2}\right\rangle_{\mu_2,S_2}\!+\!\left\vert \frac{1}{2},\frac{1}{2}\right\rangle_{\mu_1,S_1}\otimes \left\vert \frac{3}{2},\frac{3}{2}\right\rangle_{\mu_2,S_2}\right].
\label{eq14}
\end{align}
Obviously, the increasing magnetic field changes one of two entangled ferrimagnetic spin-(1/2,1) Heisenberg dimers $\vert \frac{1}{2},\frac{1}{2}\rangle_{\mu_{\gamma},S_{\gamma}}$ emergent in the ground-state eigenvector \eqref{eq12} into the fully polarized separable state $\vert \frac{3}{2},\frac{3}{2}\rangle_{\mu_{\gamma},S_{\gamma}}$ ($\gamma\!=\!1,2$).  As a result, the respective negativity saturates at non-zero, but significantly smaller value in comparison to aforementioned two cases. 
For a completeness, the last ground-state eigenvector $|3,\frac{3}{2},\frac{3}{2}\rangle$ is a pure state composed from  two fully polarized separable spin-(1/2,1) Heisenberg dimers
\begin{align}
\left\vert 3,\frac{3}{2},\frac{3}{2}\right\rangle\!=\!\left\vert \frac{3}{2},\frac{3}{2}\right\rangle_{\mu_1,S_1}\otimes \left\vert \frac{3}{2},\frac{3}{2}\right\rangle_{\mu_2,S_2},
\label{eq15}
\end{align}
which determines  zero magnitude of all measures of the bipartite entanglement ${\cal N}_{\mu_{1}|S_{1}}$. The explicit values of ${\cal N}_{\mu_{1}|S_{1}}$ within each ground state are given in the fifth column of Tab.~\ref{tab1}.
\begin{table}[t!]
\resizebox{1\textwidth}{!}{
\begin{tabular}{l || l || l | l | l | l }
  $J_1/J$ & $|\sigma^z_T,\sigma_{1},\sigma_{2}\rangle$ & ${\cal N}_{S_{1}|S_{2}}$ & ${\cal N}_{\mu_{1}|\mu_{2}}$ & ${\cal N}_{\mu_{1}|S_{1}}$ & ${\cal N}_{\mu_{1}|S_{2}}$\\\hline\hline
   $<1$ & $\vert 0,\frac{1}{2},\frac{1}{2}\rangle$ & 1/9$\approx$0.111 & 0 & 1/3$\approx$0.333  & 0\\
         & $\vert 1,\frac{1}{2},\frac{1}{2}\rangle$ & 0 & 0 & $\sqrt{2}/3\approx$0.471  & 0\\
        & $\vert 2,\frac{1}{2},\frac{3}{2}\rangle$, $\vert 2,\frac{3}{2},\frac{1}{2}\rangle$ & 0 & 0 & $\frac{1}{12}(\sqrt{17}\!-\!3)\approx$0.094 & 0\\
    & $\vert 3,\frac{3}{2},\frac{3}{2}\rangle$ & 0 & 0 & 0 & 0\\
\hline
   $=1$ & $\vert 0,\frac{1}{2},\frac{1}{2}\rangle$,  $\vert 0,\frac{3}{2},\frac{3}{2}\rangle$& 1/4$\approx$0.250  & 0 & 0 & 0\\
         & $\vert 1,\frac{1}{2},\frac{1}{2}\rangle$, $\vert 1,\frac{3}{2},\frac{3}{2}\rangle$ & $\frac{1}{60}(6\sqrt{13}\!-\!23\!-\!\sqrt{139}\cos\left( \frac{\phi}{3}\!+\!\frac{2\pi}{3}\right))\approx$0.572& 0 & 1/80$\approx$0.013 & 1/80$\approx$0.013\\
         & $\vert 1,\frac{1}{2},\frac{3}{2}\rangle$, $\vert 1,\frac{3}{2},\frac{1}{2}\rangle$ &$\phi\!=\!\arctan\left(\frac{\sqrt{p^3-q^2}}{q}\right), p\!=\!\frac{139}{(120)^2}, q\!=\!\frac{1531}{(120)^3}$  & &  & \\
        & $\vert 2,\frac{1}{2},\frac{3}{2}\rangle$, $\vert 2,\frac{3}{2},\frac{1}{2}\rangle$, $\vert 2,\frac{3}{2},\frac{3}{2}\rangle$ & $\frac{1}{18}(\sqrt{29}\!-\!5)\approx$0.021 & $\frac{1}{18}(\sqrt{17}\!-\!4)\approx$0.007 & $\frac{1}{36}(\sqrt{89}\!-\!9)\approx$0.012 & $\frac{1}{36}(\sqrt{89}\!-\!9)\approx$0.012\\
    & $\vert 3,\frac{3}{2},\frac{3}{2}\rangle$ & 0 & 0 & 0 & 0\\
	      \hline	   
	  $>1$ & $\vert 0,\frac{3}{2},\frac{3}{2}\rangle$ & 1/2$\approx$0.500 & 1/6$\approx$0.167 & 0 & 2/9$\approx$0.222\\
             & $\vert 1,\frac{3}{2},\frac{3}{2}\rangle$ & $\frac{1}{135}(3\sqrt{313}\!-\!49\!-\!\sqrt{2179}\cos(\frac{\phi}{3}\!+\!\frac{2\pi}{3}))\approx$0.301  &$ \frac{1}{90}(3\sqrt{89}\!-\!19)\approx$0.103 & $ \frac{1}{30}(\sqrt{34}\!-\!4)\approx$0.061 & $ \frac{1}{60}(3\sqrt{33}\!-\!7\approx$0.171\\ 
             & &$\phi\!=\!\arctan\left(\frac{\sqrt{p^3-q^2}}{q}\right), p\!=\!\frac{2179}{(270)^2}, q\!=\!\frac{43874}{(270)^3}$ &  & &\\
               & $\vert 2,\frac{3}{2},\frac{3}{2}\rangle$ & $-\frac{1}{9}(1\!+\!4\cos(\frac{\phi}{3}\!+\!\frac{2\pi}{3}))\approx$0.206  &$ \frac{1}{6}(\sqrt{5}\!-\!2)\approx$0.039 & $ \frac{1}{12}(\sqrt{17}\!-\!3)\approx$0.094 & $ \frac{1}{12}(\sqrt{17}\!-\!3)\approx$0.094\\ 
             & &$\phi\!=\!\arctan\left(\frac{\sqrt{p^3-q^2}}{q}\right), p\!=\!\frac{4}{(9)^2}, q\!=\!\frac{11}{2(9)^3}$ &  & &\\
  & $\vert 3,\frac{3}{2},\frac{3}{2}\rangle$ & 0 & 0 & 0 & 0\\
             \hline
\end{tabular}
}
\caption{Distribution of four bipartite quantum  negativities (${\cal N}_{S_{1}|S_{2}}$, ${\cal N}_{\mu_{1}|\mu_{2}}$, ${\cal N}_{\mu_{1}|S_{1}}$ and ${\cal N}_{\mu_{1}|S_{2}}$) calculated for all available ground states  of a mixed spin-(1/2,1) Heisenberg tetramer~\eqref{eq1}.}
\label{tab1}
\end{table}

In a weak interaction limit ($J_1/J\!<\!1$) the non-zero bipartite quantum negativity can be also ascribed to the spin pair $S_{1}\!-\!S_{2}$ within the singlet quantum ground state $\vert0,\frac{1}{2},\frac{1}{2}\rangle$ given by Eq.~\eqref{eq13}. If we rewrite the respective eigenvector into a form of tensor product of eigenvectors of a spin-1/2 Heisenberg dimer and a spin-1 Heisenberg dimer (the complete lists  of eigenvectors are given in Eqs.~\eqref{A_eq3} and \eqref{A_eq4} of Appendix~\ref{App A}) we obtain the following expression 
\begin{align}
\left\vert 0,\frac{1}{2},\frac{1}{2}\right\rangle\!=\!&\frac{\sqrt{2}}{3}\left[
\left\vert1,1\right\rangle_{\mu_1,\mu_2}\otimes \left\vert1,\!-1\right\rangle_{S_1,S_2}\!-\!\left\vert1,\!-1\right\rangle_{\mu_1,\mu_2}\otimes \left\vert1,1\right\rangle_{S_1,S_2}
\right]
\nonumber\\
\!&+\!\frac{1}{3}\left[
\sqrt{2}\left\vert1,0\right\rangle_{\mu_1,\mu_2}\otimes \left\vert1,0\right\rangle_{S_1,S_2}\!-\!\sqrt{3}\left\vert0,0\right\rangle_{\mu_1,\mu_2}\otimes \left\vert0,0\right\rangle_{S_1,S_2}
\right].
\label{eq16}
\end{align}
 The spin-1 Heisenberg dimer is in a quantum superposition of four entangled states ($\vert1,\pm1\rangle_{S_1,S_2}$, $\vert1,0\rangle_{S_1,S_2}$, $\vert0,0\rangle_{S_1,S_2}$), and therefore  ${\cal N}_{S_{1}|S_{2}}(\vert 0,\frac{1}{2},\frac{1}{2}\rangle)\!\neq\!0$. In contradiction, the spin-1/2 Heisenberg dimer  is in a quantum superposition of two entangled states ($\vert1,0\rangle_{\mu_1,\mu_2}$, $|0,0\rangle_{\mu_1,\mu_2}$) and two more probable separable states ($\vert1,\pm1\rangle_{\mu_1,\mu_2}$), which leads to a zero value of the bipartite negativity ${\cal N}_{\mu_{1}|\mu_{2}}(\vert 0,\frac{1}{2},\frac{1}{2}\rangle)\!=\!0$.
\\
\\

 The highest negativity in the strong interaction limit ($J_1/J\!>\!1$) should be expected within the spin dimers  connected through the  interaction $J_1$ with the highest sum of spin quantum numbers.  In fact, the maximal bipartite negativity in the region of $J_1/J\!>\!1$ is detected for the $S_{1}\!-\!S_{2}$ dimer  at low enough magnetic fields. The ground-state eigenvector  $\vert 0,\frac{3}{2},\frac{3}{2}\rangle$ involves  only  microstates with antiferromagnetic ($\vert1,0\rangle_{S_1,S_2}$, $\vert0,0\rangle_{S_1,S_2}$) or ferrimagnetic ($\vert1,\pm1\rangle_{S_1,S_2}$) arrangements of $S_{1}$ and $S_{2}$, which are characterized by the non-zero negativity~\cite{Ghannadan2021}
\begin{align}
\left\vert 0,\frac{3}{2},\frac{3}{2}\right\rangle\!=\!&-\frac{1}{3}\left[
\left\vert1,1\right\rangle_{\mu_1,\mu_2}\otimes \left\vert1,\!-1\right\rangle_{S_1,S_2}\!-\!\left\vert1,\!-1\right\rangle_{\mu_1,\mu_2}\otimes \left\vert1,1\right\rangle_{S_1,S_2}
\right]
\nonumber\\
\!&-\!\frac{1}{3}
\left\vert1,0\right\rangle_{\mu_1,\mu_2}\otimes \left\vert1,0\right\rangle_{S_1,S_2}\!-\!\frac{\sqrt{2}}{\sqrt{3}}\left\vert0,0\right\rangle_{\mu_1,\mu_2}\otimes \left\vert0,0\right\rangle_{S_1,S_2}
.
\label{eq17}
\end{align}
At the same time, the bipartite negativity between unlike spins is zero ${\cal N}_{\mu_{1}|S_{1}}\!=\!0$ because the respective eigenvector involves the combination of both  separable  states ($\vert \frac{3}{2},\pm\frac{3}{2}\rangle_{\mu_{\gamma},S_{\gamma}}$) with a relative high probability amplitude
\begin{align}
\left\vert 0,\frac{3}{2},\frac{3}{2}\right\rangle\!=\!&-\frac{1}{2}\left[ \left\vert \frac{3}{2},\frac{3}{2}\right\rangle_{\mu_1,S_1}\otimes  \left\vert \frac{3}{2},\!-\frac{3}{2}\right\rangle_{\mu_2,S_2}\!-\! \left\vert \frac{3}{2},\!-\frac{3}{2}\right\rangle_{\mu_1,S_1}\otimes  \left\vert \frac{3}{2},\frac{3}{2}\right\rangle_{\mu_2,S_2}\right]
\nonumber\\
\!&+\!\frac{1}{2}\left[ \left\vert \frac{3}{2},\frac{1}{2}\right\rangle_{\mu_1,S_1}\otimes \left\vert \frac{3}{2},\!-\frac{1}{2}\right\rangle_{\mu_2,S_2}\!-\!\left\vert \frac{3}{2},\!-\frac{1}{2}\right\rangle_{\mu_1,S_1}\otimes \left\vert \frac{3}{2},\frac{1}{2}\right\rangle_{\mu_2,S_2}\right].
\label{eq18}
\end{align}
One can immediately  deduce from Fig.~\ref{fig2} that the increasing magnetic field gradually reduces the quantum negativity in all three dimers coupled by the $J_1$ interaction (e.g., $S_{1}\!-\!S_{2}$, $\mu_{1}\!-\!\mu_{2}$ and $\mu_{1}\!-\!S_{2}$).  Since the coupling constant $J$ is relatively weak, an increasing magnetic field gradually suppresses (enlarges) probability amplitudes of highly entangled (separable) spin pairs within the respective ground state as is clearly evident from the transcription of eigenvectors into the more convenient form
\begin{align}
\left\vert 1,\frac{3}{2},\frac{3}{2}\right\rangle\!=\!&-\frac{\sqrt{2}}{3\sqrt{5}}\left\vert 1,\!-1\right\rangle_{\mu_1,\mu_2}\otimes\left\vert 2,2\right\rangle_{S_1,S_2}\!+\!
\frac{1}{3\sqrt{30}}\left\vert 1,1\right\rangle_{\mu_1,\mu_2}\otimes \left[
10\left\vert 0,0\right\rangle_{S_1,S_2}\!-\!\sqrt{2} \left\vert 2,0\right\rangle_{S_1,S_2}
\right]
\nonumber\\
\!&+\!\frac{1}{3\sqrt{5}}
\left\vert 1,0\right\rangle_{\mu_1,\mu_2}\otimes \left\vert 2,1\right\rangle_{S_1,S_2}\!+\!\frac{\sqrt{5}}{3}\left\vert 0,0\right\rangle_{\mu_1,\mu_2}\otimes \left\vert 1,1\right\rangle_{S_1,S_2},
\label{eq19}\\
\left\vert 2,\frac{3}{2},\frac{3}{2}\right\rangle\!&=\!-\frac{\sqrt{2}}{\sqrt{3}}
\left\vert 1,1\right\rangle_{\mu_1,\mu_2}\otimes \left\vert 1,1\right\rangle_{S_1,S_2}\!-\!\frac{1}{\sqrt{3}}\left\vert 0,0\right\rangle_{\mu_1,\mu_2}\otimes \left\vert 2,2\right\rangle_{S_1,S_2},
\label{eq20}
\end{align}
and
\begin{align}
\left\vert 0,\frac{3}{2},\frac{3}{2}\right\rangle\!=\!&-\frac{1}{6}\left[ \left\vert \frac{3}{2},\frac{3}{2}\right\rangle_{\mu_1,S_2}\otimes \left\vert \frac{3}{2},\!-\frac{3}{2}\right\rangle_{\mu_2,S_1}\!-\! \left\vert \frac{3}{2},\!-\frac{3}{2}\right\rangle_{\mu_1,S_2}\otimes \left\vert \frac{3}{2},\frac{3}{2}\right\rangle_{\mu_2,S_1}\right]
\nonumber\\
\!&-\!\frac{2}{3}\left[ \left\vert \frac{1}{2},\!-\frac{1}{2}\right\rangle_{\mu_1,S_2}\otimes\left\vert \frac{1}{2},\frac{1}{2}\right\rangle_{\mu_2,S_1}\!-\! \left\vert \frac{1}{2},\frac{1}{2}\right\rangle_{\mu_1,S_2}\otimes \left\vert \frac{1}{2},\!-\frac{1}{2}\right\rangle_{\mu_2,S_1}\right]
\nonumber\\
\!&-\!\frac{1}{6}\left[ \left\vert \frac{3}{2},\!\frac{1}{2}\right\rangle_{\mu_1,S_2}\otimes \left\vert \frac{3}{2},\frac{1}{2}\right\rangle_{\mu_2,S_1}\!-\! \left\vert \frac{3}{2},\frac{1}{2}\right\rangle_{\mu_1,S_2}\otimes \left\vert \frac{3}{2},\!-\frac{1}{2}\right\rangle_{\mu_2,S_1}\right],
\label{eq21}\\
\left\vert 1,\frac{3}{2},\frac{3}{2}\right\rangle\!=\!&-\frac{1}{3\sqrt{30}}\left\{  \left\vert \frac{3}{2},\frac{3}{2}\right\rangle_{\mu_1,S_2}\otimes \left[5\sqrt{2} \left\vert \frac{1}{2},\!-\frac{1}{2}\right\rangle_{\mu_2,S_1}\!+\!\left\vert \frac{3}{2},\!-\frac{1}{2}\right\rangle_{\mu_2,S_1}\right]\right.
\nonumber\\
\!&\hspace*{3.5cm}+\!\left.\left[5\sqrt{2}\left\vert \frac{1}{2},\!-\frac{1}{2}\right\rangle_{\mu_1,S_2}\!+\!\left\vert \frac{3}{2},\!-\frac{1}{2}\right\rangle_{\mu_1,S_2}\right]\otimes \left\vert \frac{3}{2},\frac{3}{2}\right\rangle_{\mu_2,S_1}\right\}
\nonumber\\
\!&-\!\frac{\sqrt{5}}{9}\left[2\sqrt{2} \left\vert \frac{1}{2},\frac{1}{2}\right\rangle_{\mu_1,S_2}\!+\! \left\vert \frac{3}{2},\frac{1}{2}\right\rangle_{\mu_1,S_2}\right]\otimes \left\vert \frac{1}{2},\frac{1}{2}\right\rangle_{\mu_2,S_1}
\nonumber\\
\!&-\!\frac{\sqrt{5}}{9}\left[ \left\vert \frac{1}{2},\frac{1}{2}\right\rangle_{\mu_1,S_2}\!-\! \left\vert \frac{3}{2},\frac{1}{2}\right\rangle_{\mu_1,S_2}\right]\otimes \left\vert \frac{3}{2},\frac{1}{2}\right\rangle_{\mu_2,S_1},
\label{eq22}\\
\left\vert 2,\frac{3}{2},\frac{3}{2}\right\rangle\!=\!&\frac{1}{3\sqrt{2}}\left\{  \left\vert \frac{3}{2},\frac{3}{2}\right\rangle_{\mu_1,S_2}\otimes \left[2\sqrt{2}\left\vert \frac{1}{2},\frac{1}{2}\right\rangle_{\mu_2,S_1}\!+\!\left\vert \frac{3}{2},\frac{1}{2}\right\rangle_{\mu_2,S_1}\right]\right.
\nonumber\\
\!&\hspace*{2.9cm}-\!\left.\left[2\sqrt{2}\left\vert \frac{1}{2},\frac{1}{2}\right\rangle_{\mu_1,S_2}\!+\!\left\vert \frac{3}{2},\frac{1}{2}\right\rangle_{\mu_1,S_2}\right]\otimes \left\vert \frac{3}{2},\frac{3}{2}\right\rangle_{\mu_2,S_1}\right\}.
\label{eq23}
\end{align}
The quantum negativity ${\cal N}_{\mu_{1}|S_{1}}$ contrarily increases with an increasing magnetic field due to the lower coupling within the spin pair $\mu_{1}$-$S_{1}$.  A weak interaction $J$ allows the magnetic field to progressively reorient the fully polarized spins of a  
spin-(1/2,1) Heisenberg dimer from the opposite direction towards the magnetic-field direction (see Eq.~\eqref{eq18}).
Analytically detected magnitudes for all four bipartite quantum negativities ${\cal N}_{S_{1}|S_{2}}$, ${\cal N}_{\mu_{1}|\mu_{2}}$, ${\cal N}_{\mu_{1}|S_{1}}$ and ${\cal N}_{\mu_{1}|S_{2}}$ in respective ground states are collected in Tab.~\ref{tab1}.

The most interesting observation of our study relates to the fact that  the bipartite quantum entanglement can simultaneously exist  between all possible spin-pair decompositions of a mixed spin-(1/2,1) Heisenberg tetramer for the case of $J_1/J\!>\!1$ and non-zero magnetic field $h/J\!>\!J_1/J$. This unconventional spatial distribution of the bipartite entanglement represents a useful prerequisite to presume a possible existence of a multipartite quantum entanglement in the mixed spin-(1/2,1) Heisenberg tetramer. 

 For a completeness, in Tab.~\ref{tab1} we also present analytic values of the bipartite negativities for fully spatially isotropic case with $J_1/J\!=\!1$. In general, the bipartite negativities should be reduced due to the enhancement of ground-state degeneracy, which leads to a more complex form of the ground-state eigenvectors having the character of mixed states. Only along the border of the ground states $\vert 0,\frac{1}{2},\frac{1}{2}\rangle$ and $\vert 0,\frac{3}{2},\frac{3}{2}\rangle$  the bipartite negativity ${\cal N}_{S_{1}|S_{2}}$ lies in between the values ${\cal N}_{S_{1}|S_{2}}(\vert 0,\frac{1}{2},\frac{1}{2}\rangle)$ and ${\cal N}_{S_{1}|S_{2}}(\vert 0,\frac{3}{2},\frac{3}{2}\rangle)$, see Tab.~\ref{tab1}. This peculiar finding is a direct consequence of the fact that the number of antiferromagnetically ordered entangled $S_{1}-S_{2}$ spins increases (decreases) in a new composite eigenvector in comparison to the $\vert 0,\frac{1}{2},\frac{1}{2}\rangle$ ($\vert 0,\frac{3}{2},\frac{3}{2}\rangle$), see Eqs.~\eqref{eq16} and \eqref{eq17}. 

\subsection{Bipartite thermal negativity} 
\label{thermal}
\begin{figure}[b!]
\centering
{\includegraphics[width=.48\textwidth,trim=0.5cm 7.45cm 1.cm 8cm, clip]{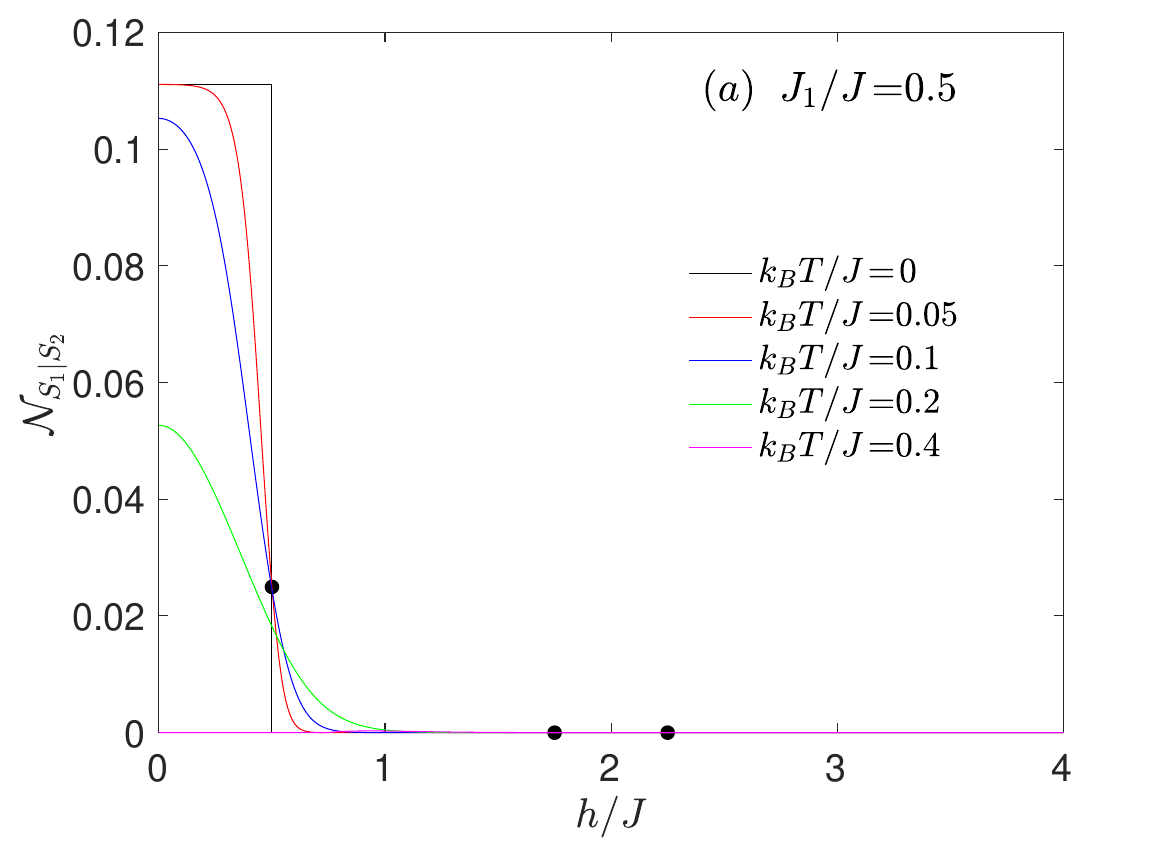}}
{\includegraphics[width=.48\textwidth,trim=0.5cm 7.45cm 1.cm 8cm, clip]{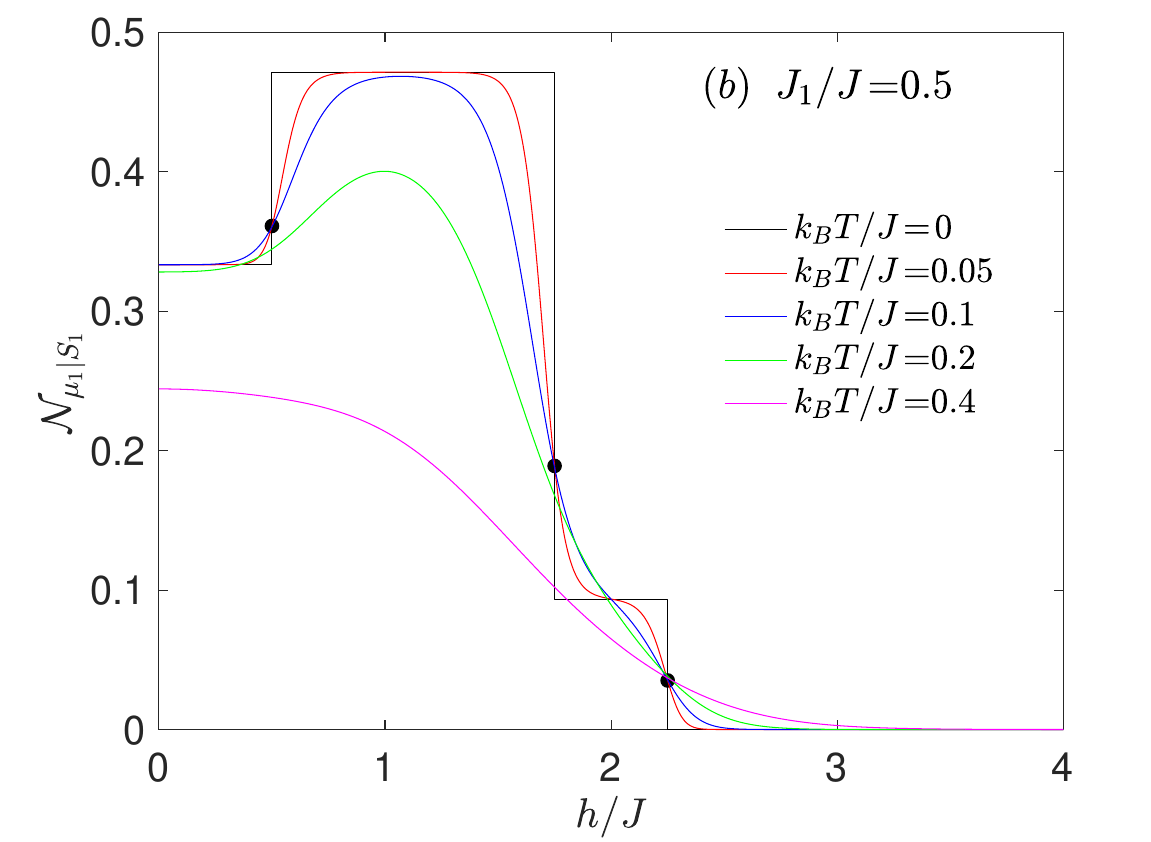}}
\caption{Magnetic-field variation of the bipartite thermal negativities   ${\cal N}_{S_{1}|S_{2}}$ ($a$) and ${\cal N}_{\mu_{1}|S_{1}}$ ($b$) of the mixed spin-(1/2,1) Heisenberg tetramer \eqref{eq1} for a few selected values of temperature and the fixed value of the  interaction ratio $J_1/J\!=\!0.5$. Black dots denote the value of the quantum negativity at the ground-state boundary.}
\label{fig3}
\end{figure}

\begin{figure}[b!]
\centering
{\includegraphics[width=.45\textwidth,trim=0.5cm 7.45cm 1.cm 8cm, clip]{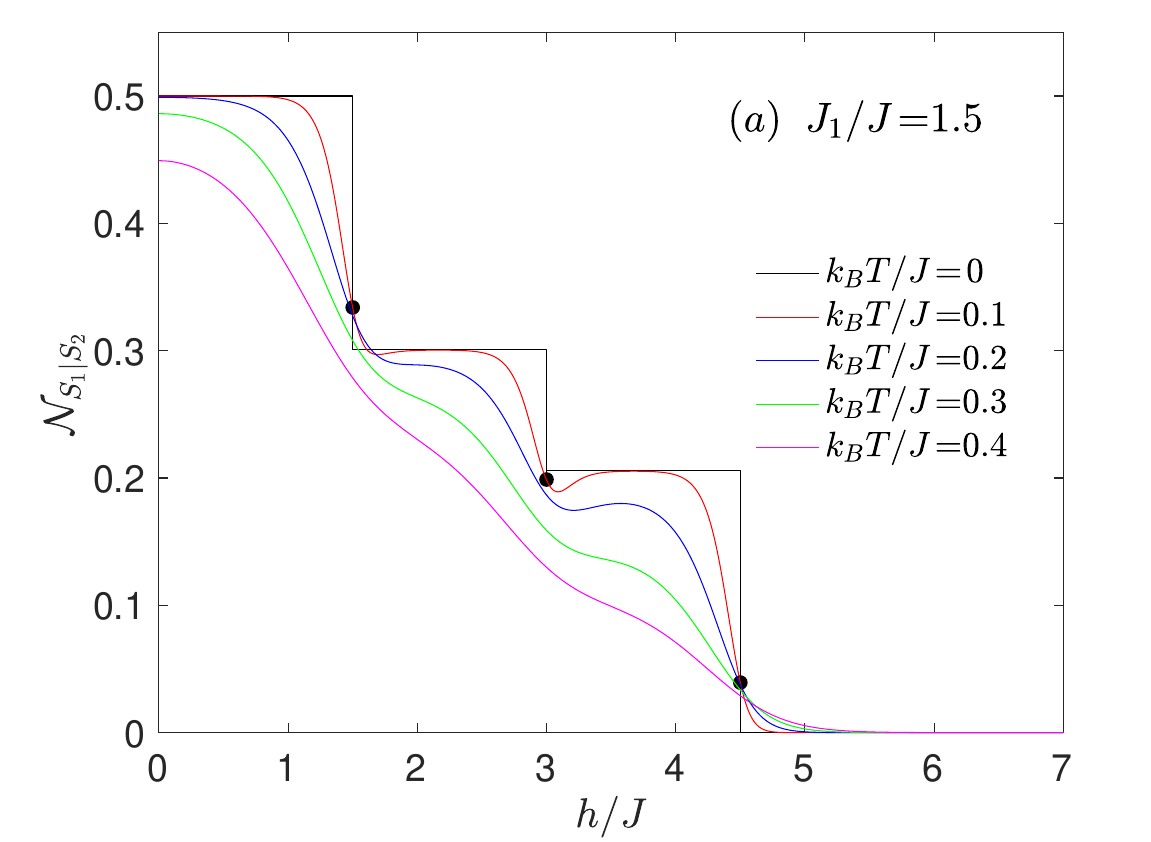}}
{\includegraphics[width=.45\textwidth,trim=0.5cm 7.45cm 1.cm 8cm, clip]{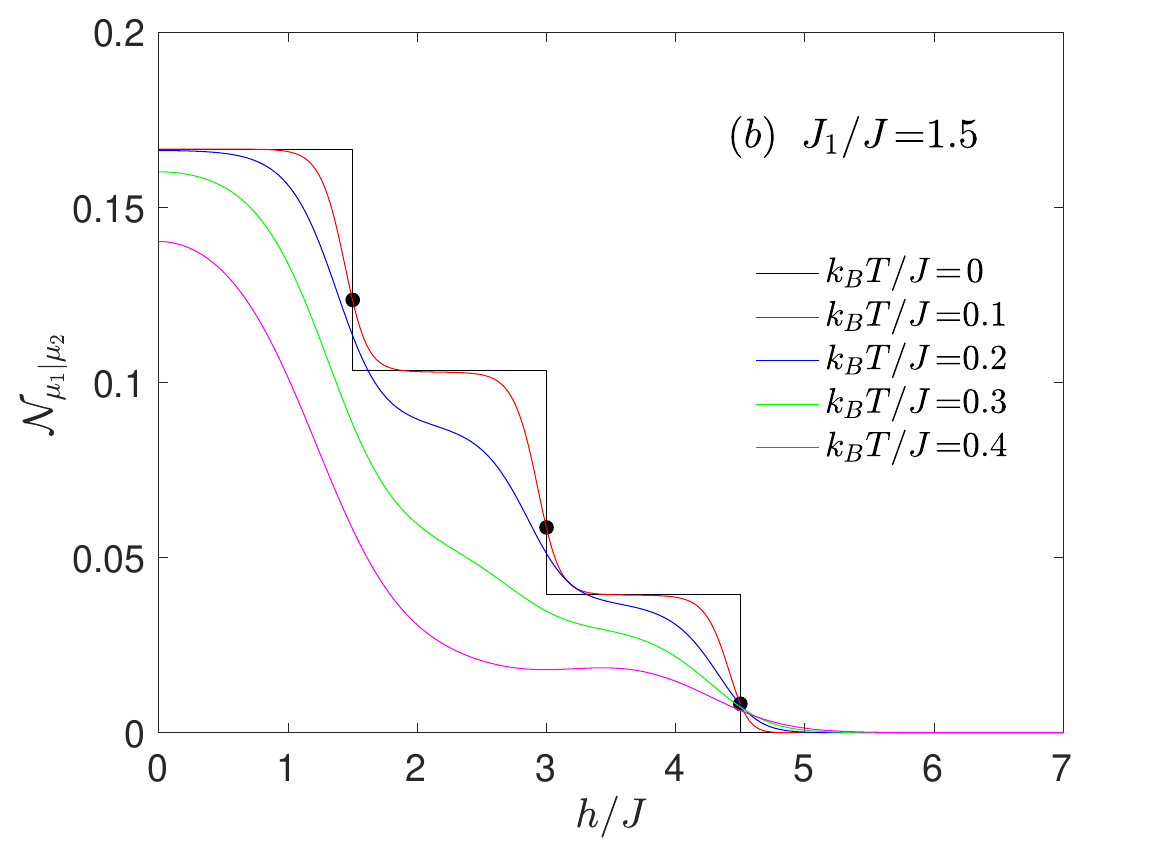}}\\
{\includegraphics[width=.45\textwidth,trim=0.5cm 7.45cm 1.cm 8cm, clip]{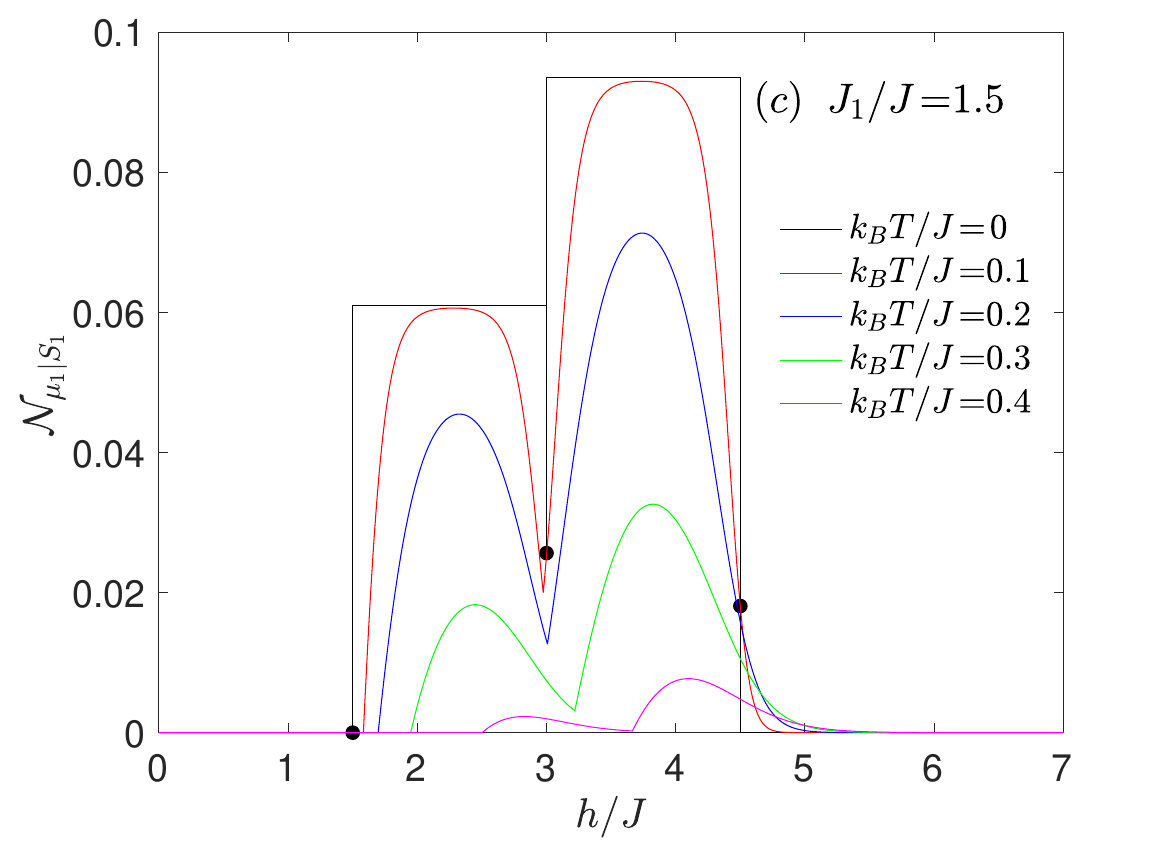}}
{\includegraphics[width=.45\textwidth,trim=0.5cm 7.45cm 1.cm 8cm, clip]{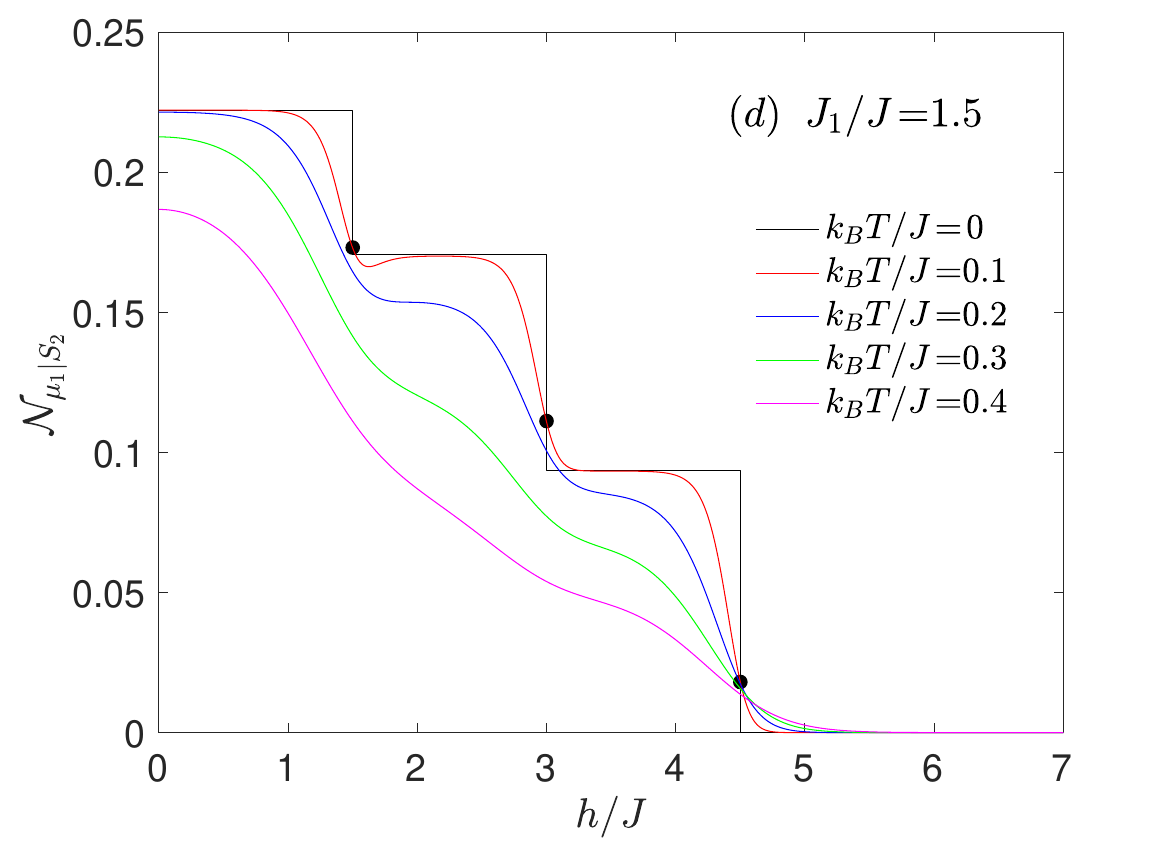}}
\caption{Magnetic-field-variation of the bipartite negativities   ${\cal N}_{S_{1}|S_{2}}$, ${\cal N}_{\mu_{1}|\mu_{2}}$,  ${\cal N}_{\mu_{1}|S_{1}}$ and ${\cal N}_{\mu_{1}|S_{2}}$  of the mixed spin-(1/2,1) Heisenberg tetramer \eqref{eq1}  for a few selected values of temperature in a strong interaction limit $J_1/J\!=\!1.5$. Black dots denote the value of the quantum negativity at the ground-state boundaries.}
\label{fig4}
\end{figure}

The definition of the bipartite negativity through the partially transposed reduced density matrix allows us to study also thermal behavior of the bipartite entanglement. The obtained results are collected in Fig.~\ref{fig3} for a weak interaction coupling $J_1/J\!=\!0.5$ and in Fig.~\ref{fig4} for a strong interaction coupling $J_1/J\!=\!1.5$ in the form of the bipartite negativity vs. magnetic field plots.

Taking into account the fact that the quantum negativity exists for $J_1/J\!<\!1$   only within the $S_{1}\!-\!S_{2}$ and $\mu_{1}\!-\!S_{1}$ dimers, one could anticipate its inherence at  finite temperatures solely for two aforementioned spin pairs as exemplified in Fig.~\ref{fig3}. In both cases the increasing temperature smears out a stepwise dependence emergent in zero-temperature curves and it reduces in general  the degree of a bipartite entanglement. However, in a close vicinity of the magnetic-field-driven phase transition  the increasing temperature can locally enhance the bipartite negativity as a consequence of thermal excitations between two states with a different level of the  quantum entanglement. It should be emphasized that the zero-temperature negativity shows discontinuity at the respective transition field(s).

\begin{figure}[t!]
\centering
{\includegraphics[width=.45\textwidth,trim=0.5cm 7.45cm 1.cm 8cm, clip]{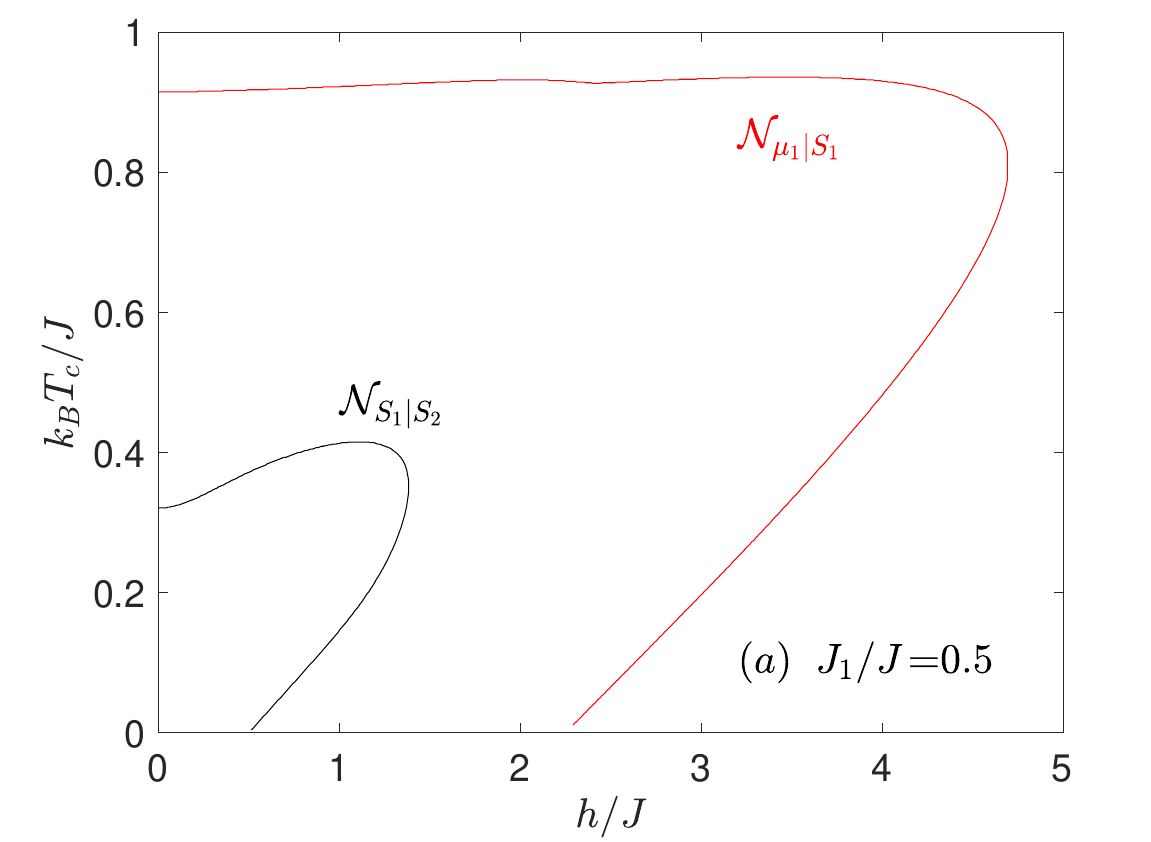}}
{\includegraphics[width=.45\textwidth,trim=0.5cm 7.45cm 1.cm 8cm, clip]{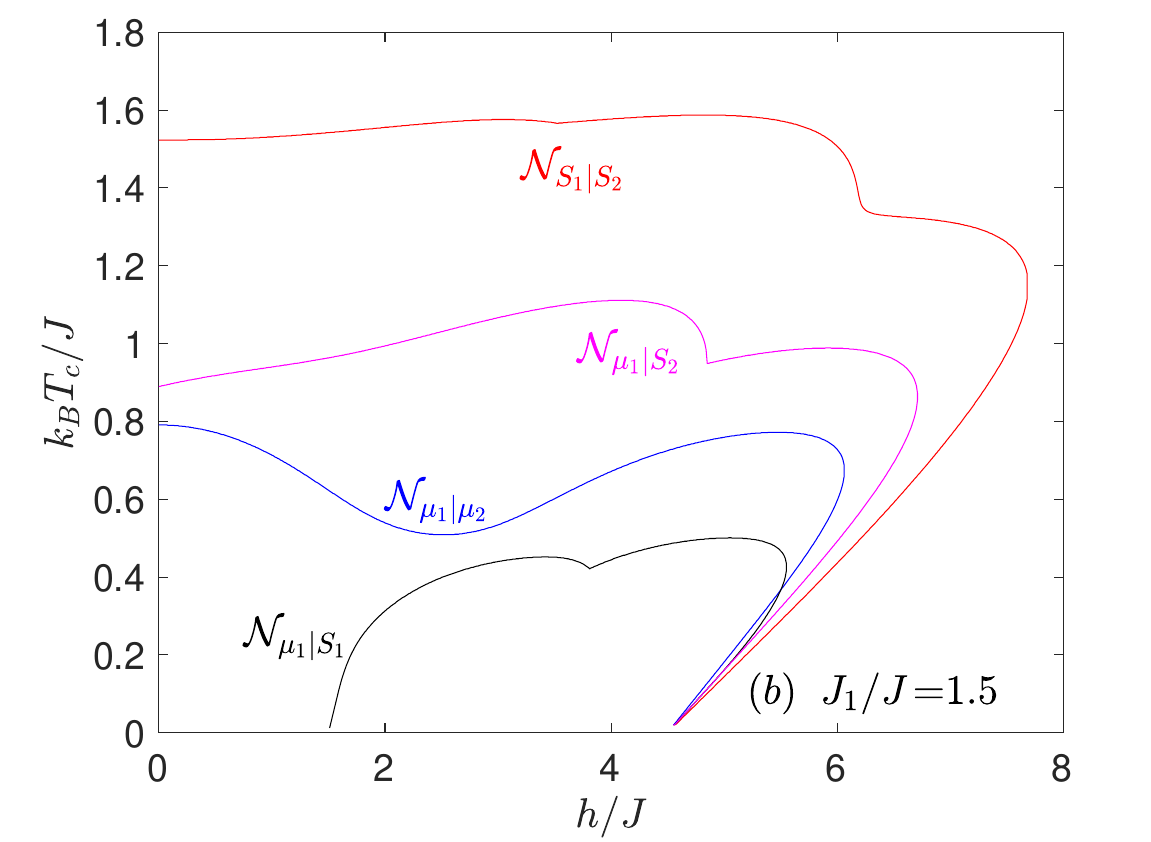}}
\caption{The threshold temperature of the bipartite negativity as a function of the magnetic field by assuming: ($a$) the weak interaction limit $J_1/J\!=\!0.5$, ($b$) the strong interaction limit $J_1/J\!=\!1.5$. Solid curves represent borders between the entangled region (inside of the 'loop') and the separable region (above the 'loop'). }
\label{fig3b}
\end{figure}

 Both coexistent phases are degenerate at the respective transition field and hence, the negativity at this specific point acquires different value due to the emergence of the eigenstate with character of a mixed state (the special points are visualized through  black dots in Fig.~\ref{fig3}). The rigorously calculated values of the bipartite quantum negativity at these specific points are collected in Tab.~\ref{tab2}. Because the spin-1 Heisenberg dimer $S_{1}\!-\!S_{2}$ is coupled via the weak interaction $J_1$, the increasing temperature more significantly destroys the respective bipartite entanglement in comparison with the mixed spin-(1/2,1) Heisenberg dimer 
$\mu_{1}\!-\!S_{1}$. In addition, the combination of a ground state and the first excited state in the mixed spin-(1/2,1) Heisenberg dimer 
$\mu_{1}\!-\!S_{1}$ (Eq.~\eqref{eq13}) makes such a spin distribution quite resistant with respect to thermal fluctuations. For low enough magnetic fields, the negativity ${\cal N}_{\mu_{1}|S_{1}}$ is almost unchanged up to relatively high temperatures. 
All the conclusions about the thermal stability of the bipartite entanglement are supported by the magnetic-field dependence of the threshold temperature, above which the bipartite entanglement disappears (see Fig.~\ref{fig3b}($a$)). The continuous curves thus delimit 
the entangled region (inside of the 'loop') from the fully separable one (above the 'loop'). In addition, it was detected that the bipartite negativity ${\cal N}_{\mu_{1}|S_{1}}$ vanishes at  almost identical threshold temperature ($k_BT_c/J\!\sim\!0.92$) for an arbitrary magnetic field being smaller than the saturation field.
\begin{table}[t!]
\resizebox{1\textwidth}{!}{
\begin{tabular}{l || l || l | l | l | l }
  $J_1/J$ & coexisting  & ${\cal N}_{S_{1}|S_{2}}$ & ${\cal N}_{\mu_{1}|\mu_{2}}$ & ${\cal N}_{\mu_{1}|S_{1}}$ & ${\cal N}_{\mu_{1}|S_{2}}$\\
   &  ground states & & &  & \\
 
  \hline\hline
   $<1$ & $\vert 0,\frac{1}{2},\frac{1}{2}\rangle$ , $\vert 1,\frac{1}{2},\frac{1}{2}\rangle$ & $\frac{1}{18}(\sqrt{6}\!-\!2) \approx$0.025  & 0 & $\frac{1}{24}\left(\sqrt{73}\!+\!\sqrt{17}\!-\!4 \right) \approx$0.361 & 0\\
         & $\vert 1,\frac{1}{2},\frac{1}{2}\rangle$, $\vert 2,\frac{1}{2}\,\frac{3}{2}\rangle$, $\vert 2,\frac{3}{2},\frac{1}{2}\rangle$ & 0 & 0 & $\frac{1}{18}(\sqrt{41}\!-\!3) \approx$0.189  & 0\\
        & $\vert 2,\frac{1}{2},\frac{3}{2}\rangle$, $\vert 2,\frac{3}{2},\frac{1}{2}\rangle$, $\vert 3,\frac{1}{2},\frac{1}{2}\rangle$ &  0& 0 & $\frac{1}{9}(\sqrt{11}\!-\!3) \approx$0.035  & 0\\
	      \hline	   
	  $>1$ & $\vert 0,\frac{3}{2},\frac{3}{2}\rangle$, $\vert 1,\frac{3}{2},\frac{3}{2}\rangle$ &$\frac{1}{270}(3\sqrt{1193}\!-\!74\!-\!\sqrt{5494}\cos\left(\frac{\phi}{3}\!+\!\frac{2\pi}{3} \right))\approx$0.334   & $ \frac{1}{180}(\sqrt{2626}\!-\!29)\approx$0.124  & 0 & $\frac{1}{720}(3\sqrt{3313}\!+\!\sqrt{19601}\!-\!188)\approx$0.173 \\
	 & &$\phi\!=\!\arctan\left(\frac{\sqrt{p^3-q^2}}{q}\right)$ &  & &\\
	  & &$ p\!=\!\frac{5494}{(540)^2}, q\!=\!\frac{110314}{(540)^3}$ &  & &\\	
             & $\vert 1,\frac{3}{2},\frac{3}{2}\rangle$, $\vert 2,\frac{3}{2},\frac{3}{2}\rangle$ &$\frac{1}{270}(12\sqrt{58}\!-\!109\!-\!\sqrt{9859}\cos\left(\frac{\phi}{3}\!+\!\frac{2\pi}{3} \right))\approx$0.199    &$ \frac{1}{180}(3\sqrt{394}\!-\!49)\approx$0.059  & $ \frac{1}{20}(3\sqrt{3}\!+\!\sqrt{11}\!-\!8)\approx$0.026  & $\frac{1}{180}(3\sqrt{323}+2\sqrt{274}\!-\!67)\approx$0.111   \\      
 	 & &$\phi\!=\!\arctan\left(\frac{\sqrt{p^3-q^2}}{q}\right)$ &  & &\\
	  & &$ p\!=\!\frac{9859}{(540)^2}, q\!=\!\frac{659429}{(540)^3}$ &  & &\\	
               & $\vert 2,\frac{3}{2},\frac{3}{2}\rangle$, $\vert 3,\frac{3}{2},\frac{3}{2}\rangle$ & $-\frac{1}{9}(2\!+\!\sqrt{19}\cos\left(\frac{\phi}{3}\!+\!\frac{2\pi}{3} \right))\approx$0.039  & $ \frac{1}{12}(\sqrt{26}\!-\!5)\approx$0.008 & $ \frac{1}{24}(\sqrt{89}\!-\!9)\approx$0.018  & $\frac{1}{24}(\sqrt{89}\!-\!9) \approx$0.018 \\ 
 	 & &$\phi\!=\!\arctan\left(\frac{\sqrt{p^3-q^2}}{q}\right)$ &  & &\\
	  & &$ p\!=\!\frac{19}{4(9)^2}, q\!=\!\frac{41}{4(9)^3}$ &  & &\\	
              \hline
\end{tabular}
}
\caption{Four bipartite quantum  negativities (${\cal N}_{S_{1}|S_{2}}$, ${\cal N}_{\mu_{1}|\mu_{2}}$, ${\cal N}_{\mu_{1}|S_{1}}$ and ${\cal N}_{\mu_{1}|S_{2}}$)  of a mixed spin-(1/2,1) Heisenberg tetramer~\eqref{eq1} calculated at the specific magnetic-field-driven phase transitions.}
\label{tab2}
\end{table}

In the strong interaction limit $J_1/J\!>\!1.0$ thermal fluctuations again smear out a sharp stepwise zero-temperature dependencies of the bipartite negativity (Fig.~\ref{fig4}) and they generally reduce a degree of the thermal entanglement.
For any non-zero temperature  the bipartite negativity ${\cal N}_{\mu_{1}|\mu_{2}}$ (Fig.~\ref{fig4}($b$)) displays a monotonous decline with increasing of an external magnetic field until the thermal entanglement is completely destroyed at
the threshold temperature. In the remaining cases the bipartite negativity as a function of magnetic field involves  round or sharp local minimum as a consequence of degeneracy at the respective transition field. Round local minima observable at ${\cal N}_{S_{1}|S_{2}}$ (Fig.~\ref{fig4}($a$)) and ${\cal N}_{\mu_{1}|S_{2}}$ (Fig.~\ref{fig4}($d$)) gradually vanish with  increasing of temperature
in contradiction to  ${\cal N}_{\mu_{1}|S_{1}}$ (Fig.~\ref{fig4}($c$)), where the pronounced cusp is detected until the thermal fluctuations completely destroy the entangled state. The  rapid changes of the negativity  in a close vicinity of the cusp may be  deduced from the respective tangents. Due to a relative strength of $J_1$ coupling, the spin arrangement within the $\mu_{1}\!-\!S_{1}$ dimer is promptly changed unlike the arrangement within the $S_{1}\!-\!S_{2}$ and $\mu_{1}\!-\!S_{2}$ dimers. In addition, the increasing temperature  significantly shifts the first transition field, below which the $\mu_{1}\!-\!S_{1}$ dimer is always in a separable state (Fig.~\ref{fig4}($c$)). Because the ground state $|0,\frac{3}{2},\frac{3}{2}\rangle$ and the first excited state $|1,\frac{3}{2},\frac{3}{2}\rangle$ are very close in energy, the first excited state is easily populated by the thermal fluctuations and the thermal negativity is consequently  dramatically reduced to the zero value. 

As it is illustrated in Fig.~\ref{fig3b}($b$), the non-zero bipartite negativity  of a mixed spin-(1/2,1) Heisenberg tetramer persists up to  relative high temperatures in a strong coupling limit. In accordance with general expectations, the thermal negativity vanishes a more rapidly within a dimer with the weak interaction coupling as well as within a dimer with a lower quantum spin numbers. Interestingly, the thermal stability of respective bipartite entanglement can be modulated by the variation of external magnetic field depending on the spin value of the dimer constituents as well as the strength of the interaction constant, see for example the local minimum in ${\cal N}_{\mu_{1}|\mu_{2}}$ around $h/J\!\approx\!2.2$. Moreover, it is worthy to notice that the bipartite negativity ${\cal N}_{\mu_{1}|S_{1}}$ is prohibited at low enough magnetic fields at the specific condition $J_1/J\!>\!1$.
\\
\section{Conclusions}
\label{Conclusions}

In this paper the distribution of four different types of bipartite entanglement in the mixed spin-(1/2,1) Heisenberg tetramer  is analyzed with the help of entanglement measure negativity. Two spin-1/2 and two spin-1 entities are localized in corners of a square plaquette in a such manner that the mixed spin-(1/2,1) Heisenberg tetramer can be viewed as being composed from two interacting mixed spin-(1/2,1) Heisenberg dimers with the intra-dimer interaction $J$ and the inter-dimer interaction $J_1$. While the coupling constant $J$ refers to the intra-dimer interaction between the nearest-neighbor spins of different magnitudes, the coupling constant $J_1$ is ascribed to the inter-dimer interaction between alike as well as different spins. To provide a deeper insight into the quantum entanglement  we have analyzed in detail the quantum negativity calculated for  all four inequivalent spin dimers emergent within the studied mixed spin-(1/2,1) Heisenberg tetramer, namely, ${\cal N}_{S_{1}|S_{2}}$, ${\cal N}_{\mu_{1}|\mu_{2}}$, ${\cal N}_{\mu_{1}|S_{1}}$ and ${\cal N}_{\mu_{1}|S_{2}}$. It was found that the distribution of bipartite quantum negativities is different in a weak ($J_1/J\!<\!1$) and strong  ($J_1/J\!>\!1$) interaction limit. In the particular case of dominant intra-dimer coupling  the bipartite entanglement exists only within the $\mu_{1}\!-\!S_{1}$ and $S_{1}\!-\!S_{2}$ dimers. Out of two available ground states the simultaneous bipartite quantum entanglement has been detected only for the  ground state $|0,\frac{1}{2},\frac{1}{2}\rangle$ realized at low enough magnetic fields. The stronger bipartite entanglement with the highest  value of the negativity has been confirmed in the other ground state $|1,\frac{1}{2},\frac{1}{2}\rangle$, which is in fact  a tensor product of two maximally entangled states of the mixed spin-(1/2,1) Heisenberg dimers. In the strong interaction limit  ($J_1/J\!>\!1$) the bipartite quantum entanglement is detected for all dimers at arbitrarily small magnetic field below the saturation field.  Only the the bipartite negativity ${\cal N}_{\mu_{1}|S_{1}}\!=\!0$ vanishes within the ground state $|0,\frac{3}{2},\frac{3}{2}\rangle$,  whose respective eigenvector is a superposition of separable and entangled states with an identical probability amplitude.

Temperature dependencies of the bipartite negativity evidenced that the increasing magnetic field can locally enhance the bipartite entanglement at non-zero temperatures depending on a character of the ground state.  In the close vicinity of magnetic-field-driven phase transition the bipartite negativity may exhibit a sharp or round minimum, which can or cannot persist until thermal fluctuations cause  breakdown of the local minimum. It was demonstrated additionally that a thermal stability of the bipartite entanglement can be almost independent of the strength of the magnetic field (e.g., ${\cal N}_{\mu_{1}|S_{1}}$ for $J_1/J\!=\!0.5$) or oppositely it can be manifested through the changes of the threshold temperature (e.g., ${\cal N}_{\mu_{1}|\mu_{2}}$ for $J_1/J\!=\!1.5$) depending on the dimer constituents as well as the coupling strength.

Finally, it should be emphasized that the deeper understanding of a multipartite entanglement  is a complex task and requires more than the analysis of all bipartite entanglements. However, the simultaneous existence of the bipartite entanglement in at least two different spin dimers points on the possibility to detect a genuinely tripartite and/or tetrapartite quantum correlations in a mixed spin-(1/2,1) Heisenberg tetramer. In the next step, we plan to perform an exhaustive analysis of a distribution of the tripartite and tetrapartite entanglement in a mixed spin-(1/2,1) Heisenberg tetramer.

This work was financially supported by the grant of  the Slovak Research and Development Agency provided under the contract No. APVV-20-0150 and by the grant of the Slovak Academy of Sciences and The Ministry of Education, Science, Research, and Sport of the Slovak Republic provided under the contract No. VEGA 1/0105/20. 
\\
\vfill
\pagebreak
\newpage
\appendix
\section{}
\label{App A}
\setcounter{equation}{0}
\renewcommand{\theequation}{\thesection.\arabic{equation}}
\renewcommand{\thetable}{\thesection.\arabic{equation}}

The basis of the mixed spin-(1/2,1) Heisenberg tetramer \eqref{eq1} involves totally  36 different states. Each of them can be defined as a  tensor product of basis states of two spin-(1/2,1) Heisenberg dimers. Namely, $\vert \varphi_i^{\pm}\rangle\!=\!\vert \mu^z_{1},S^z_{1}\rangle \otimes\vert \mu^z_{2},S^z_{2}\rangle$, where $\mu^z_{\gamma}$, $S^z_{\gamma}$ can reach one of the following state $\mu^z_{\gamma}\!=\!\pm\tfrac{1}{2}$ and $S^z_{\gamma}\!=\!\pm1,0$ ($\gamma\!=\!\{1,2\}$). The full basis, classified according to the $\sigma^z_T\!=\!S^z_{1}\!+\!S^z_{2}\!+\!\mu^z_{1}\!+\!\mu^z_{2}$, is listed below in Tab.~\ref{tab_A1}.
\begin{table}[h!]
\centering
\renewcommand\thetable{A.1} 
\begin{tabular}{l|l|| l| l }
  $\sigma_T^z$ & base states & $\sigma_T^z$ & base states \\\hline\hline
   -3 & ${\vert \varphi_{1}^-\rangle}\!=\!\vert \!-\!\tfrac{1}{2},\!-\!1\rangle\otimes\vert \!-\!\tfrac{1}{2},\!-\!1\rangle$ &
    3 & ${\vert \varphi_{1}^+\rangle}\!=\!\vert \tfrac{1}{2},1\rangle\otimes\vert \tfrac{1}{2},1\rangle$    \\
  \hline
      -2     & ${\vert \varphi_{2}^-\rangle}\!=\!\vert \!-\!\tfrac{1}{2},\!-\!1\rangle\otimes\vert \!-\!\tfrac{1}{2},0\rangle$ & 2&     ${\vert \varphi_{2}^+\rangle}\!=\!\vert \tfrac{1}{2},1\rangle\otimes \vert \tfrac{1}{2},0\rangle$\\ 
        & ${\vert \varphi_{3}^-\rangle}\!=\!\vert \!-\!\tfrac{1}{2},\!-\!1\rangle\otimes\vert \tfrac{1}{2},\!-\!1\rangle$ &  & ${\vert \varphi_{3}^+\rangle}\!=\!\vert \tfrac{1}{2},1\rangle\otimes\vert \!-\!\tfrac{1}{2},1\rangle$ \\
    & ${\vert \varphi_{4}^-\rangle}\!=\!\vert \!-\!\tfrac{1}{2},0\rangle\otimes\vert \!-\!\tfrac{1}{2},\!-\!1\rangle$ &  & ${\vert \varphi_{4}^+\rangle}\!=\!\vert \tfrac{1}{2},0\rangle\otimes\vert \tfrac{1}{2},1\rangle$\\
 & ${\vert \varphi_{5}^-\rangle}\!=\!\vert \tfrac{1}{2},\!-\!1\rangle\otimes \vert \!-\!\tfrac{1}{2},\!-\!1\rangle$ &  & ${\vert \varphi_{5}^+\rangle}\!=\!\vert \!-\!\tfrac{1}{2},1\rangle\otimes\vert \tfrac{1}{2},1\rangle$
  \\
     \hline
   -1    &  ${\vert \varphi_{6}^-\rangle}\!=\!\vert \!-\!\tfrac{1}{2},\!-\!1\rangle\otimes \vert \!-\!\tfrac{1}{2},1\rangle$&  1&  ${\vert \varphi_{6}^+\rangle}\!=\!\vert \tfrac{1}{2},1\rangle\otimes \vert \tfrac{1}{2},\!-\!1\rangle$\\
      &  ${\vert \varphi_{7}^-\rangle}\!=\!\vert \!-\!\tfrac{1}{2},\!-\!1\rangle\otimes \vert \tfrac{1}{2},0\rangle$&  & ${\vert \varphi_{7}^+\rangle}\!=\!\vert \tfrac{1}{2},1\rangle\otimes \vert \!-\!\tfrac{1}{2},0\rangle$\\ 
       &  ${\vert \varphi_{8}^-\rangle}\!=\!\vert \!-\!\tfrac{1}{2},0\rangle\otimes \vert \!-\!\tfrac{1}{2},0\rangle$&  &   ${\vert \varphi_{8}^+\rangle}\!=\!\vert \tfrac{1}{2},0\rangle\otimes \vert \tfrac{1}{2},0\rangle$\\
       &  ${\vert \varphi_{9}^-\rangle}\!=\!\vert \!-\!\tfrac{1}{2},0\rangle\otimes \vert \tfrac{1}{2},\!-\!1\rangle$&  &  ${\vert \varphi_{9}^+\rangle}\!=\!\vert \tfrac{1}{2},0\rangle\otimes \vert \!-\!\tfrac{1}{2},1\rangle$\\ 
      &  ${\vert \varphi_{10}^-\rangle}\!=\!\vert \tfrac{1}{2},\!-\!1\rangle\otimes \vert \!-\!\tfrac{1}{2},0\rangle$&  & ${\vert \varphi_{10}^+\rangle}\!=\!\vert \!-\!\tfrac{1}{2},1\rangle\otimes \vert \tfrac{1}{2},0\rangle$\\
    &  ${\vert \varphi_{11}^-\rangle}\!=\!\vert \tfrac{1}{2},\!-\!1\rangle\otimes \vert \tfrac{1}{2},\!-\!1\rangle$&  & ${\vert \varphi_{11}^+\rangle}\!=\!\vert \!-\!\tfrac{1}{2},1\rangle\otimes \vert \!-\!\tfrac{1}{2},1\rangle$\\
  &  ${\vert \varphi_{12}^-\rangle}\!=\!\vert \!-\!\tfrac{1}{2},1\rangle\otimes \vert \!-\!\tfrac{1}{2},\!-\!1\rangle$ &  &  ${\vert \varphi_{12}^+\rangle}\!=\!\vert \tfrac{1}{2},\!-\!1\rangle\otimes \vert \tfrac{1}{2},1\rangle$\\
&  ${\vert \varphi_{13}^-\rangle}\!=\!\vert \tfrac{1}{2},0\rangle\otimes \vert \!-\!\tfrac{1}{2},\!-\!1\rangle$ &&  ${\vert \varphi_{13}^+\rangle}\!=\!\vert \!-\!\tfrac{1}{2},0\rangle\otimes \vert \tfrac{1}{2},1\rangle$\\
     \hline
        0 &  ${\vert \varphi_{14}^-\rangle}\!=\!\vert \tfrac{1}{2},1\rangle\otimes \vert \!-\!\tfrac{1}{2},\!-\!1\rangle$  & 0& ${\vert \varphi_{14}^+\rangle}\!=\!\vert \!-\!\tfrac{1}{2},\!-\!1\rangle\otimes \vert \tfrac{1}{2},1\rangle$ \\
          &  ${\vert \varphi_{15}^-\rangle}\!=\!\vert \tfrac{1}{2},0\rangle\otimes \vert \tfrac{1}{2},\!-\!1\rangle$  & &  ${\vert \varphi_{15}^+\rangle}\!=\!\vert \!-\!\tfrac{1}{2},0\rangle\otimes \vert \!-\!\tfrac{1}{2},1\rangle$ \\
          &  ${\vert \varphi_{16}^-\rangle}\!=\!\vert \tfrac{1}{2},0\rangle\otimes \vert \!-\!\tfrac{1}{2},0\rangle$  & & ${\vert \varphi_{16}^+\rangle}\!=\!\vert \!-\!\tfrac{1}{2},0\rangle\otimes \vert \tfrac{1}{2},0\rangle$ \\
          &  ${\vert \varphi_{17}^-\rangle}\!=\!\vert \!-\!\tfrac{1}{2},1\rangle\otimes \vert \tfrac{1}{2},\!-\!1\rangle$  & & ${\vert \varphi_{17}^+\rangle}\!=\!\vert \tfrac{1}{2},\!-\!1\rangle\otimes \vert \!-\!\tfrac{1}{2},1\rangle$ \\
         &  ${\vert \varphi_{18}^-\rangle}\!=\!\vert \!-\!\tfrac{1}{2},1\rangle\otimes \vert \!-\!\tfrac{1}{2},0\rangle$  & &  ${\vert \varphi_{18}^+\rangle}\!=\!\vert \tfrac{1}{2},\!-\!1\rangle\otimes \vert \tfrac{1}{2},0\rangle$  \\
	      \hline	    
\end{tabular}
\caption{The complete list of basis states $\vert \varphi_i^{\pm} \rangle\!=\!\vert \mu^z_{1},S^z_{1}\rangle \otimes \vert\mu^z_{2},S^z_{2} \rangle$ of the mixed spin-(1/2,1) Heisenberg tetramer~\eqref{eq1} classified according to  the value of $\sigma^z_T\!=\!S_{1}^z\!+\!\mu_{1}^z\!+\!S_{2}^z\!+\!\mu_{2}^z$.}
\label{tab_A1}
\end{table}

The partition function ${\cal Z}$ of the the mixed spin-(1/2,1) Heisenberg tetramer~\eqref{eq1} reads
\begin{eqnarray}
{\cal Z}&=&\sum_{k=1}^{36}{\rm e}^{-\beta\varepsilon_k}
\nonumber\\
&=&2{\rm e}^{\frac{\beta}{4} J_1}\bigg\{\cosh(3\beta h){\rm e}^{-\beta (J\!+\!\frac{5}{2}J_1)}\!+\!
2\cosh(2\beta h){\rm e}^{-\frac{\beta}{4}(J\!+\!J_1)}\bigg[\cosh\left(\frac{3}{4}\beta(J\!-\!J_1)\right)+\cosh\left(\frac{3}{4}\beta(J\!+\!J_1)\right){\rm e}^{-\frac{3}{2}\beta J_1}\bigg]
\nonumber\\
&+&2\cosh(\beta h)\bigg[\cosh\left(\frac{5\beta J_1}{2}\right){\rm e}^{-\beta J}\!+\!
{\rm e}^{\frac{\beta}{2}J}\left( 2\cosh(\beta J_1)\!+\!\cosh\left(\frac{\beta}{2}(3J\!-\!J_1)\right)\right)
\bigg]\nonumber\\
&+&{\rm e}^{-\beta J}\left(\cosh\left(\frac{5\beta J_1}{2}\right)+\cosh\left(\frac{3\beta J_1}{2}\right){\rm e}^{2\beta J_1}\right)\!+\!{\rm e}^{\frac{\beta}{2}J}\left(2\cosh(\beta J_1)+\cosh\left(\frac{\beta J_1}{2}\right){\rm e}^{\frac{3}{2}\beta J}\right)\bigg\}.
\label{A_eq1}
\end{eqnarray}
\\
\begin{table}[th!]
\renewcommand\thetable{A.2} 
\begin{tabular}{l|c | c}
  $\vert\sigma_T^z,\sigma_T,\sigma_{1},\sigma_{2}\rangle$ & Eigenvalues ($\varepsilon_k$) & Eigenvectors in original base ($\vert \psi \rangle_k$)\\\hline\hline
  $\vert \pm3,3,\tfrac{3}{2},\tfrac{3}{2}\rangle$   & $J+(9/4)J_1\mp 3h$ & $\varphi_1^{\pm}$\\
   \hline
 	$\vert \pm2,3,\tfrac{3}{2},\tfrac{3}{2}\rangle$ 	  & $J+(9/4)J_1\mp 2h$& $\tfrac{1}{\sqrt{6}}\left(\sqrt{2}\varphi_2^{\pm}\!+\!\varphi_3^{\pm}\!+\!\sqrt{2}\varphi_4^{\pm}\!+\!\varphi_5^{\pm}\right)$\\
 	 $\vert \pm2,2,\tfrac{3}{2},\tfrac{3}{2}\rangle$ 	 &  $J-(3/4)J_1\mp 2h$ & $\mp\tfrac{1}{\sqrt{6}}\left(\sqrt{2}\varphi_2^{\pm}\!+\!\varphi_3^{\pm}\!-\!\sqrt{2}\varphi_4^{\pm}\!-\!\varphi_5^{\pm}\right)$\\ 
   $\vert \pm 2,2,\tfrac{3}{2},\tfrac{1}{2}\rangle$          & $-J/2+(3/4)J_1\mp 2h$ & $\mp\tfrac{1}{\sqrt{3}}\left(\varphi_2^{\pm}\!-\!\sqrt{2}\varphi_3^{\pm}\right)$\\
	   $\vert \pm 2,2,\tfrac{1}{2},\tfrac{3}{2}\rangle$         & $-J/2+(3/4)J_1\mp 2h$& $\mp\tfrac{1}{\sqrt{3}}\left(\varphi_4^{\pm}\!-\!\sqrt{2}\varphi_5^{\pm}\right)$\\
  \hline
	$\vert \pm1,3,\tfrac{3}{2},\tfrac{3}{2}\rangle$ 	  & $J+(9/4)J_1\mp h$& $\tfrac{1}{\sqrt{15}}\left(\varphi_6^{\pm}\!+\!\sqrt{2}\varphi_7^{\pm}\!+\!2\varphi_8^{\pm}\!+\!\sqrt{2}\varphi_9^{\pm}
  		  \!+\!\sqrt{2}\varphi_{10}^{\pm}\!+\!\varphi_{11}^{\pm}\!+\!\varphi_{12}^{\pm}\!+\!\sqrt{2}\varphi_{13}^{\pm}\right)$\\
  		     $\vert \pm1,2,\tfrac{3}{2},\tfrac{3}{2}\rangle$  & $J-(3/4)J_1\mp h$ & $\mp\tfrac{1}{\sqrt{6}}\left(\varphi_6^{\pm}\!+\!\sqrt{2}\varphi_7^{\pm}\!-\!\varphi_{12}^{\pm}\!-\!\sqrt{2}\varphi_{13}^{\pm}\right)$
  		     \\  
  $\vert \pm 1,2,\tfrac{3}{2},\tfrac{1}{2}\rangle$          & $-J/2+(3/4)J_1\mp h$& $\mp\tfrac{1}{2\sqrt{3}}\left(\sqrt{2}\varphi_6^{\pm}\!-\!\varphi_7^{\pm}\!+\!\sqrt{2}\varphi_8^{\pm}\!-\!2\varphi_9^{\pm}
  \!+\!\varphi_{10}^{\pm}\!-\!\sqrt{2}\varphi_{11}^{\pm}\right)$\\
		   $\vert \pm 1,2,\tfrac{1}{2},\tfrac{3}{2}\rangle$               & $-J/2+(3/4)J_1\mp h$& $\mp\tfrac{1}{2\sqrt{3}}\left(\sqrt{2}\varphi_8^{\pm}\!+\!\varphi_9^{\pm}\!-\!2\varphi_{10}^{\pm}\!-\!\sqrt{2}\varphi_{11}^{\pm}
		   \!+\!\sqrt{2}\varphi_{12}^{\pm}\!-\!\varphi_{13}^{\pm}\right)$\\	      	     
  	$\vert \pm1,1,\tfrac{3}{2},\tfrac{3}{2}\rangle$ 	  & $J-(11/4)J_1\mp h$& $\tfrac{1}{\sqrt{10}}\left(\varphi_6^{\pm}\!+\!\sqrt{2}\varphi_7^{\pm}\!-\!\tfrac{4}{3}\varphi_8^{\pm}\!-\!\tfrac{2\sqrt{2}}{3}\varphi_9^{\pm}
  	\!-\!\tfrac{2\sqrt{2}}{3}\varphi_{10}^{\pm}\!-\!\tfrac{2}{3}\varphi_{11}^{\pm}\!+\!\varphi_{12}^{\pm}\!+\!\sqrt{2}\varphi_{13}^{\pm}\right)$\\
  		     $\vert \pm 1,1,\tfrac{3}{2},\tfrac{1}{2}\rangle$               & $-J/2-(5/4)J_1\mp h$& $\tfrac{1}{6}\left(3\sqrt{2}\varphi_6^{\pm}\!-\!3\varphi_7^{\pm}\!-\!\sqrt{2}\varphi_8^{\pm}\!+\!2\varphi_9^{\pm}
  		     \!-\!\varphi_{10}^{\pm}\!+\!\sqrt{2}\varphi_{11}^{\pm}\right)$\\
		  $\vert \pm 1,1,\tfrac{1}{2},\tfrac{3}{2}\rangle$             & $-J/2-(5/4)J_1\mp h$& $\tfrac{1}{6}\left(\sqrt{2}\varphi_8^{\pm}\!+\!\varphi_9^{\pm}\!-\!2\varphi_{10}^{\pm}\!-\!\sqrt{2}\varphi_{11}^{\pm}
		  \!-\!3\sqrt{2}\varphi_{12}^{\pm}\!+\!3\varphi_{13}^{\pm}\right)$\\	  		  
  	$\vert \pm1,1,\tfrac{1}{2},\tfrac{1}{2}\rangle$ 	  & $-2J+(1/4)J_1\mp h$& $\tfrac{1}{3}\left(\varphi_8^{\pm}\!-\!\sqrt{2}\varphi_9^{\pm}\!-\!\sqrt{2}\varphi_{10}^{\pm}\!+\!2\varphi_{11}^{\pm}\right)$\\
  \hline
 	 $\vert 0,3,\tfrac{3}{2},\tfrac{3}{2}\rangle$   & $J+(9/4)J_1$& $\tfrac{1}{2\sqrt{5}}\left(\varphi_{14}^{-}\!+\!\sqrt{2}\varphi_{15}^{-}\!+\!2\varphi_{16}^{-}\!+\!\varphi_{17}^{-}\!+\!\sqrt{2}\varphi_{18}^{-}
 	 \!+\!\varphi_{14}^{+}\!+\!\sqrt{2}\varphi_{15}^{+}\!+\!2\varphi_{16}^{+}\!+\!\varphi_{17}^{+}\!+\!\sqrt{2}\varphi_{18}^{+}\right)$\\
  		         $\vert 0,2,\tfrac{3}{2},\tfrac{3}{2}\rangle$ & $J-(3/4)J_1$& $-\tfrac{1}{6}\left(3\varphi_{14}^{-}\!+\!\sqrt{2}\varphi_{15}^{-}\!+\!2\varphi_{16}^{-}\!+\!\varphi_{17}^{-}\!+\!\sqrt{2}\varphi_{18}^{-}
  		         \!-\!3\varphi_{14}^{+}\!-\!\sqrt{2}\varphi_{15}^{+}\!-\!2\varphi_{16}^{+}\!-\!\varphi_{17}^{+}\!-\!\sqrt{2}\varphi_{18}^{+}\right)$\\  	
  $\vert 0,2,\tfrac{3}{2},\tfrac{1}{2}\rangle$        & $-J/2+(3/4)J_1$&$-\tfrac{1}{3\sqrt{10}}\left(5\varphi_{15}^{-}\!-\!\sqrt{2}\varphi_{16}^{-}\!+\!\sqrt{2}\varphi_{17}^{-}\!-\!4\varphi_{18}^{-}
  \!-\!5\varphi_{15}^{+}\!+\!\sqrt{2}\varphi_{16}^{+}\!-\!\sqrt{2}\varphi_{17}^{+}\!+\!4\varphi_{18}^{+}\right)$\\	  			  $\vert 0,2,\tfrac{1}{2},\tfrac{3}{2}\rangle$     & $-J/2+(3/4)J_1$&$-\tfrac{1}{\sqrt{10}}\left(\sqrt{2}\varphi_{16}^{-}\!-\!\sqrt{2}\varphi_{17}^{-}\!-\!\varphi_{18}^{-}
  \!-\!\sqrt{2}\varphi_{16}^{+}\!+\!\sqrt{2}\varphi_{17}^{+}\!+\!\varphi_{18}^{+}\right)$\\
	  
  		 $\vert 0,1,\tfrac{3}{2},\tfrac{3}{2}\rangle$   & $J-(11/4)J_1$& $\tfrac{1}{6\sqrt{5}}\left(9\varphi_{14}^{-}\!-\!\sqrt{2}\varphi_{15}^{-}\!-\!2\varphi_{16}^{-}\!-\!\varphi_{17}^{-}\!-\!\sqrt{2}\varphi_{18}^{-}
  		 \!+\!9\varphi_{14}^{+}\!-\!\sqrt{2}\varphi_{15}^{+}\!-\!2\varphi_{16}^{+}\!-\!\varphi_{17}^{+}\!-\!\sqrt{2}\varphi_{18}^{+}\right)$\\
    $\vert 0,1,\tfrac{3}{2},\tfrac{1}{2}\rangle$      & $-J/2-(5/4)J_1$&$\tfrac{1}{3\sqrt{10}}\left(5\varphi_{15}^{-}\!-\!\sqrt{2}\varphi_{16}^{-}\!+\!\sqrt{2}\varphi_{17}^{-}\!-\!4\varphi_{18}^{-}
    \!+\!5\varphi_{15}^{+}\!-\!\sqrt{2}\varphi_{16}^{+}\!+\!\sqrt{2}\varphi_{17}^{+}\!-\!4\varphi_{18}^{+}\right)$\\	   		    $\vert 0,1,\tfrac{1}{2},\tfrac{3}{2}\rangle$        & $-J/2-(5/4)J_1$&$\tfrac{1}{\sqrt{10}}\left(\sqrt{2}\varphi_{16}^{-}\!-\!\sqrt{2}\varphi_{17}^{-}\!-\!\varphi_{18}^{-}
    \!+\!\sqrt{2}\varphi_{16}^{+}\!-\!\sqrt{2}\varphi_{17}^{+}\!-\!\varphi_{18}^{+}\right)$\\
  $\vert 0,1,\tfrac{1}{2},\tfrac{1}{2}\rangle$ 	  & $-2J+(1/4)J_1$ &$\tfrac{1}{3\sqrt{2}}\left(\sqrt{2}\varphi_{15}^{-}\!-\!\varphi_{16}^{-}\!-\!2\varphi_{17}^{-}\!+\!\sqrt{2}\varphi_{18}^{-}
  \!+\!\sqrt{2}\varphi_{15}^{+}\!-\!\varphi_{16}^{+}\!-\!2\varphi_{17}^{+}\!+\!\sqrt{2}\varphi_{18}^{+}\right)$\\
  	 	   	 $\vert 0,0,\tfrac{3}{2},\tfrac{3}{2}\rangle$ 	  & $J-(15/4)J_1$ &$-\tfrac{1}{6}\left(3\varphi_{14}^{-}\!-\!\sqrt{2}\varphi_{15}^{-}\!-\!2\varphi_{16}^{-}\!-\!\varphi_{17}^{-}\!-\!\sqrt{2}\varphi_{18}^{-}
  	 	   	 \!-\!3\varphi_{14}^{+}\!+\!\sqrt{2}\varphi_{15}^{+}\!+\!2\varphi_{16}^{+}\!+\!\varphi_{17}^{+}\!+\!\sqrt{2}\varphi_{18}^{+}\right)$\\  	
 		 $\vert 0,0,\tfrac{1}{2},\tfrac{1}{2}\rangle$  & $-2J-(3/4)J_1$&$-\tfrac{1}{3\sqrt{2}}\left(\sqrt{2}\varphi_{15}^{-}\!-\!\varphi_{16}^{-}\!-\!2\varphi_{17}^{-}\!+\!\sqrt{2}\varphi_{18}^{-}
 		 \!-\!\sqrt{2}\varphi_{15}^{+}\!+\!\varphi_{16}^{+}\!+\!2\varphi_{17}^{+}\!-\!\sqrt{2}\varphi_{18}^{+}\right)$\\    	
	      \hline	      
\end{tabular}
\caption{The complete list of eigenvalues ($\varepsilon_k$) and eigenvectors ($\vert \psi \rangle_k$) of the mixed spin-(1/2,1) Heisenberg tetramer~\eqref{eq1}.}
\label{tab_A2}
\end{table}
\\\\
The complete list of eigenvalues and respective eigenvectors of a  spin-1/2 Heisenberg dimer.  Each eigenvector $\vert M^z_T, M_T\rangle$ is expressed in terms of original basis $\vert \mu_1^z,\mu_2^z\rangle$, where two quantum numbers $M^z_T$, $M_T$  are associated with a new composite operator $\hat{M}_T\!=\!\hat{\boldsymbol\mu}_{1}\!+\!\hat{\boldsymbol\mu}_{2}$. 
\begin{equation}
\begin{array}{lll}
E_{\vert 0,0\rangle}\!=\!-\frac{3}{4}J, & &\vert 0,0\rangle\!=\!\frac{1}{\sqrt{2}}\left[ \vert \tfrac{1}{2},\!-\!\tfrac{1}{2}\rangle\!-\!\vert \!-\!\tfrac{1}{2},\tfrac{1}{2}\rangle \right],\\
E_{\vert 1,1\rangle}\!=\!\frac{J}{4}\!-\!h, & &\vert 1,1\rangle\!=\!\vert \tfrac{1}{2},\tfrac{1}{2}\rangle,\\E_{\vert 1,0\rangle}\!=\!\frac{J}{4}, & &\vert 1,0\rangle\!=\!\frac{1}{\sqrt{2}}\left[ \vert \tfrac{1}{2},\!-\!\tfrac{1}{2}\rangle\!+\!\vert \!-\!\tfrac{1}{2},\tfrac{1}{2}\rangle \right],\\
E_{\vert 1,\!-\!1\rangle}\!=\!\frac{J}{4}\!+\!h, & &\vert 1,\!-\!1\rangle\!=\!\vert \!-\!\tfrac{1}{2},\!-\!\tfrac{1}{2}\rangle.
\end{array}
\label{A_eq3}
\end{equation}
\\\\
The complete list of eigenvalues and respective eigenvectors of a mixed spin-(1/2,1) Heisenberg dimer.  Each eigenvector $\vert N^z_T, N_T\rangle$ is expressed in terms of original basis $\vert \mu^z,S^z\rangle$, where two quantum numbers $N^z_T$, $N_T$  are associated with a new composite operator $\hat{N}_T\!=\!\hat{\boldsymbol\mu}\!+\!\hat{\boldsymbol S}$. 
\begin{equation}
\begin{array}{lll}
E_{\vert\frac{1}{2},\frac{1}{2}\rangle}\!=\!-J\!-\!\frac{h}{2}, & &\vert\frac{1}{2},\frac{1}{2}\rangle\!=\!\frac{1}{\sqrt{3}}\left[ \vert \tfrac{1}{2},0\rangle\!-\!\sqrt{2}\vert \!-\!\tfrac{1}{2},1\rangle \right],\\
E_{\vert\frac{1}{2},\!-\!\frac{1}{2}\rangle}\!=\!-J\!+\!\frac{h}{2}, & &\vert\frac{1}{2},\!-\!\frac{1}{2}\rangle\!=\!\frac{1}{\sqrt{3}}\left[ \vert \!-\!\tfrac{1}{2},0\rangle\!-\!\sqrt{2}\vert \tfrac{1}{2},\!-\!1\rangle \right],\\
E_{\vert\frac{3}{2},\frac{3}{2}\rangle}\!=\!\frac{J}{2}\!-\!\frac{3}{2}h, & &\vert\frac{3}{2},\frac{3}{2}\rangle\!=\!\vert \tfrac{1}{2},1\rangle,\\
E_{\vert\frac{3}{2},\frac{1}{2}\rangle}\!=\!\frac{J}{2}\!-\!\frac{h}{2}, & &\vert\frac{3}{2},\frac{1}{2}\rangle\!=\!\frac{1}{\sqrt{3}}\left[\sqrt{2} \vert \tfrac{1}{2},0\rangle\!+\!\vert \!-\!\tfrac{1}{2},1\rangle \right],\\
E_{\vert\frac{3}{2},\!-\!\frac{1}{2}\rangle}\!=\!\frac{J}{2}\!+\!\frac{h}{2}, & &\vert\frac{3}{2},\!-\!\frac{1}{2}\rangle\!=\!\frac{1}{\sqrt{3}}\left[\sqrt{2} \vert \!-\!\tfrac{1}{2},0\rangle\!+\!\vert \tfrac{1}{2},\!-\!1\rangle \right],\\
E_{\vert\frac{3}{2},\!-\!\frac{3}{2}\rangle}\!=\!\frac{J}{2}\!+\!\frac{3}{2}h, & &\vert\frac{3}{2},\!-\!\frac{3}{2}\rangle\!=\!\vert \!-\!\tfrac{1}{2},\!-\!1\rangle.
\end{array}
\label{A_eq2}
\end{equation}
\\\\
The complete list of eigenvalues and respective eigenvectors of a  spin-1 Heisenberg dimer.  Each eigenvector $\vert \Xi^z_T, \Xi_T\rangle$ is expressed in terms of original basis $\vert S_1^z,S_2^z\rangle$, where two quantum numbers $\Xi^z_T$, $\Xi_T$  are associated with a new composite operator $\hat{\Xi}_T\!=\!\hat{\boldsymbol S}_1\!+\!\hat{\boldsymbol S}_2$. 
\begin{equation}
\begin{array}{lll}
E_{\vert 0,0\rangle}\!=\!-2J, & &{\vert 0,0\rangle}\!=\!\frac{1}{\sqrt{3}}\left[ \vert 1,\!-\!1\rangle\!+\!\vert \!-\!1,1\rangle\!-\!\vert 0,0\rangle \right],\\
E_{\vert 1,1\rangle}\!=\!-J\!-\!h, & &\vert 1,1\rangle\!=\!\frac{1}{\sqrt{2}}\left[ \vert 1,0\rangle\!-\!\vert 0,1\rangle \right],\\
E_{\vert 1,0\rangle}\!=\!-J, & &\vert 1,0\rangle\!=\!\frac{1}{\sqrt{2}}\left[ \vert 1,\!-\!1\rangle\!-\!\vert \!-\!1,1\rangle \right],\\
E_{\vert 1,\!-\!1\rangle}\!=\!-J\!+\!h, & &\vert 1,\!-\!1\rangle\!=\!\frac{1}{\sqrt{2}}\left[ \vert \!-\!1,0\rangle\!-\!\vert 0,\!-\!1\rangle \right],\\
E_{\vert 2,2\rangle}\!=\!J\!-\!2h, & &\vert 2,2\rangle\!=\!\vert 1,1\rangle,\\
E_{\vert 2,1\rangle}\!=\!J\!-\!h, & &\vert 2,1\rangle\!=\!\frac{1}{\sqrt{2}}\left[ \vert 1,0\rangle\!+\!\vert 0,1\rangle \right],\\
E_{\vert 2,0\rangle}\!=\!J, & &{\vert 2,0\rangle}\!=\!\frac{1}{\sqrt{6}}\left[ \vert 1,\!-\!1\rangle\!+\!\vert \!-\!1,1\rangle\!+\!2\vert 0,0\rangle \right],\\
E_{\vert 2,\!-\!1\rangle}\!=\!J\!+\!h, & &\vert 2,\!-\!1\rangle\!=\!\frac{1}{\sqrt{2}}\left[ \vert \!-\!1,0\rangle\!+\!\vert 0,\!-\!1\rangle \right],\\
E_{\vert 2,\!-\!2\rangle}\!=\!J\!+\!2h, & &\vert 2,\!-\!2\rangle\!=\!\vert \!-\!1,\!-\!1\rangle.
\end{array}
\label{A_eq4}
\end{equation}
\vfill
\section{Bipartite negativity ${\cal N}_{S_{1}|S_{2}}$ }
\label{App B}
The reduced density operator $\hat{\rho}_{S_{1}|S_{2}}$ is defined after tracing out degree of freedom of two spins $\mu_{1}$ and $\mu_{2}$. Thus,
\begin{align}
\hat{\rho}_{S_{1}|S_{2}}&\!=\!\sum_{\mu_{1}^z}\sum_{\mu_{2}^z} \langle \mu_{1}^z,\mu_{2}^z\vert\hat{\rho}\vert \mu_{1}^z,\mu_{2}^z\rangle\!=\!\frac{1}{\cal Z}\sum_{k=1}^{36}  {\rm e}^{-\beta \varepsilon_k}\left( \sum_{\mu_1^z}\sum_{\mu_2^z} \langle \mu_1^z,\mu_2^z\vert\psi_k\rangle \langle \psi_k\vert \mu_1^z,\mu_2^z\rangle\right).
\label{d14}
\end{align}
The corresponding reduced density matrix ${\hat{\rho}_{S_{1}|S_{2}}}$ in a basis of $\vert S_{1}^z,S^z_{2}\rangle$  reads as follows%
\begin{align}
\allowdisplaybreaks
{\hat{\rho}_{S_{1}|S_{2}}}\!=\!
\begin{blockarray}{r ccc ccc ccc}
 & \vert 1,1\rangle & \vert 1,0\rangle &\vert 1,\!-\!1\rangle  & \vert 0,1\rangle & \vert 0,0\rangle &\vert 0,\!-\!1\rangle  &\vert \!-\!1,1\rangle&\vert \!-\!1,0\rangle&\vert \!-\!1,\!-\!1\rangle\\
\begin{block}{r(ccc ccc ccc)}
\langle 1,1\vert \;\;\;&\rho_{11}(h) & 0 & 0 & 0 & 0 & 0 & 0 & 0 & 0 \\
\langle 1,0\vert \;\;\;&0 & \rho_{22}(h) & 0 & \rho_{24}(h) & 0 & 0 & 0 & 0 & 0\\
\langle 1,\!-\!1\vert\;\;\; & 0 & 0 & \rho_{33}(h) &  0 &\rho_{35}(h)& 0 & \rho_{37}(h) & 0 & 0\\
\langle 0,1\vert \;\;\;&0&\rho_{42}(h) & 0 & \rho_{44}(h) & 0 & 0 & 0 & 0 & 0  \\
\langle 0,0\vert \;\;\;&0 & 0 & \rho_{53}(h) & 0 & \rho_{55}(h)& 0 & \rho_{57}(h) & 0 & 0\\
\langle 0,\!-\!1\vert\;\;\; & 0 & 0 & 0 &  0 &0& \rho_{66}(h)& 0 & \rho_{68}(h)& 0\\
 \langle \!-\!1,1\vert\; \;\;& 0 & 0 & \rho_{73}(h)& 0& \rho_{75}(h) &0 & \rho_{77}(h) & 0 & 0\\
\langle \!-\!1,0\vert\;\;\;& 0 & 0 & 0&0 &0 & \rho_{86}(h) &0  &  \rho_{88}(h) & 0 \\
 \langle \!-\!1,\!-\!1\vert \;\;\;& 0 & 0 & 0& 0 & 0 & 0 & 0 & 0 & \rho_{99}(h) \\
\end{block}
\end{blockarray}\;\;.
\label{d15}
\end{align}
Here
\begin{flalign}
\rho_{11}(h)\!&=\!\frac{{\rm e}^{ \frac{\beta }{4}J_1}}{45{\cal Z}}\bigg\{
\cosh\left(\frac{\beta h}{2}\right){\rm e}^{\frac{5}{2}\beta h}\left[54{\rm e}^{-\beta(J+\frac{5}{2}J_1)} \right]
\nonumber\\
\!&+\!\cosh\left(\frac{\beta h}{2}\right){\rm e}^{\frac{3}{2}\beta h}\left[30{\rm e}^{-\beta(J-\frac{J_1}{2})}\!+\!9{\rm e}^{-\beta(J+\frac{5}{2}J_1)}\!+\!75{\rm e}^{-\beta(-\frac{J}{2}+J_1)} \right]
\nonumber\\
\!&+\!\sinh\left(\frac{\beta h}{2}\right){\rm e}^{\frac{5}{2}\beta h}\left[36{\rm e}^{-\beta(J+\frac{5}{2}J_1)} \right]
\nonumber\\
\!&+\!\sinh\left(\frac{\beta h}{2}\right){\rm e}^{\frac{3}{2}\beta h}\left[3{\rm e}^{-\beta(J+\frac{5}{2}J_1)}\!+\!45{\rm e}^{-\beta(-\frac{J}{2}+J_1)} \right]
\nonumber\\
\!&+\!5{\rm e}^{-\beta(-\frac{5}{4}J-\frac{J_1}{4})}{\rm e}^{\beta h}\left[ 
5\cosh\left(\frac{3\beta}{4}(J\!-\!J_1)\right)
\!+\!3\sinh\left(\frac{3\beta}{4}(J\!-\!J_1)\right)\right]
\nonumber\\
\!&-\!{\rm e}^{-\beta(J-\frac{3}{2}J_1)}{\rm e}^{\beta h}\left[ 
13\cosh\left(\beta J_1\right)
\!-\!17\sinh\left(\beta J_1\right)\right]
\bigg\},
\label{d16}\\
\rho_{99}(h)\!&=\!\rho_{11}(-h);
\label{d21}\\
\rho_{22}(h)\!&=\!\rho_{44}(h)\!=\!\frac{{\rm e}^{ \frac{\beta }{4}J_1}}{90{\cal Z}}\bigg\{
\cosh\left(\frac{\beta h}{2}\right){\rm e}^{\frac{3}{2}\beta h}\left[45{\rm e}^{-\beta(J-\frac{J_1}{2})}\!+\!42{\rm e}^{-\beta(J+\frac{5}{2}J_1)}\!+\!50{\rm e}^{-\beta(-\frac{J}{2}+J_1)} \right]
\nonumber\\
\!&+\!\cosh\left(\frac{\beta h}{2}\right){\rm e}^{\frac{\beta}{2} h}\left[30{\rm e}^{-\beta(-2J+\frac{J_1}{2})}\!+\!20{\rm e}^{-\beta(J-\frac{J_1}{2})} \!+\!21{\rm e}^{-\beta(J+\frac{5}{2}J_1)}\!+\!27{\rm e}^{-\beta(J-\frac{5}{2}J_1)}\!+\!50{\rm e}^{-\beta(-\frac{J}{2}+J_1)}\right.
\nonumber\\
\!&+\!\left.60{\rm e}^{-\beta(-\frac{J}{2}-J_1)}\right]
\nonumber\\
\!&+\!\sinh\left(\frac{\beta h}{2}\right){\rm e}^{\frac{3}{2}\beta h}\left[15{\rm e}^{-\beta(J-\frac{J_1}{2})}\!+\!18{\rm e}^{-\beta(J+\frac{5}{2}J_1)}\!+\!10{\rm e}^{-\beta(-\frac{J}{2}+J_1)} \right]
\nonumber\\
\!&+\!\sinh\left(\frac{\beta h}{2}\right){\rm e}^{\frac{\beta}{2} h}\left[10{\rm e}^{-\beta(-2J+\frac{J_1}{2})}\!+\!10{\rm e}^{-\beta(J-\frac{J_1}{2})}\!+\!3{\rm e}^{-\beta(J+\frac{5}{2}J_1)}\!+\!25{\rm e}^{-\beta(J-\frac{5}{2}J_1)}\!+\!10{\rm e}^{-\beta(-\frac{J}{2}-J_1)}\right]
\nonumber\\
\!&+\!5{\rm e}^{-\beta(-\frac{J}{2}-2J_1)}
\left[3\cosh\left(\frac{3\beta}{2}(J\!-\!J_1)\right)\!+\!\sinh\left(\frac{3\beta}{2}(J\!-\!J_1)\right)\right]\bigg\},
\label{d17}\\
\rho_{66}(h)\!&=\!\rho_{88}(h)\!=\!\rho_{22}(-h),
\label{d20}\\
\rho_{33}(h)\!&=\!\rho_{77}(h)\!=\!\frac{{\rm e}^{ \frac{\beta }{4}J_1}}{90{\cal Z}}\bigg\{
\cosh^2\left(\frac{\beta h}{2}\right)\left[55{\rm e}^{-\beta(J-\frac{J_1}{2})}\!+\!21{\rm e}^{-\beta(J+\frac{5}{2}J_1)}\!+\!59{\rm e}^{-\beta(J-\frac{5}{2}J_1)}\!+\!50{\rm e}^{-\beta(-\frac{J}{2}+J_1)}\!+\!110{\rm e}^{-\beta(-\frac{J}{2}-J_1)} \right]
\nonumber\\
\!&+\!\sinh^2\left(\frac{\beta h}{2}\right)\left[5{\rm e}^{-\beta(J-\frac{J_1}{2})}\!+\!3{\rm e}^{-\beta(J+\frac{5}{2}J_1)}\!-\!23{\rm e}^{-\beta(J-\frac{5}{2}J_1)}\!+\!10{\rm e}^{-\beta(-\frac{J}{2}+J_1)}\!+\!70{\rm e}^{-\beta(-\frac{J}{2}-J_1)} \right]
\nonumber\\
\!&+\!5{\rm e}^{-\beta(-\frac{J}{2}-2J_1)}
\left[7\cosh\left(\frac{3\beta}{2}(J\!-\!J_1)\right)\!-\!3\sinh\left(\frac{3\beta}{2}(J\!-\!J_1)\right)\right]
\nonumber\\
\!&+\!10{\rm e}^{2\beta J}
\left[3\cosh\left(\frac{\beta J_1}{2}\right)\!-\!\sinh\left(\frac{\beta J_1}{2}\right)\right]
\bigg\},
\label{d18}\\
\rho_{55}(h)\!&=\!\frac{{\rm e}^{ \frac{\beta }{4}J_1}}{45{\cal Z}}\bigg\{
\cosh^2\left(\frac{\beta h}{2}\right)\left[15{\rm e}^{-\beta(-2J+\frac{J_1}{2})}\!+\!42{\rm e}^{-\beta(J+\frac{5}{2}J_1)}\!+\!18{\rm e}^{-\beta(J-\frac{5}{2}J_1)}\!+\!50{\rm e}^{-\beta(-\frac{J}{2}+J_1)}\!+\!30{\rm e}^{-\beta(-\frac{J}{2}-J_1)} \right]
\nonumber\\
\!&+\!\sinh^2\left(\frac{\beta h}{2}\right)\left[5{\rm e}^{-\beta(-2J+\frac{J_1}{2})}\!+\!6{\rm e}^{-\beta(J+\frac{5}{2}J_1)}\!+\!14{\rm e}^{-\beta(J-\frac{5}{2}J_1)}\!+\!10{\rm e}^{-\beta(-\frac{J}{2}+J_1)}\!-\!10{\rm e}^{-\beta(-\frac{J}{2}-J_1)} \right]
\nonumber\\
\!&+\!5{\rm e}^{-\beta(J-2J_1)}
\left[3\cosh\left(\frac{3\beta J_1}{2}\right)\!-\!\sinh\left(\frac{3\beta J_1}{2}\right)\right]
\nonumber\\
\!&+\!10{\rm e}^{-\beta(-\frac{J}{2}-2J_1)}
\cosh\left(\frac{3\beta}{2}(J\!-\!J_1)\right)\bigg\},
\label{d19}\\
\rho_{24}(h)\!&=\!\rho_{42}(h)\!=\!\frac{{\rm e}^{ \frac{\beta }{4}J_1}}{90{\cal Z}}\bigg\{
\cosh\left(\frac{\beta h}{2}\right){\rm e}^{\frac{3}{2}\beta h}\left[42{\rm e}^{-\beta(J+\frac{5}{2}J_1)}\right]
\nonumber\\
\!&+\!\cosh\left(\beta h\right){\rm e}^{\beta h}\left[-35{\rm e}^{-\beta(J-\frac{J_1}{2})}\right]
\nonumber\\
\!&+\!\cosh\left(\frac{\beta h}{2}\right){\rm e}^{\frac{\beta}{2} h}\left[
21{\rm e}^{-\beta(J+\frac{5}{2}J_1)}\!-\!23{\rm e}^{-\beta(J-\frac{5}{2}J_1)}\!+\! 50{\rm e}^{-\beta(-\frac{J}{2}+J_1)}\!-\! 50{\rm e}^{-\beta(-\frac{J}{2}-J_1)}\right]
\nonumber\\
\!&+\!\sinh\left(\frac{\beta h}{2}\right){\rm e}^{\frac{3}{2}\beta h}\left[18{\rm e}^{-\beta(J+\frac{5}{2}J_1)} \right]
\nonumber\\
\!&+\!\sinh\left(\beta h\right){\rm e}^{\beta h}\left[-25{\rm e}^{-\beta(J-\frac{J_1}{2})} \right]
\nonumber\\
\!&+\!\sinh\left(\frac{\beta h}{2}\right){\rm e}^{\frac{\beta}{2} h}\left[3{\rm e}^{-\beta(J+\frac{5}{2}J_1)}\!-\!25{\rm e}^{-\beta(J-\frac{5}{2}J_1)}\!+\!10{\rm e}^{-\beta(-\frac{J}{2}+J_1)}\!-\!10{\rm e}^{-\beta(-\frac{J}{2}-J_1)}\right]
\nonumber\\
\!&+\!5{\rm e}^{2\beta J}
\left[\cosh\left(\frac{\beta J_1}{2}\right)\!-\!5\sinh\left(\frac{\beta J_1}{2}\right)\right]
\nonumber\\
\!&-\!10{\rm e}^{-\beta(-\frac{J}{2}-\frac{3}{2}J_1)}
\cosh\left(\frac{\beta}{2}(3J\!-\!4J_1)\right)\bigg\},
\label{d22}\\
\rho_{68}(h)\!&=\!\rho_{86}(h)\!=\!\rho_{24}(-h),
\label{d25}\\
\rho_{35}(h)\!&=\!\rho_{53}(h)\!=\!\rho_{57}(h)\!=\!\rho_{75}(h)\!=\!\frac{{\rm e}^{ \frac{\beta }{4}J_1}}{45{\cal Z}}\bigg\{
\cosh^2\left(\frac{\beta h}{2}\right)\left[21{\rm e}^{-\beta(J+\frac{5}{2}J_1)}\!-\!16{\rm e}^{-\beta(J-\frac{5}{2}J_1)} \!+\! 25{\rm e}^{-\beta(-\frac{J}{2}+J_1)}\!-\! 25{\rm e}^{-\beta(-\frac{J}{2}-J_1)}\right]
\nonumber\\
\!&+\!\sinh^2\left(\frac{\beta h}{2}\right)\left[3{\rm e}^{-\beta(J+\frac{5}{2}J_1)}\!-\!8{\rm e}^{-\beta(J-\frac{5}{2}J_1)} \!+\! 5{\rm e}^{-\beta(-\frac{J}{2}+J_1)}\!-\! 5{\rm e}^{-\beta(-\frac{J}{2}-J_1)}\right]
\nonumber\\
\!&-\!10\sinh\left(\frac{\beta J_1}{2}\right){\rm e}^{2\beta J}
\nonumber\\
\!&-\!5{\rm e}^{-\beta(J-2J_1)}
\left[\cosh\left(\frac{3\beta J_1}{2}\right)\!+\!3\sinh\left(\frac{3\beta J_1}{2}\right)\right]
\bigg\},
\label{d23}\\
\rho_{37}(h)\!&=\!\rho_{73}(h)\!=\!\frac{{\rm e}^{ \frac{\beta }{4}J_1}}{30{\cal Z}}\bigg\{
\cosh^2\left(\frac{\beta h}{2}\right)\left[-15{\rm e}^{-\beta(J-\frac{J_1}{2})}\!+\!7{\rm e}^{-\beta(J+\frac{5}{2}J_1)} \right]
\nonumber\\
\!&+\!\sinh^2\left(\frac{\beta h}{2}\right)\left[-5{\rm e}^{-\beta(J-\frac{J_1}{2})}\!+\!{\rm e}^{-\beta(J+\frac{5}{2}J_1)}\!+\!12{\rm e}^{-\beta(J-\frac{5}{2}J_1)}\right]
\nonumber\\
\!&+\!2{\rm e}^{-\beta(J-3J_1)}
\left[4\cosh\left(\frac{\beta J_1}{2}\right)\!+\!\sinh\left(\frac{\beta J_1}{2}\right)\right]
\bigg\}.
\label{d24}
\end{flalign}
Subsequently,  the partial transposition of the reduced density matrix  $\hat{\rho}_{S_{1}|S_{2}}^{T_{S_{1}}}$,  transposed with respect the $S_{1}$ spin, has the following block-diagonal structure
\begin{align}
\allowdisplaybreaks
{\hat{\rho}_{S_{1}|S_{2}}^{T_{S_{1}}}}\!=\;
\begin{blockarray}{ (ccc ccc ccc)}
\rho_{33}(h) & 0 & 0 & 0 & 0 & 0& 0 & 0 & 0 \\
0 & \rho_{77}(h) & 0 & 0 & 0 & 0& 0 & 0 & 0 \\
 0 & 0 &\rho_{22}(h) & \rho_{53}(h) & 0 & 0 & 0 & 0 &0\\
 0 & 0 & \rho_{35}(h) & \rho_{66}(h) & 0 & 0 & 0 & 0 &0\\
 0 & 0 & 0 & 0 & \rho_{44}(h) &\rho_{75}(h) &  0 & 0 &0 \\
 0 & 0 & 0& 0 &\rho_{57}(h) & \rho_{88}(h) & 0 & 0 & 0\\
 0 & 0 & 0&0 &0 & 0  &  \rho_{11}(h) & \rho_{42}(h) & \rho_{73}(h) \\
 0 & 0 & 0 & 0 & 0& 0 & \rho_{24}(h) & \rho_{55}(h) & \rho_{86}(h) \\
 0 & 0 & 0 & 0 & 0& 0 & \rho_{37}(h) & \rho_{68}(h) & \rho_{99}(h) \\
\end{blockarray}
\;\;,
\label{d26}
\end{align}
which directly results to the respective eigenvalues

\begin{flalign}
\lambda_1\!&=\!\rho_{33}(h),
\nonumber\\
\lambda_2\!&=\!\rho_{77} (h),
\nonumber\\
\lambda_3^{\pm}\!&=\!\frac{1}{2}\left(\rho_{22}(h)\!+\!\rho_{66}(h)\!\pm\!\sqrt{(\rho_{22}(h)\!-\!\rho_{66}(h))^2\!+\!4\rho_{35}(h)\rho_{53}(h)} \right),
\nonumber\\
\lambda_4^{\pm}\!&=\!\frac{1}{2}\left(\rho_{44}(h)\!+\!\rho_{88}(h)\!\pm\!\sqrt{(\rho_{44}(h)\!-\!\rho_{88}(h))^2\!+\!4\rho_{57}(h)\rho_{75}(h)} \right),
\nonumber\\
\lambda_{4+n}\!&=\!\frac{a}{3}\!+\!2{\rm sgn}(q)\sqrt{p}\cos\left(\frac{\phi}{3}\!+\!\frac{2\pi n}{3}\right),\;\;\;n\!=\!1,2,3,
\label{d27}\\
a\!&=\!\rho_{11}(h)\!+\!\rho_{55}(h)\!+\!\rho_{99}(h), 
\nonumber\\
b\!&=\!\rho_{11}(h)\rho_{55}(h)\!+\!\rho_{11}(h)\rho_{99}(h)\!+\!\rho_{55}(h)\rho_{99}(h)\!-\!\left( \rho_{24}(h)\rho_{42}(h)\!+\!\rho_{37}(h)\rho_{73}(h)\!+\!\rho_{68}(h)\rho_{86}(h)\right), 
\nonumber\\
c\!&=\!\rho_{11}(h)\rho_{68}(h)\rho_{86}(h)\!+\!\rho_{55}(h)\rho_{37}(h)\rho_{73}(h)\!+\!\rho_{99}(h)\rho_{24}(h)\rho_{42}(h)
\nonumber\\
\!&-\!\rho_{11}(h)\rho_{55}(h)\rho_{99}(h)\!-\!\rho_{24}(h)\rho_{37}(h)\rho_{68}(h)\!-\!\rho_{42}(h)\rho_{73}(h)\rho_{86}(h),
\nonumber\\
p\!&=\!\left(\frac{a}{3}\right)^2\!-\!\frac{b}{3},
\hspace*{2cm}
q\!=\!\left(\frac{a}{3}\right)^3\!-\!\frac{a}{3}\frac{b}{2}\!-\!\frac{c}{2},
\hspace*{2cm}
\phi\!=\!\arctan\left(\frac{\sqrt{p^3\!-\!q^2}}{q} \right).
\nonumber
\end{flalign}
It should be emphasized that the eigenvalues $\lambda_{5,6,7}$ are the solutions of a cubic equation $\lambda^3\!-\!a\lambda^2\!+\!b\lambda\!+\!c\!=\!0$ with a standard substitution $\lambda\!=\!x\!+\!a/3$ in Ferrari's solution (see, for instance, Ref~\cite{Rektorys}).
\section{Bipartite negativity ${\cal N}_{\mu_{1}|\mu_{2}}$ }
\label{App C}

The reduced density operator $\hat{\rho}_{\mu_{1}|\mu_{2}}$ is defined after tracing out degree of freedom of two spins $S_{1}$ and $S_{2}$. Thus,
\begin{align}
\hat{\rho}_{\mu_{1}|\mu_{2}}&\!=\!\sum_{S_{1}^z}\sum_{S_{2}^z} \langle S_{1}^z,S_{2}^z\vert\hat{\rho}\vert S_{1}^z,S_{2}^z\rangle\!=\!\frac{1}{\cal Z}\sum_{k=1}^{36}  {\rm e}^{-\beta \varepsilon_k}\left( \sum_{S_1^z}\sum_{S_2^z} \langle S_1^z,S_2^z\vert\psi_k\rangle \langle \psi_k\vert S_1^z,S_2^z\rangle\right).
\label{c11}
\end{align}
The corresponding reduced density matrix ${\hat{\rho}_{\mu_{1}|\mu_{2}}}$ in a basis of $\vert\mu^z_{1},\mu^z_{2}\rangle$ reads as follows
\begin{align}
\allowdisplaybreaks
{\hat{\rho}_{\mu_{1}|\mu_{2}}}\!=\!
\begin{blockarray}{r cc cc }
 & \vert \tfrac{1}{2},\tfrac{1}{2}\rangle & \vert \tfrac{1}{2},\!-\!\tfrac{1}{2}\rangle &\vert \!-\!\tfrac{1}{2},\tfrac{1}{2}\rangle  &\vert \!-\!\tfrac{1}{2},\!-\!\tfrac{1}{2}\rangle\\
\begin{block}{r(cc cc )}
\langle \tfrac{1}{2},\tfrac{1}{2}\vert \;\;\;&\rho_{11}(h) & 0 & 0 & 0  \\
\langle \tfrac{1}{2},\!-\!\tfrac{1}{2}\vert \;\;\;&0 & \rho_{22}(h) &  \rho_{23}(h)  & 0 \\
\langle \!-\!\tfrac{1}{2},\tfrac{1}{2}\vert\;\;\; & 0 & \rho_{32}(h) & \rho_{33}(h)& 0 \\
 \langle \!-\!\tfrac{1}{2},\!-\!\tfrac{1}{2}\vert \;\;\; & 0 & 0 & 0 & \rho_{44}(h) \\
\end{block}
\end{blockarray}\;\;.
\label{c12}
\end{align}
Here
\begin{flalign}
\rho_{11}(h)\!&=\!\frac{{\rm e}^{ \frac{\beta }{4}J_1}}{90{\cal Z}}\bigg\{
\cosh\left(\frac{\beta h}{2}\right){\rm e}^{\frac{5}{2}\beta h}\left[150{\rm e}^{-\beta(J+\frac{5}{2}J_1)} \right]
\nonumber\\
\!&+\!15\cosh\left(\frac{\beta h}{2}\right){\rm e}^{\frac{3}{2}\beta h}\left[
5{\rm e}^{-\beta(J-\frac{J_1}{2})}\!+\!8{\rm e}^{\beta(-\frac{J}{2}+J_1)} \right]
\nonumber\\
\!&+\!5\cosh\left(\frac{\beta h}{2}\right){\rm e}^{\frac{\beta}{2} h}\left[4{\rm e}^{-\beta(-2J+\frac{J_1}{2})}\!+\!5{\rm e}^{-\beta(J-\frac{J_1}{2})}\!+\!9{\rm e}^{-\beta(J+\frac{5}{2}J_1)}\!+\!7{\rm e}^{-\beta(J-\frac{5}{2}J_1)}\!+\!25{\rm e}^{-\beta(-\frac{J}{2}-J_1)} \right]
\nonumber\\
\!&+\!5\cosh\left(\frac{\beta h}{2}\right){\rm e}^{-\frac{\beta}{2} h}\left[10{\rm e}^{-\beta(-2J+\frac{J_1}{2})}\!+\!3{\rm e}^{-\beta(J+\frac{5}{2}J_1)}\!+\!{\rm e}^{-\beta(J-\frac{5}{2}J_1)}\!+\!16{\rm e}^{-\beta(-\frac{J}{2}+J_1)} \!+\!7{\rm e}^{-\beta(-\frac{J}{2}-J_1)} \right]
\nonumber\\
\!&+\!\sinh\left(\frac{\beta h}{2}\right){\rm e}^{\frac{5}{2}\beta h}\left[30{\rm e}^{-\beta(J+\frac{5}{2}J_1)} \right]
\nonumber\\
\!&+\!\sinh\left(\frac{\beta h}{2}\right){\rm e}^{\frac{3}{2}\beta h}\left[45{\rm e}^{-\beta(J-\frac{J_1}{2})} \right]
\nonumber\\
\!&+\!\sinh\left(\frac{\beta h}{2}\right){\rm e}^{\frac{\beta}{2} h}\left[5{\rm e}^{-\beta(J-\frac{J_1}{2})}\!+\!27{\rm e}^{-\beta(J+\frac{5}{2}J_1)}\!+\!33{\rm e}^{-\beta(J-\frac{5}{2}J_1)}\!+\! 75{\rm e}^{-\beta(-\frac{J}{2}-J_1)}\right]
\nonumber\\
\!&+\!\sinh\left(\frac{\beta h}{2}\right){\rm e}^{-\frac{\beta}{2} h}\left[-30{\rm e}^{-\beta(-2J+\frac{J_1}{2})}\!+\!3{\rm e}^{-\beta(J+\frac{5}{2}J_1)}\!-\!3{\rm e}^{-\beta(J-\frac{5}{2}J_1)}\!+\!20{\rm e}^{-\beta(-\frac{J}{2}+J_1)} \!+\! 15{\rm e}^{-\beta(-\frac{J}{2}-J_1)}\right]
\nonumber\\
\!&+\!10{\rm e}^{-\beta(-\frac{J}{2}-2J_1)}\left[ 
3\cosh\left(\frac{3\beta}{2}(J\!-\!J_1)\right)
\!+\!\sinh\left(\frac{3\beta}{2}(J\!-\!J_1)\right)\right]\bigg\},
\label{c13}\\
\rho_{44}(h)\!&=\!\rho_{11}(-h),
\label{c16}\\
\rho_{22}(h)\!&=\!\rho_{33}(h)\!=\!\frac{{\rm e}^{ \frac{\beta }{4}J_1}}{90{\cal Z}}\bigg\{
3\cosh\left(\frac{\beta h}{2}\right)\cosh\left(\frac{3\beta h}{2}\right)\left[20{\rm e}^{-\beta(J-\frac{J_1}{2})}\!+\!18{\rm e}^{-\beta(J+\frac{5}{2}J_1)}\!+\!55{\rm e}^{-\beta(-\frac{J}{2}+J_1)} \right]
\nonumber\\
\!&+\!\cosh^2\left(\frac{\beta h}{2}\right)\left[65{\rm e}^{-\beta(-2J+\frac{J_1}{2})}\!+\!65{\rm e}^{-\beta(J-\frac{J_1}{2})} \!+\!51{\rm e}^{-\beta(J+\frac{5}{2}J_1)}\!+\!95{\rm e}^{-\beta(J-\frac{5}{2}J_1)}\!+\!85{\rm e}^{-\beta(-\frac{J}{2}+J_1)}\right.
\nonumber\\
\!&+\!\left.110{\rm e}^{-\beta(-\frac{J}{2}-J_1)}\right]
\nonumber\\
\!&+\!3\sinh\left(\frac{\beta h}{2}\right)\sinh\left(\frac{3\beta h}{2}\right)\left[2{\rm e}^{-\beta(J+\frac{5}{2}J_1)}\!+\!25{\rm e}^{-\beta(-\frac{J}{2}+J_1)} \right]
\nonumber\\
\!&+\!\sinh^2\left(\frac{\beta h}{2}\right)\left[15{\rm e}^{-\beta(-2J+\frac{J_1}{2})}\!-\!5{\rm e}^{-\beta(J-\frac{J_1}{2})} \!-\!3{\rm e}^{-\beta(J+\frac{5}{2}J_1)}\!+\!9{\rm e}^{-\beta(J-\frac{5}{2}J_1)}\!+\!5{\rm e}^{-\beta(-\frac{J}{2}+J_1)}\right.
\nonumber\\
\!&+\!\left.30{\rm e}^{-\beta(-\frac{J}{2}-J_1)}\right]
\nonumber\\
\!&+\!10{\rm e}^{-\beta(-\frac{J}{2}-2J_1)}\left[
6\cosh\left(\frac{3\beta}{2}(J\!-\!J_1)\right)\!-\!\sinh\left(\frac{3\beta}{2}(J\!-\!J_1)\right)\right]
\bigg\},
\label{c14}\\
\rho_{23}(h)\!&=\!\rho_{32}(h)\!=\!\frac{{\rm e}^{ \frac{\beta }{4}J_1}}{90{\cal Z}}\bigg\{
10\cosh\left(\frac{\beta h}{2}\right)\cosh\left(\frac{3\beta h}{2}\right)\left[-5{\rm e}^{-\beta(J-\frac{J_1}{2})}\!+\!6{\rm e}^{-\beta(J+\frac{5}{2}J_1)} \right]
\nonumber\\
\!&+\!\cosh^2\left(\frac{\beta h}{2}\right)\left[25{\rm e}^{-\beta(J-\frac{J_1}{2})} \!+\!45{\rm e}^{-\beta(J+\frac{5}{2}J_1)}\!-\!51{\rm e}^{-\beta(J-\frac{5}{2}J_1)}\!-\!50{\rm e}^{-\beta(-\frac{J}{2}+J_1)}
\!+\!50{\rm e}^{-\beta(-\frac{J}{2}-J_1)}\right]
\nonumber\\
\!&+\!\sinh\left(\frac{\beta h}{2}\right)\sinh\left(\frac{3\beta h}{2}\right)\left[-10{\rm e}^{-\beta(J-\frac{J_1}{2})} \right]
\nonumber\\
\!&+\!\sinh^2\left(\frac{\beta h}{2}\right)\left[15{\rm e}^{-\beta(J-\frac{J_1}{2})} \!-\!9{\rm e}^{-\beta(J+\frac{5}{2}J_1)}\!-\!45{\rm e}^{-\beta(J-\frac{5}{2}J_1)}\!-\!10{\rm e}^{-\beta(-\frac{J}{2}+J_1)}
\!+\!10{\rm e}^{-\beta(-\frac{J}{2}-J_1)}\right]
\nonumber\\
\!&-\!5{\rm e}^{-\beta(-\frac{J}{2}-2J_1)}\left[
7\cosh\left(\frac{3\beta}{2}(J\!-\!J_1)\right)\!-\!5\sinh\left(\frac{3\beta}{2}(J\!-\!J_1)\right)\right]
\nonumber\\
\!&+\!10{\rm e}^{-\beta(-\frac{J}{2}-\frac{3}{2}J_1)}
\cosh\left(\frac{\beta}{2}(3J\!-\!4J_1)\right)
\bigg\}.
\label{c17}
\end{flalign}

Subsequently,  the partial transposition of the reduced density matrix  $\hat{\rho}_{\mu_{1}
|\mu_{2}}^{T_{\mu_{1}}}$,  transposed with respect the $\mu_{1}$ spin, has the following block-diagonal structure
\begin{align}
\allowdisplaybreaks
{\hat{\rho}_{\mu_{1}|\mu_{2}}^{T_{\mu_{1}}}}\!=\;
\begin{blockarray}{( cc cc)}
\rho_{11}(h) & \rho_{32}(h)  & 0 & 0 \\
 \rho_{23} & \rho_{44} & 0 & 0 \\
 0 & 0 & \rho_{33}(h) &  0  \\
  0 & 0 & 0 & \rho_{22}(h) \\
\end{blockarray}\;\;,
\label{c18}
\end{align}
which directly results to the respective eigenvalues
\begin{flalign}
\lambda_1\!&=\!\rho_{33}(h),\hspace*{1cm} \lambda_2\!=\!\rho_{22}(h),
\hspace*{1cm}
\lambda_3^{\pm}\!=\!\frac{1}{2}\left(\rho_{11}(h)\!+\!\rho_{44}(h)\!\pm\!\sqrt{(\rho_{11}(h)\!-\!\rho_{44}(h))^2\!+\!4\rho_{23}(h)\rho_{32}(h)} \right).
\label{c19}
\end{flalign}
\section{Bipartite negativity ${\cal N}_{\mu_{1}|S_{1}}$ }
\label{App D}

The reduced density operator $\hat{\rho}_{\mu_{1}|S_{1}}$ is defined after tracing out degree of freedom of two spins $\mu_{2}$ and $S_{2}$. Thus,
\begin{align}
\hat{\rho}_{\mu_{1}|S_{1}}&\!=\!\sum_{\mu_{2}^z}\sum_{S_{2}^z} \langle \mu_{2}^z,S_{2}^z\vert\hat{\rho}\vert \mu_{2}^z,S_{2}^z\rangle\!=\!\frac{1}{\cal Z}\sum_{k=1}^{36}  {\rm e}^{-\beta \varepsilon_k}\left( \sum_{\mu_2^z}\sum_{S_2^z} \langle \mu_2^z,S_2^z\vert\psi_k\rangle \langle \psi_k\vert \mu_2^z,S_2^z\rangle\right).
\label{b11}
\end{align}
The corresponding reduced density matrix in a basis of $\vert \mu_{1}^z,S^z_{1}\rangle$ reads as follows
\begin{align}
\allowdisplaybreaks
{\hat{\rho}_{\mu_{1}|S_{1}}}\!=\!
\begin{blockarray}{r cc cc cc}
 & \vert \frac{1}{2},1\rangle & \vert \frac{1}{2},0\rangle &\vert \frac{1}{2},\!-\!1\rangle  &\vert \!-\!\frac{1}{2},1\rangle&\vert \!-\!\frac{1}{2},0\rangle&\vert \!-\!\frac{1}{2},\!-\!1\rangle\\
\begin{block}{r(cc cc cc)}
\langle \frac{1}{2},1\vert \;\;\;&\rho_{11}(h) & 0 & 0 & 0 & 0 & 0 \\
\langle \frac{1}{2},0\vert \;\;\;&0 & \rho_{22}(h) & 0 & \rho_{24}(h) & 0 & 0 \\
\langle \frac{1}{2},\!-\!1\vert\;\;\; & 0 & 0 & \rho_{33}(h) &  0 &\rho_{35}(h)& 0 \\
 \langle \!-\!\frac{1}{2},1\vert\; \;\;& 0& \rho_{42}(h) &0 & \rho_{44}(h) & 0 & 0\\
\langle \!-\!\frac{1}{2},0\vert\;\;\;&0 &0 & \rho_{53}(h) &0  &  \rho_{55}(h) & 0 \\
 \langle \!-\!\frac{1}{2},\!-\!1\vert \;\;\;& 0 & 0 & 0 & 0 & 0 & \rho_{66}(h) \\
\end{block}
\end{blockarray}\;\;.
\label{b12}
\end{align}
Here
\begin{flalign}
\rho_{11}(h)\!&=\!\frac{{\rm e}^{ \frac{\beta }{4}J_1}}{20{\cal Z}}\bigg\{
\cosh\left(\frac{\beta h}{2}\right){\rm e}^{\frac{5}{2}\beta h}\left[30{\rm e}^{-\beta(J+\frac{5}{2}J_1)} \right]
\nonumber\\
\!&+\!5\cosh\left(\frac{\beta h}{2}\right){\rm e}^{\frac{3}{2}\beta h}\left[3{\rm e}^{-\beta(J-\frac{J_1}{2})}\!+\!5{\rm e}^{-\beta(-\frac{J}{2}+J_1)} \right]
\nonumber\\
\!&+\!5\cosh\left(\frac{\beta h}{2}\right){\rm e}^{\frac{\beta}{2} h}\left[2{\rm e}^{-\beta(J-\frac{J_1}{2})}\!+\!{\rm e}^{-\beta(J+\frac{5}{2}J_1)}\!+\!3{\rm e}^{-\beta(J-\frac{5}{2}J_1)}\!+\!4{\rm e}^{-\beta(-\frac{J}{2}-J_1)} \right]
\nonumber\\
\!&+\!\sinh\left(\frac{\beta h}{2}\right){\rm e}^{\frac{5}{2}\beta h}\left[10{\rm e}^{-\beta(J+\frac{5}{2}J_1)} \right]
\nonumber\\
\!&+\!5\sinh\left(\frac{\beta h}{2}\right){\rm e}^{\frac{3}{2}\beta h}\left[{\rm e}^{-\beta(J-\frac{J_1}{2})}\!+\!3{\rm e}^{-\beta(-\frac{J}{2}+J_1)}\right]
\nonumber\\
\!&+\!\sinh\left(\frac{\beta h}{2}\right){\rm e}^{\frac{\beta}{2} h}\left[3{\rm e}^{-\beta(J+\frac{5}{2}J_1)}\!-\!3{\rm e}^{-\beta(J-\frac{5}{2}J_1)}\!+\! 10{\rm e}^{-\beta(-\frac{J}{2}-J_1)}\right]
\nonumber\\
\!&-\!10{\rm e}^{-\beta(-\frac{J}{4}-\frac{9J_1}{4})}\sinh\left(\frac{\beta}{4}(3J\!-\!5J_1)\right)\bigg\},
\label{b13}\\
\rho_{66}(h)\!&=\!\rho_{11}(-h),
\label{b18}\\
\rho_{22}(h)\!&=\!\frac{{\rm e}^{ \frac{\beta }{4}J_1}}{60{\cal Z}}\bigg\{
\cosh\left(\frac{\beta h}{2}\right){\rm e}^{\frac{3}{2}\beta h}\left[30{\rm e}^{-\beta(J-\frac{J_1}{2})}\!+\!44{\rm e}^{-\beta(J+\frac{5}{2}J_1)}\!+\!65{\rm e}^{-\beta(-\frac{J}{2}+J_1)} \right]
\nonumber\\
\!&+\!\cosh\left(\frac{\beta h}{2}\right){\rm e}^{\frac{\beta}{2} h}\left[30{\rm e}^{-\beta(-2J+\frac{J_1}{2})}\!-\!20{\rm e}^{-\beta(J-\frac{J_1}{2})} \!+\!17{\rm e}^{-\beta(-\frac{J}{2}+J_1)}\!+\!30{\rm e}^{-\beta(-\frac{J}{2}-J_1)}\right]
\nonumber\\
\!&+\!\cosh\left(\frac{\beta h}{2}\right){\rm e}^{-\frac{\beta}{2} h}\left[
40{\rm e}^{-\beta(J-\frac{J_1}{2})}\!+\!26{\rm e}^{-\beta(J+\frac{5}{2}J_1)}\!+\!13{\rm e}^{-\beta(J-\frac{5}{2}J_1)}\!+\! 35{\rm e}^{-\beta(-\frac{J}{2}+J_1)}\!+\! 30{\rm e}^{-\beta(-\frac{J}{2}-J_1)}\right]
\nonumber\\
\!&+\!\sinh\left(\frac{\beta h}{2}\right){\rm e}^{\frac{3}{2}\beta h}\left[10{\rm e}^{-\beta(J-\frac{J_1}{2})}\!-\!4{\rm e}^{-\beta(J+\frac{5}{2}J_1)}\!-\!25{\rm e}^{-\beta(-\frac{J}{2}+J_1)} \right]
\nonumber\\
\!&+\!\sinh\left(\frac{\beta h}{2}\right){\rm e}^{\frac{\beta}{2} h}\left[10{\rm e}^{-\beta(-2J+\frac{J_1}{2})}\!+\! 15{\rm e}^{-\beta(-\frac{J}{2}+J_1)}\right]
\nonumber\\
\!&+\!\sinh\left(\frac{\beta h}{2}\right){\rm e}^{-\frac{\beta}{2} h}\left[10{\rm e}^{-\beta(J+\frac{5}{2}J_1)}\!-\!11{\rm e}^{-\beta(J-\frac{5}{2}J_1)} \!+\!25{\rm e}^{-\beta(-\frac{J}{2}+J_1)}\right]
\nonumber\\
\!&+\!20{\rm e}^{-\beta(-\frac{J}{2}-2J_1)}
\cosh\left(\frac{3\beta}{2}(J\!-\!J_1)\right)\bigg\},
\label{b14}\\
\rho_{55}(h)\!&=\!\rho_{22}(-h),
\label{b17}\\
\rho_{33}(h)\!&=\!\rho_{44}(-h),
\label{b15}\\
\rho_{44}(h)\!&=\!\frac{{\rm e}^{ \frac{\beta }{4}J_1}}{60{\cal Z}}\bigg\{
\cosh\left(\frac{\beta h}{2}\right){\rm e}^{\frac{3}{2}\beta h}\left[20{\rm e}^{-\beta(J-\frac{J_1}{2})}\!+\!22{\rm e}^{-\beta(J+\frac{5}{2}J_1)}\!+\!85{\rm e}^{-\beta(-\frac{J}{2}+J_1)} \right]
\nonumber\\
\!&+\!\cosh\left(\frac{\beta h}{2}\right){\rm e}^{\frac{\beta}{2} h}\left[60{\rm e}^{-\beta(-2J+\frac{J_1}{2})} \!-\!15{\rm e}^{-\beta(J-\frac{J_1}{2})}\!+\!14{\rm e}^{-\beta(J-\frac{5}{2}J_1)}\!+\!30{\rm e}^{-\beta(-\frac{J}{2}-J_1)}\right]
\nonumber\\
\!&+\!\cosh\left(\frac{\beta h}{2}\right){\rm e}^{-\frac{\beta}{2} h}\left[
20{\rm e}^{-\beta(J-\frac{J_1}{2})}\!+\!13{\rm e}^{-\beta(J+\frac{5}{2}J_1)}\!+\!{\rm e}^{-\beta(J-\frac{5}{2}J_1)}\!+\!40{\rm e}^{-\beta(-\frac{J}{2}+J_1)}\!+\! 45{\rm e}^{-\beta(-\frac{J}{2}-J_1)}\right]
\nonumber\\
\!&-\!\sinh\left(\frac{\beta h}{2}\right){\rm e}^{\frac{3}{2}\beta h}\left[2{\rm e}^{-\beta(J+\frac{5}{2}J_1)}\!+\!5{\rm e}^{-\beta(-\frac{J}{2}+J_1)} \right]
\nonumber\\
\!&+\!\sinh\left(\frac{\beta h}{2}\right){\rm e}^{\frac{\beta}{2} h}\left[20{\rm e}^{-\beta(-2J+\frac{J_1}{2})}\!-\!5{\rm e}^{-\beta(J-\frac{J_1}{2})}\!+\!2{\rm e}^{-\beta(J+\frac{5}{2}J_1)}\right]
\nonumber\\
\!&+\!\sinh\left(\frac{\beta h}{2}\right){\rm e}^{-\frac{\beta}{2} h}\left[5{\rm e}^{-\beta(J+\frac{5}{2}J_1)}\!-\!11{\rm e}^{-\beta(J-\frac{5}{2}J_1)}\!+\!20{\rm e}^{-\beta(-\frac{J}{2}+J_1)}\!-\!15{\rm e}^{-\beta(-\frac{J}{2}-J_1)}\right]
\nonumber\\
\!&+\!5{\rm e}^{-\beta(-\frac{J}{2}-2J_1)}
\left[5\cosh\left(\frac{3\beta}{2}(J\!-\!J_1)\right)\!+\!3\sinh\left(\frac{3\beta}{2}(J\!-\!J_1)\right)\right]\bigg\},
\label{b16}\\
\rho_{24}(h)\!&=\!\rho_{42}(h)\!=\!\frac{\sqrt{2}{\rm e}^{ \frac{\beta }{4}J_1}}{60{\cal Z}}\bigg\{
\cosh\left(\frac{\beta h}{2}\right){\rm e}^{\frac{3}{2}\beta h}\left[20{\rm e}^{-\beta(J-\frac{J_1}{2})}\!+\!22{\rm e}^{-\beta(J+\frac{5}{2}J_1)}\!-\! 30{\rm e}^{-\beta(-\frac{J}{2}+J_1)}\right]
\nonumber\\
\!&+\!\cosh\left(\frac{\beta h}{2}\right){\rm e}^{\frac{\beta}{2} h}\left[-15{\rm e}^{-\beta(J-\frac{J_1}{2})}\!-\!30{\rm e}^{-\beta(-2J+\frac{J_1}{2})}\!+\!14{\rm e}^{-\beta(J-\frac{5}{2}J_1)}\!+\! 20{\rm e}^{-\beta(-\frac{J}{2}+J_1)}\right]
\nonumber\\
\!&+\!\cosh\left(\frac{\beta h}{2}\right){\rm e}^{-\frac{\beta}{2} h}\left[
20{\rm e}^{-\beta(J-\frac{J_1}{2})}\!+\!13{\rm e}^{-\beta(J+\frac{5}{2}J_1)}\!+\!{\rm e}^{-\beta(J-\frac{5}{2}J_1)}\!-\! 15{\rm e}^{-\beta(-\frac{J}{2}+J_1)}\!-\! 30{\rm e}^{-\beta(-\frac{J}{2}-J_1)}\right]
\nonumber\\
\!&-\!2\sinh\left(\frac{\beta h}{2}\right){\rm e}^{\frac{3}{2}\beta h}\left[{\rm e}^{-\beta(J+\frac{5}{2}J_1)} \!+\! 5{\rm e}^{-\beta(-\frac{J}{2}+J_1)}\right]
\nonumber\\
\!&+\!\sinh\left(\frac{\beta h}{2}\right){\rm e}^{\frac{\beta}{2} h}\left[-5{\rm e}^{-\beta(J-\frac{J_1}{2})}\!-\!10{\rm e}^{-\beta(-2J+\frac{J_1}{2})} \!+\!2{\rm e}^{-\beta(J-\frac{5}{2}J_1)}\right]
\nonumber\\
\!&+\!\sinh\left(\frac{\beta h}{2}\right){\rm e}^{-\frac{\beta}{2} h}\left[5{\rm e}^{-\beta(J+\frac{5}{2}J_1)}\!-\!11{\rm e}^{-\beta(J-\frac{5}{2}J_1)}\!-\!5{\rm e}^{-\beta(-\frac{J}{2}+J_1)}\right]
\nonumber\\
\!&+\!10{\rm e}^{-\beta(-\frac{J}{2}-2J_1)}
\cosh\left(\frac{3\beta}{2}(J\!-\!J_1)\right)
\nonumber\\
\!&-\!30{\rm e}^{-\beta(-\frac{5J}{4}-\frac{3J_1}{4})}
\sinh\left(\frac{\beta}{4}(3J\!-\!J_1)\right)\bigg\},
\label{b19}\\
\rho_{35}(h)\!&=\!\rho_{53}(h)\!=\!\rho_{24}(-h).
\label{b20}
\end{flalign}
Subsequently, the partial transposition  of the reduced density matrix  $\hat{\rho}_{\mu_{1}|S_{1}}^{T_{\mu_{1}}}$, transposed with respect the $\mu_{1}$ spin has the following block-diagonal structure
\begin{align}
\allowdisplaybreaks
{\hat{\rho}_{\mu_{1}|S_{1}}^{T_{\mu_{1}}}}\!=\;
\begin{blockarray}{( cc cc cc)}
\rho_{11}(h) & \rho_{42}(h) & 0 & 0 & 0 & 0 \\
 \rho_{24}(h) & \rho_{55}(h) & 0 & 0 & 0 & 0 \\
 0 & 0 & \rho_{33}(h) &  0 & 0 & 0 \\
 0& 0 &0 & \rho_{44}(h) & 0 & 0\\
0 &0 & 0 &0  &  \rho_{22}(h) & \rho_{53}(h) \\
 0 & 0 & 0 & 0 & \rho_{35}(h) & \rho_{66}(h) \\
\end{blockarray}\;\;,
\label{b21}
\end{align}
which directly results to the respective eigenvalues
\begin{flalign}
\lambda_1\!&=\!\rho_{33}(h),
\nonumber\\
 \lambda_2\!&=\!\rho_{44}(h),
\nonumber\\
\lambda_3^{\pm}\!&=\!\frac{1}{2}\left(\rho_{11}(h)\!+\!\rho_{55}(h)\!\pm\!\sqrt{(\rho_{11}(h)\!-\!\rho_{55}(h))^2\!+\!4\rho_{24}(h)\rho_{42}(h)} \right),
\nonumber\\
\lambda_4^{\pm}\!&=\!\frac{1}{2}\left(\rho_{22}(h)\!+\!\rho_{66}(h)\!\pm\!\sqrt{(\rho_{22}(h)\!-\!\rho_{66}(h))^2\!+\!4\rho_{35}(h)\rho_{53}(h)} \right).
\label{b22}
\end{flalign}
%
%


\section{Bipartite entanglement ${\cal N}_{\mu_{1}|S_{2}}$ }
\label{App E}
The reduced density operator $\hat{\rho}_{\mu_{1}|S_{2}}$ is defined after tracing out degree of freedom of two spins $\mu_{2}$ and $S_{1}$. Thus,
\begin{align}
\hat{\rho}_{\mu_{1}|S_{2}}&\!=\!\sum_{\mu_{2}^z}\sum_{S_{1}^z} \langle \mu_{2}^z,S_{1}^z\vert\hat{\rho}\vert \mu_{2}^z,S_{1}^z\rangle\!=\!\frac{1}{\cal Z}\sum_{k=1}^{36}  {\rm e}^{-\beta \varepsilon_k}\left( \sum_{\mu_2^z}\sum_{S_1^z} \langle \mu_2^z,S_1^z\vert\psi_k\rangle \langle \psi_k\vert \mu_2^z,1_2^z\rangle\right).
\label{e11}
\end{align}
The corresponding reduced density matrix in a basis of $\vert \mu^z_{1},S^z_{2}\rangle$  reads as follows
\begin{align}
\allowdisplaybreaks
{\hat{\rho}_{\mu_{1}|S_{2}}}\!=\!
\begin{blockarray}{r cc cc cc}
 & \vert \frac{1}{2},1\rangle & \vert \frac{1}{2},0\rangle &\vert \frac{1}{2},\!-\!1\rangle  &\vert \!-\!\frac{1}{2},1\rangle&\vert \!-\!\frac{1}{2},0\rangle&\vert \!-\!\frac{1}{2},\!-\!1\rangle\\
\begin{block}{r(cc cc cc)}
\langle \frac{1}{2},1\vert \;\;\;&\rho_{11}(h) & 0 & 0 & 0 & 0 & 0 \\
\langle \frac{1}{2},0\vert \;\;\;&0 & \rho_{22}(h) & 0 & \rho_{24}(h) & 0 & 0 \\
\langle \frac{1}{2},\!-\!1\vert\;\;\; & 0 & 0 & \rho_{33}(h) &  0 &\rho_{35}(h)& 0 \\
 \langle \!-\!\frac{1}{2},1\vert\; \;\;& 0& \rho_{42}(h) &0 & \rho_{44}(h) & 0 & 0\\
\langle \!-\!\frac{1}{2},0\vert\;\;\;&0 &0 & \rho_{53}(h) &0  &  \rho_{55}(h) & 0 \\
 \langle \!-\!\frac{1}{2},\!-\!1\vert \;\;\;& 0 & 0 & 0 & 0 & 0 & \rho_{66}(h) \\
\end{block}
\end{blockarray}\;\;.
\label{e12}
\end{align}
Here
\begin{flalign}
\rho_{11}(h)\!&=\!\frac{{\rm e}^{ \frac{\beta }{4}J_1}}{180{\cal Z}}\bigg\{
\cosh\left(\frac{\beta h}{2}\right){\rm e}^{\frac{5}{2}\beta h}\left[270{\rm e}^{-\beta(J+\frac{5}{2}J_1)} \right]
\nonumber\\
\!&+\!5\cosh\left(\frac{\beta h}{2}\right){\rm e}^{\frac{3}{2}\beta h}\left[23{\rm e}^{-\beta(J-\frac{J_1}{2})}\!+\!49{\rm e}^{-\beta(-\frac{J}{2}+J_1)} \right]
\nonumber\\
\!&+\!5\cosh\left(\frac{\beta h}{2}\right){\rm e}^{\frac{\beta}{2} h}\left[16{\rm e}^{-\beta(-2J+\frac{J_1}{2})}\!+\!2{\rm e}^{-\beta(J-\frac{J_1}{2})}\!+\!9{\rm e}^{-\beta(J+\frac{5}{2}J_1)}\!+\!7{\rm e}^{-\beta(J-\frac{5}{2}J_1)}\!+\! 16{\rm e}^{-\beta(-\frac{J}{2}+J_1)}\right.
\nonumber\\
\!&+\!\left. 31{\rm e}^{-\beta(-\frac{J}{2}-J_1)} \right]
\nonumber\\
\!&+\!\sinh\left(\frac{\beta h}{2}\right){\rm e}^{\frac{5}{2}\beta h}\left[90{\rm e}^{-\beta(J+\frac{5}{2}J_1)} \right]
\nonumber\\
\!&+\!5\sinh\left(\frac{\beta h}{2}\right){\rm e}^{\frac{3}{2}\beta h}\left[23{\rm e}^{-\beta(-\frac{J}{2}+J_1)}\!+\!13{\rm e}^{-\beta(J-\frac{J_1}{2})} \right]
\nonumber\\
\!&+\!3\sinh\left(\frac{\beta h}{2}\right){\rm e}^{\frac{\beta}{2} h}\left[9{\rm e}^{-\beta(J+\frac{5}{2}J_1)}\!+\!11{\rm e}^{-\beta(J-\frac{5}{2}J_1)}\!+\! 25{\rm e}^{-\beta(-\frac{J}{2}-J_1)}\right]
\nonumber\\
\!&+\!5{\rm e}^{-\beta(-\frac{J}{2}-2J_1)}\left[ 
9\cosh\left(\frac{3\beta}{2}(J\!-\!J_1)\right)\right.
\!+\!\left.7\sinh\left(\frac{3\beta}{2}(J\!-\!J_1)\right)\right]\bigg\},
\label{e13}\\
\rho_{66}(h)\!&=\!\rho_{11}(-h),
\label{e18}\\
\rho_{22}(h)\!&=\!\frac{{\rm e}^{ \frac{\beta }{4}J_1}}{180{\cal Z}}\bigg\{
3\cosh\left(\frac{\beta h}{2}\right){\rm e}^{\frac{3}{2}\beta h}\left[30{\rm e}^{-\beta(J-\frac{J_1}{2})}\!+\!44{\rm e}^{-\beta(J+\frac{5}{2}J_1)}\!+\!45{\rm e}^{-\beta(-\frac{J}{2}+J_1)} \right]
\nonumber\\
\!&+\!2\cosh\left(\frac{\beta h}{2}\right){\rm e}^{\frac{\beta}{2} h}\left[15{\rm e}^{-\beta(-2J+\frac{J_1}{2})}\!+\!30{\rm e}^{-\beta(J-\frac{J_1}{2})} \!+\!36{\rm e}^{-\beta(J-\frac{5}{2}J_1)}\!+\!65{\rm e}^{-\beta(-\frac{J}{2}-J_1)}\right]
\nonumber\\
\!&+\!\cosh\left(\frac{\beta h}{2}\right){\rm e}^{-\frac{\beta}{2} h}\left[
60{\rm e}^{-\beta(-2J+\frac{J_1}{2})}\!+\!78{\rm e}^{-\beta(J+\frac{5}{2}J_1)}\!+\!18{\rm e}^{-\beta(J-\frac{5}{2}J_1)}\!+\! 165{\rm e}^{-\beta(-\frac{J}{2}+J_1)}\!+\! 50{\rm e}^{-\beta(-\frac{J}{2}-J_1)}\right]
\nonumber\\
\!&+\!3\sinh\left(\frac{\beta h}{2}\right){\rm e}^{\frac{3}{2}\beta h}\left[10{\rm e}^{-\beta(J-\frac{J_1}{2})}\!-\!4{\rm e}^{-\beta(J+\frac{5}{2}J_1)}\!-\!5{\rm e}^{-\beta(-\frac{J}{2}+J_1)} \right]
\nonumber\\
\!&+\!2\sinh\left(\frac{\beta h}{2}\right){\rm e}^{\frac{\beta}{2} h}\left[5{\rm e}^{-\beta(-2J+\frac{J_1}{2})}\!+\!32{\rm e}^{-\beta(J-\frac{5}{2}J_1)}\right]
\nonumber\\
\!&+\!\sinh\left(\frac{\beta h}{2}\right){\rm e}^{-\frac{\beta}{2} h}\left[-20{\rm e}^{-\beta(-2J+\frac{J_1}{2})}\!+\!30{\rm e}^{-\beta(J+\frac{5}{2}J_1)}\!-\!14{\rm e}^{-\beta(J-\frac{5}{2}J_1)} \!+\!15{\rm e}^{-\beta(-\frac{J}{2}+J_1)}\right]
\nonumber\\
\!&+\!60{\rm e}^{-\beta(-\frac{J}{2}-2J_1)}
\cosh\left(\frac{3\beta}{2}(J\!-\!J_1)\right)\bigg\},
\label{e14}\\
\rho_{55}(h)\!&=\!\rho_{22}(-h),
\label{e17}\\
\rho_{33}(h)\!&=\!\rho_{44}(-h),
\label{e15}\\
\rho_{44}(h)\!&=\!\frac{{\rm e}^{ \frac{\beta }{4}J_1}}{180{\cal Z}}\bigg\{
3\cosh\left(\frac{\beta h}{2}\right){\rm e}^{\frac{3}{2}\beta h}\left[30{\rm e}^{-\beta(J-\frac{J_1}{2})}\!+\!22{\rm e}^{-\beta(J+\frac{5}{2}J_1)}\!+\!65{\rm e}^{-\beta(-\frac{J}{2}+J_1)} \right]
\nonumber\\
\!&+\!\cosh\left(\frac{\beta h}{2}\right){\rm e}^{\frac{\beta}{2} h}\left[100{\rm e}^{-\beta(-2J+\frac{J_1}{2})} \!+\!109{\rm e}^{-\beta(J-\frac{5}{2}J_1)}\!+\!86{\rm e}^{-\beta(-\frac{J}{2}-J_1)}\right]
\nonumber\\
\!&+\!\cosh\left(\frac{\beta h}{2}\right){\rm e}^{-\frac{\beta}{2} h}\left[
85{\rm e}^{-\beta(J-\frac{J_1}{2})}\!+\!39{\rm e}^{-\beta(J+\frac{5}{2}J_1)}\!+\!36{\rm e}^{-\beta(J-\frac{5}{2}J_1)}\!+\! 14{\rm e}^{-\beta(-\frac{J}{2}+J_1)}\!+\! 111{\rm e}^{-\beta(-\frac{J}{2}-J_1)}\right]
\nonumber\\
\!&-\!3\sinh\left(\frac{\beta h}{2}\right){\rm e}^{\frac{3}{2}\beta h}\left[10{\rm e}^{-\beta(J-\frac{J_1}{2})}\!+\!2{\rm e}^{-\beta(J+\frac{5}{2}J_1)}\!-\!15{\rm e}^{-\beta(-\frac{J}{2}+J_1)} \right]
\nonumber\\
\!&-\!\sinh\left(\frac{\beta h}{2}\right){\rm e}^{\frac{\beta}{2} h}\left[-60{\rm e}^{-\beta(-2J+\frac{J_1}{2})}\!+\!21{\rm e}^{-\beta(J-\frac{5}{2}J_1)}\!-\!44{\rm e}^{-\beta(-\frac{J}{2}-J_1)} \right]
\nonumber\\
\!&-\!\sinh\left(\frac{\beta h}{2}\right){\rm e}^{-\frac{\beta}{2} h}\left[-25{\rm e}^{-\beta(J-\frac{J_1}{2})}\!-\!15{\rm e}^{-\beta(J+\frac{5}{2}J_1)}\!-\!12{\rm e}^{-\beta(-\frac{J}{2}+J_1)}\!+\!69{\rm e}^{-\beta(-\frac{J}{2}-J_1)}\right]
\nonumber\\
\!&+\!5{\rm e}^{-\beta(-\frac{J}{2}-2J_1)}
\left[15\cosh\left(\frac{3\beta}{2}(J\!-\!J_1)\right)\!-\!7\sinh\left(\frac{3\beta}{2}(J\!-\!J_1)\right)\right]\bigg\},
\label{e16}\\
\rho_{24}(h)\!&=\!\rho_{42}(h)\!=\!\frac{\sqrt{2}{\rm e}^{ \frac{\beta }{4}J_1}}{180{\cal Z}}\bigg\{
\cosh\left(\frac{\beta h}{2}\right){\rm e}^{\frac{3}{2}\beta h}\left[66{\rm e}^{-\beta(J+\frac{5}{2}J_1)}\right]
\nonumber\\
\!&-\!\cosh\left(\beta h\right){\rm e}^{\beta h}\left[25{\rm e}^{-\beta(J-\frac{J_1}{2})}\right]
\nonumber\\
\!&-\!\cosh\left(\frac{\beta h}{2}\right){\rm e}^{\frac{\beta}{2} h}\left[39{\rm e}^{-\beta(J-\frac{5}{2}J_1)}\right]
\nonumber\\
\!&+\!\cosh\left(\frac{\beta h}{2}\right){\rm e}^{-\frac{\beta}{2} h}\left[
39{\rm e}^{-\beta(J+\frac{5}{2}J_1)}\!-\!16{\rm e}^{-\beta(J-\frac{5}{2}J_1)}\!+\! 25{\rm e}^{-\beta(-\frac{J}{2}+J_1)}\!-\! 25{\rm e}^{-\beta(-\frac{J}{2}-J_1)}\right]
\nonumber\\
\!&-\!\sinh\left(\frac{\beta h}{2}\right){\rm e}^{\frac{3}{2}\beta h}\left[6{\rm e}^{-\beta(J+\frac{5}{2}J_1)} \right]
\nonumber\\
\!&-\!\sinh\left(\beta h\right){\rm e}^{\beta h}\left[35{\rm e}^{-\beta(J-\frac{J_1}{2})} \right]
\nonumber\\
\!&-\!\sinh\left(\frac{\beta h}{2}\right){\rm e}^{\frac{\beta}{2} h}\left[33{\rm e}^{-\beta(J-\frac{5}{2}J_1)} \right]
\nonumber\\
\!&+\!\sinh\left(\frac{\beta h}{2}\right){\rm e}^{-\frac{\beta}{2} h}\left[15{\rm e}^{-\beta(J+\frac{5}{2}J_1)}\!+\!8{\rm e}^{-\beta(J-\frac{5}{2}J_1)}\!-\!5{\rm e}^{-\beta(-\frac{J}{2}+J_1)}\!+\!5{\rm e}^{-\beta(-\frac{J}{2}-J_1)}\right]
\nonumber\\
\!&+\!5{\rm e}^{-\beta(-\frac{J}{2}-2J_1)}
\left[3\cosh\left(\frac{3\beta}{2}(J\!-\!J_1)\right)\!+\!\sinh\left(\frac{3\beta}{2}(J\!-\!J_1)\right)\right]
\nonumber\\
\!&-\!20{\rm e}^{-\beta(-\frac{J}{2}-\frac{3}{2}J_1)}
\left[2\cosh\left(\frac{\beta}{2}(3J\!-\!4J_1)\right)\!-\!\sinh\left(\frac{\beta}{2}(3J\!-\!4J_1)\right)\right]\bigg\},
\label{e19}\\
\rho_{35}(h)\!&=\rho_{53}(h)\!=\!\!\rho_{24}(-h).
\label{e20}
\end{flalign}
Subsequently, the partial transposition of the reduced density matrix  $\hat{\rho}_{\mu_{1}|S_{2}}^{T_{\mu_{1}}}$,  transposed with respect the $\mu_{1}$ spin  has the following block-diagonal structure
\begin{align}
\allowdisplaybreaks
{\hat{\rho}_{\mu_{1}|S_{2}}^{T_{\mu_{1}}}}\!=\;
\begin{blockarray}{( cc cc cc)}
\rho_{11}(h) & \rho_{42}(h) & 0 & 0 & 0 & 0 \\
 \rho_{24}(h) & \rho_{55}(h) & 0 & 0 & 0 & 0 \\
 0 & 0 & \rho_{33}(h) &  0 & 0 & 0 \\
 0& 0 &0 & \rho_{44}(h) & 0 & 0\\
0 &0 & 0 &0  &  \rho_{22}(h) & \rho_{53}(h) \\
 0 & 0 & 0 & 0 & \rho_{35}(h) & \rho_{66}(h) \\
\end{blockarray}\;\;,
\label{e21}
\end{align}
which directly results to the respective eigenvalues
\begin{flalign}
\lambda_1\!&=\!\rho_{33}(h),
\nonumber\\
 \lambda_2\!&=\!\rho_{44}(h),
\nonumber\\
\lambda_3^{\pm}\!&=\!\frac{1}{2}\left(\rho_{11}(h)\!+\!\rho_{55}(h)\!\pm\!\sqrt{(\rho_{11}(h)\!-\!\rho_{55}(h))^2\!+\!4\rho_{24}(h)\rho_{42}(h)} \right),
\nonumber\\
\lambda_4^{\pm}\!&=\!\frac{1}{2}\left(\rho_{22}(h)\!+\!\rho_{66}(h)\!\pm\!\sqrt{(\rho_{22}(h)\!-\!\rho_{66}(h))^2\!+\!4\rho_{35}(h)\rho_{53}(h)} \right).
\label{e22}
\end{flalign}
%



\begin{thebibliography}{60}
\bibitem{Amico} L. Amico, R. Fazio, A. Osterloh, and V. Vedral, Rev. Mod. Phys. {\bf 80} (2008) 517. 
\bibitem{Bell} J.S. Bell, Physics {\bf 1} (1964) 195.
\bibitem{Benneta} Ch.H. Bennett, D.P. DiVincenzo, J. A. Smolin, and W.K. Wootters, Phys. Rev. A {\bf 54} (1996) 3824.
\bibitem{Nielsen} M.A. Nielsen and I. Chuang, 2000, {\it Quantum Computation
and Quantum Communication}, Cambridge University Press, Cambridge, England.
\bibitem{Furusawa} A. Furusawa, J.L. Sørensen, S.L. Braunstein, C.A. Fuchs, H.J. Kimble, and E. S. Polzik, Science {\bf 282} (1998) 706.
\bibitem{Loss} D. Loss and D.P. DiVincenzo, Phys. Rev. A {\bf 57} (1998) 120.
\bibitem{Hayashi} T. Hayashi, T. Fujisawa, H.D. Cheong, Y.H. Jeong, and Y. Hirayama, Phys. Rev. Lett. {\bf 91} (2003) 226804. 
\bibitem{Deutsch} D. Deutsch, A. Ekert, R. Jozsa, Ch. Macchiavello, S. Popescu, and A. Sanpera, Phys Rev. Lett. {\bf 77} (1996) 2818.
\bibitem{Bechmann} H. Bechmann-Pasquinucci and A. Peres,  Phys. Rev. Lett. {\bf 85}, (2000) 3313.
\bibitem{Bennet2000} C.H. Bennett and D.P. Divincenzo, Nature {\bf 404} (2000) 247.
\bibitem{Horodecki2009}  R. Horodecki, P. Horodecki, M. Horodecki, and K. Horodecki, Rev. Mod. Phys.
{\bf 81} (2009) 865.
\bibitem{Leuenberger} M.N. Leuenberger and D. Loss, Nature {\bf 410} (2001) 789.
\bibitem{Gaita} A. Gaita-Arino, F. Luis, S. Hill, and E. Coronado, Nat. Chem. {\bf 11} (2019) 301.
\bibitem{Wootters} W. Wootters, Phys. Rev. Lett. {\bf 80} (1998) 2245.
\bibitem{Vidal} G. Vidal and R.~F. Werner, Phys. Rev. A {\bf 65} (2002) 032314.
\bibitem{Bennett} Ch.H. Bennett, H.J. Bernstein, S. Popescu, and B. Schumacher, Phys. Rev. A {\bf 53} (1996) 2046.
\bibitem{Wang2001} X. Wang, H. Fu, and A.I. Solomon,  J. Phys. A: Math. Gen. {\bf 34} (2001) 11307. 
\bibitem{Bose} I. Bose and A. Tribedi, Phys. Rev. A {\bf 72} (2005) 022314.
\bibitem{Tribedi} A. Tribedi and I. Bose, Phys. Rev. A {\bf 74} (2006) 012314.
\bibitem{Pal2009} A.K. Pal and I. Bose, J. Phys.: Condens. Matter {\bf 22} (2009) 016004.
\bibitem{Pal2011} A.K. Pal and I. Bose, J. Phys. B: At. Mol. Opt. Phys. {\bf 44} (2011) 045101.
\bibitem{Cima} O.M.D. Cima, D.H.T. Franco, and S.L.L. da Silva, Quantum Stud.: Math. Found. {\bf 3} (2016) 57.
\bibitem{Liu2015} Ch.-Ch. Liu, S. Xu, J. He, and L. Ye, J. Mod. Phys. B {\bf 29} (2015) 1550005.
\bibitem{Deniz} H. S. Deniz and A. Ekrem, Chin. Phys. B {\bf 23} (2014) 050305.
\bibitem{Ahami} N. Ahami and M. E. Baz, Int. J. Quantum Inf. {\bf 19} (2021) 2150021.
\bibitem{Zhang2021} P.-P. Zhang, J. Wang, Y.-L. Xu, Ch.-Y. Wang, and X.-M. Kong, Physica A {\bf 566} (2021) 125643.
\bibitem{Zheng} Y.-D. Zheng and B. Zhou, Physica A {\bf 603} (2022) 127753.
\bibitem{Najarbashi} G. Najarbashi, L. Balazadeh, and A. Tavana,  Int. J. Theor. Phys. {\bf 57} (2018) 95.
\bibitem{Zad2017} A. Zad, J. Kor. Phys. Soc. {\bf 70} (2017) 835.
\bibitem{Zad2016} A. Zad, Chin. Phys. Lett. {\bf 33} (2016) 090302.
\bibitem{Sun} Z. Sun, X.G. Wang, A.Z. Hu, and Y.Q. Li, Physica A {\bf 370} (2006) 483.
\bibitem{Shawish} S. El Shawish, A. Ram\v{s}ak, and J. Bon\v{c}a, Phys. Rev. B {\bf 75} (2007) 205442.
\bibitem{Ma} X.-S. Ma, B. Dakic, W. Naylor, A. Zeilinger, and P. Walther, Nat. Phys. {\bf 7} (2011) 399.
\bibitem{Ananikian} N. S. Ananikian, L.N. Ananikyan, L. A. Chakhmakhchyan, and O. Rojas, J. Phys.: Cond. Matter {\bf 24} (2012) 256001. 
\bibitem{Rojas} O. Rojas, M. Rojas, N. S. Ananikian, and S.M. de Souza, Phys. Rev. A {\bf 86} (2012) 0423330.
\bibitem{Torrico} J. Torrico, M. Rojas, S. M. de Souza, O. Rojas, and N. S. Ananikian, EPL {\bf 108} (2014) 50007.
\bibitem{Irons} H.R. Irons, J. Quintanilla, T.G. Perring, L. Amico, and G. Aeppli,  Phys. Rev. B {\bf 96} (2017) 224408
\bibitem{Karlova} K. Karlova and J. Stre\v{c}ka, Acta Phys. Pol. A {\bf 137} (2020) 5.
\bibitem{Szalowski} K. Szalowski, J. Magn. Magn. Mater. {\bf 546} (2022) 168782.
\bibitem{Benabdallah} F. Benabdallah, S. Haddadi, H.A. Zad, M. R. Pourkarimi, M. Daoud, and N. Ananikian, Sci. Rep. {\bf 12} (2022) 6406.
\bibitem{Zhad2022} A. Zhad, A. Zoshki, N. Ananinikian, and M. Ja\v{s}\v{c}ur, J. Magn. Magn. Mater. {\bf 559} (2022) 169533.
\bibitem{Zheng2017} Y.-D. Zheng, Z. Mao, and B. Zhou, Chin. Phys. B {\bf 26} (2017) 070302.
\bibitem{Ghannadan2021} A. Ghannadan and J. Stre\v{c}ka,  Molecules {\bf 26} (2021) 3420.
\bibitem{Li} S.B. Li, Z.X. Xu, J.-H. Dai, and J.-B. Xu, Phys. Rev. B {\bf 73} (2006) 184411.
\bibitem{Abgaryan2015} V.S. Abgaryan, N.S. Ananikian, L.N. Ananikyan, and V. Hovhannisyan, Sol. Stat. Commun. {\bf 203} (2015) 5.
\bibitem{Hu} M.L. Hu, Mod. Phys. Lett. B {\bf 31} (2008) 3067. 
\bibitem{Xu2016} Y.-L. Xu, X.-M. Kong, Z.-Q. Liu, and Ch.-Y. Wang, Physica A {\bf 446} (2016) 217.
\bibitem{Deb} M. Deb and A. K. Ghosh, Eur. Phys. J. D {\bf 71} (2017) 173. 
\bibitem{Zhad2018} A. Zhad and N. Ananikian, Sol. Stat. Commun. {\bf 276} (2018) 24.
\bibitem{Sadiek2} G. Sadiek and M. AlQasimi, Entropy {\bf 23} (2021) 1066.
\bibitem{Sadiek} G. Sadiek and M. AlQasimi, Results in Physics {\bf 35} (2022) 105403.
\bibitem{Wang2002} X. Wang,  Phys. Rev. A {\bf 66} (2002) 044305.

\bibitem{Zhang2011}  G.-F. Zhang, Y.-C. Hou, and  A.-L. Ji, Sol. Stat. Commun. {\bf 151} (2011) 790. 
\bibitem{Cenci2020} H. \v{C}en\v{c}arikov\'a and J. Stre\v{c}ka, Phys. Rev. B {\bf 102} (2020) 184419.
\bibitem{Vargova2022} H. Vargov\'a, J. Stre\v{c}ka, and N. Toma\v{s}ovi\v{c}ov\'a, J. Magn. Magn. Mater. {\bf 546} (2022) 168799.
\bibitem{Vargova2022a} H. Vargov\'a and J. Stre\v{c}ka, Nanomatrials {\bf 11} (2022) 3096.

\bibitem{Guo} J.-L. Guo, X.-L. Huang, and H.-S. Song, Phys. Scr. {\bf 76} (2007) 327.
\bibitem{Yang} G.-H. Yang and L. Zhou, Phys. Scr. {\bf 78} (2008) 025703.
\bibitem{Wang2009} F. Wang, H. Jia, H. Zhang, X. Zhang, and S. Chang, Sci. China Ser. G {\bf 52} (2009) 1919.
\bibitem{Zhu} G.-Q. Zhu, Cent. Eur. J. Phys. {\bf 7} (2009) 135.
\bibitem{Xu} S. Xu, X. Song, and L. Ye, Quantum Inf. Process. {\bf 13} (2014) 1013.

\bibitem{Ribas} J. Ribas, C. Diaz, R. Costa, J. Tercero, X. Solans, M. Font-Bard\'{i}a, and H. Stoeckli-Evans, Inorg. Chem. {\bf 37} (1998) 233.
\bibitem{Yonemura} M. Yonemura, H. Okawa, M. Ohba, D. E. Fentonb, and L. K. Thompson, Chem. Commun. (2000) 817.
\bibitem{Park} K. Park and S. M. Holmes, Phys. Rev. B  {\bf 74} (2006) 224440.
\bibitem{Zhuang2015} P.-F. Zhuang, Y.-J. Zhang, H. Zheng, C.-Q. Jiao, L. Zhao, J.-L. Wang, C. He,  Ch.-Y. Duan, and T. Liu,  Dalton Trans. {\bf 44} (2015) 3393.

\bibitem{Matlab} MATLAB, 2022. version 9.13.0.2166757 (R2022b), Natick, Massachusetts: The MathWorks Inc.
\bibitem{Peres} A. Peres, Phys. Rev. Lett. {\bf 77} (1996) 1413.
\bibitem{Horodecki} M. Horodecki, P. Horodecki, and R. Horodecki, Phys. Lett. A {\bf 223} (1996) 1.
\bibitem{Hu2008} M.L. Hu and D.P. Tian, Chinese Phys. C {\bf 32} (2008) 303.
\bibitem{Rektorys}  K. Rektorys, {\it Survey of Applicable Mathematics} (The M. I. T. Press, Cambridge, MA, 1969).
 \end{thebibliography}
\end{document}